\documentclass[letterpaper]{JHEP3}
\usepackage{epsfig,amsmath,xspace}

\newcommand{\K}{K\"ahler\xspace}
\newcommand{\bi}{{\bar\imath}}
\newcommand{\bj}{{\bar\jmath}}
\newcommand{\bz}{{\bar z}}
\newcommand{\Rc}{\mathcal{R}}
\newcommand{\Z}{\mathbf{Z}}
\newcommand{\R}{\mathbf{R}}
\newcommand{\C}{\mathbf{C}}
\DeclareMathOperator{\I}{Im}
\DeclareMathOperator{\Rl}{Re}

\title{Numerical Ricci-flat metrics on K3}

\author{Matthew Headrick \\
Center for Theoretical Physics, Massachusetts Institute of Technology \\
77 Massachusetts Ave., Cambridge MA 02139, USA \\
E-mail: \email{headrick@mit.edu}}

\author{Toby Wiseman \\
Jefferson Physical Laboratory, Harvard University \\
Cambridge MA 02138, USA \\
E-mail: \email{twiseman@fas.harvard.edu}}

\abstract{ We develop numerical algorithms for solving the Einstein
equation on Calabi-Yau manifolds at arbitrary values of their complex
structure and \K parameters. We show that \K geometry can be exploited for significant gains in computational efficiency. As a proof of
principle, we apply our methods to a one-parameter family of K3
surfaces constructed as blow-ups of the $T^4/\Z_2$ orbifold with many
discrete symmetries. High-resolution metrics may be obtained on a
time scale of days using a desktop computer.  We compute various
geometric and spectral quantities from our numerical metrics.  Using
similar resources we expect our methods to practically extend to
Calabi-Yau three-folds with a high degree of discrete symmetry,
although we expect the general three-fold to remain a challenge due to
memory requirements.  }

\preprint{\hepth{0506129}\\HUTP-05/A0028\\MIT-CTP 3647}

\begin{document}

\section{Introduction}\label{intro}

In 1977, S.-T. Yau \cite{Yau:1977ms} proved E. Calabi's conjecture
\cite{Calabi} that a compact \K manifold with vanishing first Chern
class admits a Ricci-flat metric in each \K class. Interest in
Calabi-Yau manifolds was subsequently generated among physicists by
the discovery that they can serve as supersymmetry-preserving
compactification manifolds in string theory \cite{Candelas:1985en}. In
the three decades since the proof of Yau's theorem, many examples of
Calabi-Yau manifolds have been constructed and studied, principally
using methods of algebraic geometry, and much has been learned about
their mathematical properties and physical applications.  Nonetheless,
a major gap in our knowledge about Calabi-Yau manifolds has persisted
during this time, namely the Ricci-flat metrics themselves. Yau's
proof is not constructive, and no example of a smooth Ricci-flat
metric is known explicitly for any Calabi-Yau manifold. Indeed,
perhaps we should not expect there to exist such metrics in closed
form \cite{Morrison}.

The question thus arises as to whether it is possible to solve the
Einstein equation numerically on a Calabi-Yau manifold. The purpose of
this paper it to show that it is. We describe a method for doing so,
and display the metrics obtained by applying that method to the
smallest-dimensional Calabi-Yau, the K3 surface.

The algorithms we developed rely in an essential way on the underlying
complex and \K geometry of Calabi-Yau manifolds. In fact, one of the
main points of this paper is to show that those properties are as
powerful for numerical work as they have already proven to be for
analytical calculations and proving theorems. To explain this, let us
consider the challenges faced by someone attempting to numerically
solve the Euclidean Einstein equation on a general real
four-manifold. In some sense this is a problem in numerical
relativity, and one can get a sense of its scale by considering that
four-dimensional problems in numerical relativity, such as black hole
collisions, can be solved (if at all) only with the investment of
extremely large computing resources, typically supercomputers. (Note
that Calabi-Yau manifolds admit no continuous isometries that could
reduce the effective dimensionality of the problem.) At a more
fundamental level, whereas conventional numerical relativity deals
with solving the Einstein equation as an initial-value problem on a
Lorentzian spacetime, here we wish to solve it on a Euclidean
manifold, a problem for which general algorithms are lacking. The
challenges for creating an algorithm include the usual issues of gauge
fixing and coordinate singularities, as well as finding a way to fix
any moduli the solutions might have. An even more difficult challenge
would be to avoid (or deal with) the curvature singularities that
generically form under relaxation schemes such as Ricci flow.

Let us now see how the framework of complex and \K differential
geometry allows one to naturally solve, or greatly ameliorate, each of
these problems in turn. Firstly, the \K formulation of geometry can be
employed to vastly reduce the scale of the problem compared to the
language of real differential geometry. The metric can be encoded in a
single scalar function, the \K potential (as reviewed in subsection
\ref{Kgeom}). In terms of the \K potential the Einstein equation takes
the relatively simple form of a Monge-Amp\`ere equation (as reviewed
in subsection \ref{MAeqn}). So we have a large simplication of the
many degrees of freedom required to locally parameterize a general
metric, and of the complicated set of differential equations in the
usual form of the Einstein equation.

Secondly, complex coordinates offer a naturally adapted gauge choice,
since with respect to any \K metric they satisfy the harmonic gauge
condition, a well-known gauge choice for numerical
relativity. Furthermore the gauge fixing is nearly complete: there are
only a finite number of (continuous) residual gauge transformations,
since compact complex manifolds admit only a finite number of
holomorphic vector fields. In fact, for Calabi-Yaus this number is
zero, so the gauge fixing is complete. Since there are no gauge
transformations, no coordinate singularities can appear.

Thirdly, the moduli of Calabi-Yau manifolds, which are divided into
complex structure and \K moduli, may be fixed at any desired
values---before solving the Einstein equation---in the following
way. With the manifold defined topologically by an atlas of patches,
the complex structure is fixed by fixing the holomorphic coordinate
transition functions on the patch overlaps, while the \K moduli are
fixed by fixing the \K transformations on them (which then serve as
the boundary conditions for the \K potential). (Details of the above
are given in subsection \ref{Kgeom}.)

Finally, there is a kind of stability that appears to be inherent in
\K geometry, miraculously eliminating the problem of spontaneous
formation of curvature singularities. Ricci flow, for example,
contrary to its behaviour on real manifolds, is extremely robust on \K
manifolds, as shown by Cao's long-time existence theorems
\cite{Cao}. On Calabi-Yau manifolds in particular, starting from
\emph{any} \K metric it converges to the Ricci-flat metric in the same
class. In principle therefore it provides a general algorithm for
solving the Einstein equation on Calabi-Yaus. However, since it is a
rather inefficient method from a computational viewpoint (rather like
solving the Laplace equation by simulating diffusion), we developed
and used instead a Gauss-Seidel-type relaxation algorithm for the
Monge-Amp\`ere equation. While we lack a convergence theorem for our
algorithm, we found that in practice it was just as robust as Ricci
flow, presumably for the same underlying reasons. (In fact, in one
sense our algorithm was even more robust than Ricci flow, since
remarkably it converged even when the initial \K potential didn't
define a [positive-definite] metric.) Subsection \ref{methods}
contains a discussion of Ricci flow and a description of our
algorithm.

As a proof of principle for our methods, we applied them to a class of
K3's known as Kummer surfaces, which are blow-ups of the orbifold
$T^4/\Z_2$. For simplicity, we considered the most symmetrical Kummer
surfaces, namely those for which the torus is cubical and all 16 fixed
points are blown up identically. This symmetry leaves only one modulus
(not counting the trivial volume modulus), namely the ratio of the
size of the blow-ups to the size of the $T^4$. In fact there is a
limit to how large this ratio may be, since at some point certain
holomorphic curves shrink to zero size, signalling the appearance of
new orbifold singularities---on this wall of the \K cone the manifold
is an orbifold of another, smooth K3. (Details of the construction are
given in subsection \ref{construction}.) Using our algorithm, we
computed the Ricci-flat metric at various points over the full range
of this modulus. Each point required a few days on a garden-variety
desktop computer for the highest resolutions. Subsection \ref{results}
is devoted to an exploration of the resulting geometries as a function
of the modulus. Various curvature invariants are plotted, as well as a
low-lying eigenvalue of the scalar Laplacian.

The success we had with these highly symmetrical K3 surfaces leads us
to consider possible generalizations. Using our methods, could we
solve the Einstein equation on generic K3's, which lack such discrete
symmetries, or on Calabi-Yau three-folds? What about on other \K
manifolds such as del Pezzo surfaces, or on Calabi-Yaus with matter
such as fluxes and branes? The estimates we make in subsection
\ref{generalizations} suggest that the answer to these questions is
yes, but that due to memory limitations in the three-fold case we
would need to be helped by a high degree of discrete symmetry.

In subsection \ref{euclidean} we return to the problem of solving the
Einstein equation in the real Euclidean context, and explore what this
work has taught us that might generalize to non-\K geometries. We
conclude in subsection \ref{applications} by mentioning some possible
mathematical and physical applications of these numerical metrics.

The C code for our simulations, as well as animated versions of the
plots shown in this paper, are available at the website
\href{http://schwinger.harvard.edu/~wiseman/K3/}{\tt
http://schwinger.harvard.edu/~wiseman/K3/}.

\section{General method}\label{generalmethod}

The purpose of this section is to discuss in a general way the problem
of solving the Einstein equation numerically on a Calabi-Yau manifold
at a given point in its moduli space. We first briefly review the
essentials of \K geometry,\footnote{A detailed review of \K geometry
and Calabi-Yau manifolds may be found in the excellent set of lecture
notes \cite{Candelas:1987is}.} describing how the geometry is encoded
in the \K potential and how the moduli are fixed. We then explain how,
in \K geometry, the Einstein equation reduces to a Monge-Amp\`ere
equation. Finally, we discuss in general terms the numerical
algorithms we applied to solving that equation.

\subsection{\K geometry}\label{Kgeom}

We work on a manifold with a fixed complex structure, that is, on each
coordinate patch $U_\alpha$ we have a set of complex coordinates
$\{z_\alpha^i,\bz_\alpha^i\}$ such that on each overlap $U_\alpha\cap
U_\beta$ the transition functions $z^i_\beta(z_\alpha)$ are
holomorphic ($\alpha,\beta$ index the patch and $i,j$ the complex
coordinate). Fixing the complex structure restricts the allowed gauge
transformations to holomorphic diffeomorphisms. Continuous holomorphic
diffeomorphisms are generated by holomorphic vector fields. In
contrast to the infinite number of vector fields generating general
diffeomorphisms, it can be shown that a compact manifold admits only a
finite number of holomorphic vector fields. Even better, Calabi-Yau
manifolds have none at all, for the following reason. Every Calabi-Yau
is equipped with a nowhere-vanishing holomorphic $(n,0)$ form $\Omega$
(which depends on the complex structure but not on the metric). If
$v^i$ is non-vanishing with holomorphic components, then the same is
true of the $(n-1,0)$ form $v^{i_1}\Omega_{i_1i_2\dots
i_n}dz^{i_2}\wedge\dots\wedge dz^{i_n}$. Since the Hodge number
$h^{n-1,0}$ vanishes on a Calabi-Yau, such a form must be
$\bar\partial$ exact, which is impossible for an $(n-1,0)$
form. Hence, on a Calabi-Yau, fixing the transition functions for the
complex coordinates amounts to a full gauge fixing. As mentioned in
the introduction, this is a great advantage for numerical work.

A metric is called \K with respect to a given complex
structure if it is Hermitian, $g_{ij}=g_{\bi\bj}=0$, and if the
associated \K form $J = ig_{i\bj}dz^i\wedge d\bz^j$ is closed,
$dJ=0$. $J$ is obviously a positive $(1,1)$ form, since the metric is
everywhere positive definite. The cohomology class of $J$ is called
the \K class. The set of potential \K classes, i.e.\ $(1,1)$
cohomology classes containing at least one positive form, is called
the \K cone, or \K moduli space.

The fact that the \K form is closed implies that it, and the metric,
are locally expressible as the matrix of second derivatives of a
scalar,
\begin{equation}\label{potential}
\left.g_{i\bj}\right|_{U_\alpha} = \partial_i\partial_\bj K_\alpha,\qquad
\left.J\right|_{U_\alpha} = i\partial\bar\partial K_\alpha,
\end{equation}
where $K_\alpha$ is a real function, the \K potential, defined
on the patch $U_\alpha$.  Given $g_{i\bj}$, the \K potential is not
unique but can be changed by adding the real part of any holomorphic
function, a so-called \K transformation.  From a numerical
point of view, one advantage of using the \K potential to encode the
geometry is obvious: it reduces the number of functions to store from
$n(2n+1)$ for the full metric (or $n^2$ if the hermiticity condition
is imposed) to a single one.

The volume of the manifold is given in terms of the \K form by
\begin{equation}\label{volume}
V = \frac1{n!}\int J^n.
\end{equation}
Therefore on a compact manifold $J$ cannot be exact, or else the
manifold would have zero volume. So the \K potentials obtained by
integrating \eqref{potential} must disagree on the patch overlaps by
\K transformations:
\begin{equation}\label{transformation}
K_\alpha-K_\beta=u_{\alpha\beta}.
\end{equation}
These \K transformations serve as boundary conditions for the \K
potential (the only boundary conditions if the manifold is compact). In doing so,
they perform two important tasks. First, they fix the \K class. To see
this, note that two \K potentials $\{K_\alpha\}$ and $\{K'_\alpha\}$
which have the same \K transformations,
$K_\alpha-K_\beta=K'_\alpha-K'_\beta$, differ by a globally defined
real function $\phi$. Therefore the difference between the
corresponding \K forms, $J-J'=i\partial\bar\partial\phi$, is an exact
form. Conversely, given a representative of some particular \K class,
a corresponding set of $u_{\alpha\beta}$'s may easily be found by
solving \eqref{potential} separately on each patch. (The
representative need not be positive. This is useful particularly near
the edge of the \K cone, where it may not be easy to find positive
representatives.) The second task performed by the \K transformations
is to (almost) eliminate gauge-equivalent \K potentials, that is,
different \K potentials that give rise to the same metric: for each \K
form $J$ in the class defined by a set $\{u_{\alpha\beta}\}$, there is
a unique solution to \eqref{potential} and \eqref{transformation}, up
to constant shifts of all the $K_\alpha$.

When the metric is \K, only the purely holomorphic and antiholomorphic
components $\Gamma^i_{jk}$ and $\Gamma^{\bi}_{\bj\bar k}$ of the
Christoffel symbol are non-zero. Together with the hermiticity of the
metric, it follows that the coordinates $z^i$ and $\bz^i$ are
harmonic, a well-known gauge fixing condition for numerical
relativity. However, {\K}ity is a stronger condition than harmonicity;
whereas a given metric always admits a harmonic coordinate system, at
least locally, a generic metric is not \K with respect to any
coordinates, even locally.\footnote{Nor are \K coordinates necessarily
unique for a given metric. For example, a Ricci-flat metric on a
hyperk\"ahler manifold, such as K3, is \K with respect to a continuous
family of different complex structures. We will not make use of this
additional structure, but will be content to fix a particular complex
structure at the outset.}

To summarize: We fix the complex structure by specifying the
coordinate transition functions on the patch overlaps. Using the \K
potential to encode the geometry, we fix the \K class by specifying
the \K transformations on the overlaps.  With this set-up in hand, we
now turn to the question of what equation to solve for the \K
potential.

\subsection{The Monge-Amp\`ere equation}\label{MAeqn}

There exists a simple expression for the Ricci tensor of a \K metric:
\begin{equation}\label{ricci}
R_{kl}=R_{\bar k\bar l}=0, \qquad R_{k\bar l} = R_{\bar l k}=-\partial_k\partial_{\bar l}\ln\det g_{i\bj}
\end{equation}
(note that $\det g_{i\bj}=\sqrt{|g|}$). Just as we defined the \K form
$J$, we can define the Ricci form $\Rc=iR_{k\bar l}dz^k\wedge
d\bz^l$. By virtue of \eqref{ricci}, $\Rc$ is closed (but not
necessarily exact, since $\det g_{i\bj}$ is not in general a globally
defined function). One can show that its cohomology class, the
first Chern class $c_1$, is a topological invariant: two
different \K metrics will give rise to Ricci forms that differ by an
exact $(1,1)$ form. Calabi-Yau manifolds are defined by the condition
that $c_1$ vanishes, in other words that the Ricci form is always
exact. According to Yau's theorem \cite{Yau:1977ms}, on a Calabi-Yau
manifold each \K class (i.e.\ $(1,1)$ cohomology class containing at
least one \K form) contains a unique Ricci-flat \K form.

From \eqref{ricci}, the Einstein equation for a \K metric reads
\begin{equation}\label{ricciflat}
\partial_k\partial_{\bar l}\ln\det g_{i\bj}=0.
\end{equation}
In terms of the metric this is a second-order PDE, but in terms of the
\K potential it is fourth order. It might seem then that working with
the \K potential is not so advantageous after all. However, as we will
now explain, by the magic of complex analysis it can be reduced to a
second-order PDE, specifically a Monge-Amp\`ere equation (a PDE in
which the derivatives appear in the form of a Hessian).

For simplicity, let us assume first that the coordinates have been
arranged in such a way that the Jacobians $\det_{ij}(\partial
z^i_\alpha/\partial z^j_\beta)$ of the transition functions are 1 on
all the overlaps. In that case $\det g_{i\bj}$ is a globally defined
function, and on a compact manifold the Einstein equation
\eqref{ricciflat} is equivalent to it being constant:
\begin{equation}\label{ma2}
\det g_{i\bj} = \lambda.
\end{equation}
This is a non-linear Monge-Amp\`ere equation for the \K potential. The
constant $\lambda$ is related to the volume of the manifold,
given by \eqref{volume}, which depends only on the
\K class and may therefore be calculated a priori from the \K
transformations on the overlaps (see Appendix \ref{homology} for details).

The coordinate system used for the Kummer surface in Section
\ref{kummer} happens to satisfy the condition of unit coordinate
Jacobians assumed in the previous paragraph. It is instructive
nonetheless to consider what one would do in the more general
situation. Clearly we need to replace the right-hand side of
\eqref{ma2} with something that has the correct transformation law on
the overlaps, and also that implies the Einstein equation
\eqref{ricciflat}. The holomorphic $(n,0)$ form $\Omega$ once again
comes to the rescue; the generalization we seek is:
\begin{equation}\label{ma3}
\det g_{i\bj} = \lambda|\Omega_{1\dots n}|^2
\end{equation}
(this can also be written in the coordinate-invariant form $J^n =
(-i)^nn!\,\lambda\,\Omega\wedge\bar\Omega$). If $\Omega$ is not known
explicitly, then $|\Omega_{1\dots n}|^2$ can be found as follows in
terms of an arbitrary \K metric $\tilde g_{i\bj}$ (which need not be
in the same \K class as the desired solution, since $\Omega$ depends
only on the complex structure). Since the manifold is Calabi-Yau, its
Ricci form is exact, $\tilde R_{i\bj} = \partial_i\partial_\bj\tilde
F$, where $\tilde F$ is a globally defined function which can be
calculated explicitly (either analytically or by standard numerical
methods). We then have
\begin{equation}\label{omega}
|\Omega_{1\dots n}|^2 = e^{\tilde F}\det\tilde g_{i\bj},
\end{equation}
and equation \eqref{ma3} may be applied.

To get a feeling for the Monge-Amp\`ere equation, it is useful to
consider equations \eqref{ma3} and \eqref{omega} in the case where
$\tilde g_{i\bj}$ \emph{is} in the desired \K class. Then we can write
\begin{equation}
K_\alpha = \tilde K_\alpha + \phi,
\end{equation}
where $\phi$ is a globally defined scalar. In terms of $\phi$ the
Monge-Amp\`ere equation is
\begin{equation}\label{ma4}
\det(\delta_i^k+\tilde g^{k\bj}\partial_i\partial_{\bj}\phi) = \lambda
e^{\tilde F}.
\end{equation}
The left-hand side is a non-linear operator acting on $\phi$. If
$\partial_i\partial_\bj\phi$ is small (i.e.\ if $\tilde g_{i\bj}$ is
almost Ricci-flat) then we can linearize, yielding a Poisson
equation:
\begin{equation}\label{poisson}
\frac12\tilde\nabla^2\phi +
 \mathcal{O}\left(\partial_i\partial_\bj\phi\right)^2 = \lambda
 e^{\tilde F}-1.
\end{equation}
In this sense Yau's theorem can be understood as a generalization of
the existence theorem for the Poisson equation.

\subsection{Methods}\label{methods}

The similarity to the Poisson equation observed in the last subsection
suggests that methods for solving it might be generalized to the
Monge-Amp\`ere equation.  Among the standard methods for the Poisson
equation, some are local and others are non-local. We have restricted
ourselves to local schemes, for two reasons. First, the Monge-Amp\`ere
equation is local and non-linear. Second, for most manifolds (even
highly symmetrical ones), the individual patches $U_\alpha$ (or, more
precisely, their images in $\C^n$, i.e.\ the ranges of the coordinates
$z^i_\alpha$), have rather irregular shapes. This motivated the use of
a lattice discretization, rather than non-local spectral
representations such as Fourier modes or wavelets.

One simple (but inefficient) method for solving the Poisson equation
is to simulate diffusion. The analogous geometric relaxation equation
is Ricci flow, which is defined by a first order equation in an
auxiliary time dimension: $\dot g_{i\bj} = -R_{i\bj}$. This flow is
well studied mathematically \cite{MR2061425, MR2112626}, largely
because of its application to geometrization. It is also of some
interest in physics as the one-loop renormalization-group evolution for
the target space geometry of a sigma model. In terms of the \K potential the
flow is governed by
\begin{equation}\label{ricciflow}
\dot K_\alpha = \ln\left(\frac{\det g_{i\bj}}{\lambda|\Omega_{1\dots
n}|^2}\right).
\end{equation}
Note that on the overlaps $\dot K_\alpha-\dot K_\beta=0$, so the \K
transformations, and hence the \K moduli, are conserved. Cao
\cite{Cao} has shown that Ricci flow starting from an arbitrary \K
metric converges to the Ricci-flat metric in its class.

Some work has been done on numerical simulation of Ricci flow on real
geometries \cite{Garfinkle:2003an,Rubinstein,MR2115754}, but not (as
far as we know) in the \K case.  We experimented with simple lattice
implementations of such a scheme.  However, if one is only interested
in the endpoint of the flow, namely the Ricci-flat metric, as opposed
to the whole flow, this method is clearly very inefficient, as it
requires solving an equation in one higher dimension. Furthermore,
stability of the diffusion problem typically requires implicit finite
differencing schemes which are rather inconvenient (particularly in
several dimensions). Another more subtle drawback which nonetheless is
rather serious in practice is that it requires an initial \K (i.e.\
positive) form in the desired class, as seen above explicitly from the
logarithm in \eqref{ricciflow}. As mentioned in subsection
\ref{Kgeom}, it is not always easy to find a positive representative
of a given class, especially near the edge of the \K cone.

The prototype local method for solving the Poisson equation is the
Gauss-Seidel method. Here the discretized Poisson equation is solved
at each lattice point in turn. After a suitable number of iterations
over the whole lattice, the discretized equations will be solved to a
given accuracy. This is very robust and simple to
implement. Furthermore whilst Gauss-Seidel is slow compared to
spectral or multigrid methods in low dimension, scaling as $N^{1+2/d}$
rather than $N \log N$ in $d$ real dimensions (where $N$ is the number
of lattice points), in our K3 case of 4 real dimensions, and more so
for still larger dimensions, the advantage is not so great. Of course
multigrid could be implemented relatively simply to improve our
Gauss-Seidel method if speed became a crucial issue.

Our analog of the Gauss-Seidel method for the Monge-Amp\`ere equation
is as follows. On a lattice the metric is determined from the \K
potential by taking discrete derivatives. Thus the value of $\det
g_{i\bj}$ at a given site is a function of the values of $K_\alpha$ at
that site and its neighbors, out to some distance depending on the
order of finite differencing used. Our algorithm directs one to go
through each site of the lattice in turn, changing $K_\alpha$ at that
site to the value which solves \eqref{ma3} given its values at the
neighboring sites. Although we do not have a convergence theorem for
this algorithm, we found that in practice it converged on the full
range of \K parameters studied. This included cases where the initial
\K potential did \emph{not} define a positive $(1,1)$ form.

As noted earlier, the constant $\lambda$ which relates the coordinate
volume to the proper volume can be computed analytically from the \K
class. Now consider solving the equation
\begin{equation}\label{manum}
\det g_{i\bj} = \tilde\lambda.
\end{equation}
We will \emph{only} find a solution to this new Monge-Amp\`ere
equation when we set $\tilde\lambda = \lambda$. However, there is a
subtlety, as when we implement this equation numerically, the errors
involved in finite differencing will mean that the value of
$\tilde\lambda$ that solves the finite differenced equation will
actually differ slightly from the true continuum value $\lambda$. This
implies the finite differenced equation should have no solution when
$\tilde\lambda = \lambda$, which sounds mildly disastrous. However, we
can see in detail how this discretization error manifests itself if we
consider the Ricci flow equation \eqref{ricciflow} with $\lambda$
replaced by $\tilde\lambda$. Then, since the Monge-Amp\'ere equation
has no solution for $\tilde\lambda \ne \lambda$, the flow will never
reach a fixed point. However, the flow does asymptote to one with a
simple time dependence, namely,
\begin{equation}
K_{\tilde\lambda}(t) = K_{\lambda} + t v , \qquad v = \ln \frac{\lambda}{\tilde\lambda},
\end{equation}
where $K_{\lambda}$ is the solution of the Monge-Amp\`ere equation for
the true value of $\lambda$. The constant $v$, as determined above,
then gives the asymptotic time dependence of the flow. By analogy with
the Ricci flow above, this implies that under Gauss-Seidel iteration,
our finite differenced Monge-Amp\`ere equation \eqref{manum} will have
an asymptotic solution that drifts in iteration time by a constant
mode. This simply corresponds to a drifting \K transformation;
therefore the real metric does indeed tend to a static solution, and
disaster is avoided. However, since we would rather have a fixed
endpoint to our Gauss-Seidel method, we ``cure" this drifting due to
discretization error by solving the Monge-Amp\`ere equation
\eqref{manum}, and dynamically determine $\tilde\lambda$ by averaging
$\det g_{i\bj}$ as we perform the Gauss-Seidel iterations. This
procedure then yields a static end solution, and the value
$\tilde\lambda$ should approach the true analytic $\lambda$ as the
continuum is approached by increasing the lattice resolution.

It is important to check that the solution to the lattice version of
the Monge-Amp\`ere equation converges to the continuum solution as the
lattice resolution is improved. For this purpose it is useful to have
some quantities that can be calculated from the lattice solution, and
for which the exact, continuum value is also known. Here we will
mention three such quantities.  The first is that mentioned directly
above, namely the total volume of the manifold, in other words how
close $\tilde\lambda$ is to $\lambda$. The agreement between the
analytic and numerical values tests not only the quality of the
solution, but also the error in fixing the \K class. A second quantity
that is known exactly is the Euler number $\chi$ of the manifold
(which of course does not depend on the \K class). This can be
calculated numerically in terms of the \K potential via a Gauss-Bonnet
theorem. Finally, there is the difference between \K potentials
$K_\alpha-K_\beta$ on the overlap $U_\alpha\cap U_\beta$. In
subsection \ref{Kgeom} it was argued that this difference should be
set equal to the fixed \K transformation $u_{\alpha\beta}$, as a
boundary condition for the $K_\alpha$'s. More precisely, however, the
boundary condition is imposed only on the \emph{edge} of each
patch. For the continuum equation, this is enough to guarantee
$K_\alpha-K_\beta=u_{\alpha\beta}$ also in the interior of
$U_\alpha\cap U_\beta$, by the uniqueness of the solution to the
Monge-Amp\`ere equation. The lattice will introduce an error into this
equality; conversely, that error measures how well the lattice
solution approximates the continuum one.

\section{Application to Kummer surfaces}\label{kummer}

For a first application of the techniques described in the previous
section, we turned to the lowest-dimensional Calabi-Yau manifold, the
K3 surface. K3 has played important roles in algebraic and
differential geometry, as well as in string theory. (The lecture notes
\cite{Aspinwall:1996mn} provide an excellent review of the role of K3
in string theory. See also the more mathematically-oriented review
\cite{Nahm:1999ps}.)  The moduli space of Ricci-flat metrics on K3 is
58-dimensional: 40 complex structure moduli and 20 \K moduli, minus 2
for the hyperk\"ahler identifications. One of the moduli is the
overall volume of the manifold; the other 57 are non-trivial in the
sense that they do not act on the Ricci-flat metric by any
straightforward transformation. We studied a particular one-parameter
family of K3's which admit a simple construction and have a high
degree of discrete symmetry, which serves to reduce the number of
lattice points to be simulated at any given resolution. In the first
subsection below, we explain the construction of these K3's. In the
next subsection we describe the results obtained in our simulations,
giving several examples of the kind of concrete geometrical
information that is available from the explicit form of the metric.

\subsection{Construction}\label{construction}

Among the simplest K3 surfaces to describe explicitly are the
so-called Kummer surfaces, which are the orbifold $T^4/\Z_2$ with its
16 singular points blown up. After explaining this construction, we
will specialize to the ones with the largest discrete symmetry group,
namely those constructed from a cubical $T^4$ with all 16 singular
points blown up to the same size. The only free parameters are the
size of the $T^4$ and the size of the blow-up. We will see that the \K
cone defines a finite range of values for their ratio.

As a complex manifold, the torus $T^4$ is parametrized by a pair of
complex coordinates $(z^1,z^2)\in\C^2$, identified under translations
by a set of four linearly independent vectors $(v_a^1,v_a^2)\in\C^2$
($a=1,\dots,4$),
\begin{equation}\label{periodicity}
(z^1,z^2) \sim (z^1,z^2)+(v_a^1,v_a^2).
\end{equation}
Choosing the vectors $(v_a^1,v_a^2)$ fixes the complex structure of
the torus (there are equivalences; the complex structure moduli space
is actually only 8 real dimensional). The parity map
\begin{equation}\label{parity}
(z^1,z^2) \mapsto (-z^1,-z^2)
\end{equation}
is compatible with the equivalences \eqref{periodicity}, so we can
quotient the torus by it. Furthermore, it acts holomorphically, so the
result, $T^4/\Z_2$, is a complex manifold. More correctly, it's an
orbifold, since the parity map has fixed points. There are 16 of them,
located at $(z^1,z^2) = \frac12\sum_an^a(v^1_a,v^2_a)$ with $n^a=0$ or
1, and each carries an $A_1$ (or $\C^2/\Z_2$) type singularity.

We obtain a smooth complex manifold, known as a Kummer surface, by
blowing up the fixed points. Consider for example the fixed point at
the origin. To blow it up, remove that point from the manifold and add
two new patches, with coordinates $(y,w)$ and $(y',w')$ respectively,
and transition functions (which are of course holomorphic)
\begin{eqnarray}
(y,w) &=& \left(\frac1{y'},w'y^{\prime2}\right) =
\left(\frac{z^1}{z^2},\frac12(z^2)^2\right), \label{trans1} \\ (y',w')
&=& \left(\frac1y,wy^2\right) =
\left(\frac{z^2}{z^1},\frac12(z^1)^2\right), \label{trans2} \\
(z^1,z^2) &=& \pm\sqrt{2w}\,(y,1) =
\pm\sqrt{2w'}(1,y'). \label{trans3}
\end{eqnarray}
To avoid complications, the ranges of the new coordinates should be
bounded in such a way that they do not include the other fixed
points. Each of the 16 fixed points of the orbifold is given its own
$(y,w)$ and $(y',w')$ patches, with the same transition functions
(\ref{trans1}--\ref{trans3}) except that $(z^1,z^2)$ is replaced by
$(z^1,z^2) - \frac12\sum_an^a(v^1_a,v^2_a)$. Hence we have a total of
33 patches. There are three important points to note about the new
$(y,w)$ and $(y',w')$ coordinate systems. First, the identification
under the orbifold action \eqref{parity} is automatic in them. Second,
the origin has been replaced by the surface $w=w'=0$, parametrized by
$y=1/y'$. This is a $\C P^1$ (or $S^2$), and is homologically
non-trivial; it is called the exceptional divisor. Finally, the
transition functions (\ref{trans1}--\ref{trans3}) all have unit
Jacobian. Hence it is quite easy to write down the holomorphic $(2,0)$
form:
\begin{equation}
\Omega = dz^1\wedge dz^2 = dy\wedge dw = dw'\wedge dy'.
\end{equation}

Kummer surfaces have 8 complex structure moduli (inherited from the
$T^4$) and 20 \K moduli (the 4 inherited from the $T^4$, plus the size of
each of the 16 exceptional divisors). It can be shown that they are
special cases of K3 surfaces. (The missing 32 complex structure moduli
are due to the fact that we blew up, rather than deformed, the
orbifold fixed points.)

Both for simplicity and in order to reduce the number of lattice
points simulated, it was advantageous for us to choose highly
symmetrical Kummer surfaces. The $T^4$ was taken to be cubical;
in other words, the periodicities \eqref{periodicity} were given by
\begin{equation}\label{cube}
z^1\sim z^1+1 \sim z^1+i, \qquad z^2\sim z^2+1 \sim z^2+i,
\end{equation}
while the \K transformations are those obtained for the flat metric
on a cubical $T^4$ of side length $b$:
\begin{eqnarray}
\label{toruskt1} K(z^1+1,z^2) - K(z^1,z^2) &=& b^2\left(\Rl z^1+\frac12\right), \\
\label{toruskt2} K(z^1+i,z^2) - K(z^1,z^2) &=& b^2\left(\I z^1+\frac12\right), \\
\label{toruskt3} K(z^1,z^2+1) - K(z^1,z^2) &=& b^2\left(\Rl z^2+\frac12\right), \\
\label{toruskt4} K(z^1,z^2+i) - K(z^1,z^2) &=& b^2\left(\I z^2+\frac12\right).
\end{eqnarray}
The coefficients on the four right-hand sides correspond to the four
\K parameters of $T^4$; here since the $T^4$ is cubical, they are set
equal to a common constant $b^2$. Each blown up fixed point has only
one \K modulus; without loss of generality the \K transformations may
be taken as follows:
\begin{eqnarray}
\label{ehkt1} K_{(z^1,z^2)} - K_{(y,w)} &=& a^2\ln|z^2|, \\
\label{ehkt2} K_{(z^1,z^2)} - K_{(y',w')} &=& a^2\ln|z^1|, \\
\label{ehkt3} K_{(y,w)} - K_{(y',w')} &=& a^2\ln|y|.
\end{eqnarray}
All 16 fixed points were blown up to the same value of the modulus $a^2$.

How much discrete symmetry do these Kummer surfaces possess? Our choice of
complex structure admits a holomorphic diffeomorphism group
of order $2^8$, all of which is respected by our choice of \K
class. The generators of this group are as follows: the translations
by the vectors $(\frac12,0)$, $(\frac i2,0)$, $(0,\frac12)$, and
$(0,\frac i2)$, which generate a $\Z_2^4$ group that maps the origin
to each of the other fixed points; the rotation $z^1\mapsto iz^1$,
which generates a $\Z_4$ group; the diagonal rotation
$(z^1,z^2)\mapsto(iz^1,iz^2)$, which (in view of the identification
under parity) generates a $\Z_2$ group; and the exchange
$(z^1,z^2)\mapsto(z^2,z^1)$, which generates another $\Z_2$. Of course,
these generators don't commute with each other, so the full
holomorphic diffeomorphism group is complicated. In addition, the
anti-holomorphic diffeomorphism $(z^1,z^2)\mapsto(\bz^1,\bz^2)$ is a
symmetry of the real metric. Uniqueness of the solution to the
Monge-Amp\`ere equation guarantees that every symmetry of the \K class
is an isometry of the Ricci-flat metric in that class. Therefore it is
sufficient for us to simulate only the fundamental domain of the
symmetry group, a reduction potentially by a factor of $2^9$.

As explained in Appendix \ref{homology}, the volume is easily
calculated from the intersection matrix of the second de Rham cohomology group
$H_2(\R)$:
\begin{equation}\label{k3volume}
V = \frac12b^4 - 4\pi^2a^4.
\end{equation}
As discussed in subsection \ref{methods}, the volume calculated
numerically may be compared against this analytic result to give a
measure of how accurately we are fixing our \K class. More details are
given in Appendix \ref{appnum}.

Equation \eqref{k3volume} shows that, as the size of the blow-up
increases, it eats away at the volume of the manifold. This indicates
an upper bound $a^2/b^2<(8\pi^2)^{-1/2}$. In fact, the real bound is
slightly lower: $a^2/b^2<(4\pi)^{-1}$. We first noticed this as an
empirical fact in our numerical trials, but the reason is not hard to
understand. The volume of a holomorphic submanifold depends only on
the \K class, not the metric (the total volume being a special case of
this). Therefore, a necessary condition for being inside the \K cone
is that all the holomorphic submanifolds have positive area. Our
symmetric Kummer surfaces have three types of holomorphic curves. The
first type are the 16 exceptional divisors. From the \K transformation
\eqref{ehkt3} restricted to the $w=0$ surface, one may show that each
has area $A=\pi a^2$. We thus have the condition $a^2>0$. Another
holomorphic submanifold is the curve $\{z^1=C\}\cup\{z^1=-C\}$, which
represents the same two-cycle for all values of $C$ other than the
special values $C=0,\frac12,\frac i2,\frac12+\frac i2$. Using
(\ref{toruskt3},\ref{toruskt4}), its area is $b^2$, so we have
$b^2>0$. In the special cases $C=0,\frac12,\frac i2,\frac12+\frac i2$
the curve passes through points which are outside of the $z$ patch, so
one has to be more careful. As we show in Appendix \ref{homology},
these represent 4 different two-cycles; each has area
\begin{equation}\label{neweh}
\hat A = \frac 12b^2-2\pi a^2,
\end{equation}
accounting for the above upper limit on $a^2/b^2$. Obviously, the same
results apply to the curves of constant $z^2$, so there are a total of
8 curves of this type. In fact they are rational curves, since they
have topology $T^2/\Z_2\approx S^2$. Each of them intersects each
exceptional divisor at a single point, so what is happening as they
shrink to zero size is that the exceptional divisors are so large that
they touch each other and therefore can't be blown up any
larger.

An isolated rational curve shrinking to zero size implies the
formation of an $A_1$ orbifold singularity. In the next subsection we
will explicitly confirm the formation of these orbifold singularities
from our numerical solutions. The simultaneous formation of 8 $A_1$
singularities naturally leads to the idea that the manifold in this
limit is globally a $\Z_2$ orbifold. By counting the Euler number, one
finds that it must be an orbifold of another, smooth K3. Indeed, it
can be shown \cite{Inose,Wendland} that it is a rather well-known K3,
namely the Fermat quartic surface $x_1^4+x_2^4+x_3^4+x_4^4=0$ in $\C
P^3$, in the \K class induced from the Fubini-Study metric on $\C
P^3$.\footnote{It was shown in \cite{Inose} that (the blow-up of) the
orbifold of the Fermat surface by the $\Z_2$ action
$(x_1,x_2,x_3,x_4)\mapsto(x_1,x_2,-x_3,-x_4)$ has the same complex
structure as the Kummer surfaces we consider. This action has 8 fixed
points, and the resulting 8 exceptional divisors are identified with
our ``shrinking rational curves". It can furthermore be shown that the
\K class on the Kummer surface at $a^2/b^2=(4\pi)^{-1}$ lifts to the
\K class on the Fermat surface induced from the Fubini-Study metric on
$\C P^3$ \cite{Wendland}. Amusingly, it has also been shown
\cite{Nahm:1999ps,Wendland:2003ma,Wendland} that the sigma model with
this Kummer surface (or equivalently the orbifold of the Fermat
surface) as its target space, at a particular volume and equipped with
a particular $B$-field, is dual to one whose target space is the
Kummer surface at the opposite edge of the \K cone, $a^2/b^2=0$! We
are grateful to K. Wendland for very helpful discussions on these
issues.} Hence our one-parameter moduli space interpolates between the
orbifold of $T^4$ at $a^2/b^2 = 0$ and the orbifold of the Fermat
quartic at $a^2/b^2 = (4\pi)^{-1}$. For our fixed complex structure,
these endpoints of our modulus $a^2/b^2$ represent the edges of the \K
cone.

It is worth remarking that an approximate analytical solution to the
Einstein equation can be constructed in the limit $a^2/b^2\ll1$
by smoothly joining the Eguchi-Hanson metric (a Ricci-flat and
asymptotically flat metric on the blow-up of $\C^2/\Z_2$) onto the
flat torus metric \cite{Gibbons:1979xn,Bozhkov}. As we will see in the next subsection,
the numerical method does not perform as well in this
regime of parameters because the manifold contains a region of high curvature, namely the vicinity of the exceptional divisors. Thus the numerical and analytic approaches are effective in
complementary regimes.

\subsection{Results}\label{results}

We applied the methods described in Section \ref{generalmethod} to
find the Ricci-flat metrics on the symmetrical Kummer surfaces
constructed in the previous subsection, at 9 different values of the
modulus $a^2/b^2$. In terms of the combination
\begin{equation}
\alpha = 4\pi\frac{a^2}{b^2},
\end{equation}
which ranges from 0 to 1, the points for which the metrics were calculated were
\begin{equation}\label{alphavals}
\alpha = 0.03, 0.13, 0.28,0.50,0.61,0.72,0.79,0.85,0.92.
\end{equation}
(In figures \ref{fig:iso1}, \ref{fig:iso2}, \ref{fig:EHsphere}, and
\ref{fig:Torussphere} we have left out $\alpha = 0.79$ for typesetting
elegance.) Without loss of generality, the volume of the manifold was
fixed to be 1. The computational aspects of the problem---lattice
discretization, convergence, etc.---are discussed in detail in
Appendix \ref{appnum}. Let us simply mention here that, at each of the
above values of $\alpha$, the metric was computed at four different
lattice resolutions, labelled A, B, C, D; B has twice the linear
resolution (or 16 times the total number of points) of A, and so on.

In this subsection, we will illustrate the kind of concrete geometric
information that is available once the Ricci-flat metric is known. We
will focus on three ways to characterize the geometry: the
distribution of the Euler density; the induced geometry on the
exceptional divisors and other rational curves; and the low-lying
spectrum of the Laplacian. If unspecified, results presented are
generated from the highest resolution metric, D.

\FIGURE{
\centerline{\psfig{file=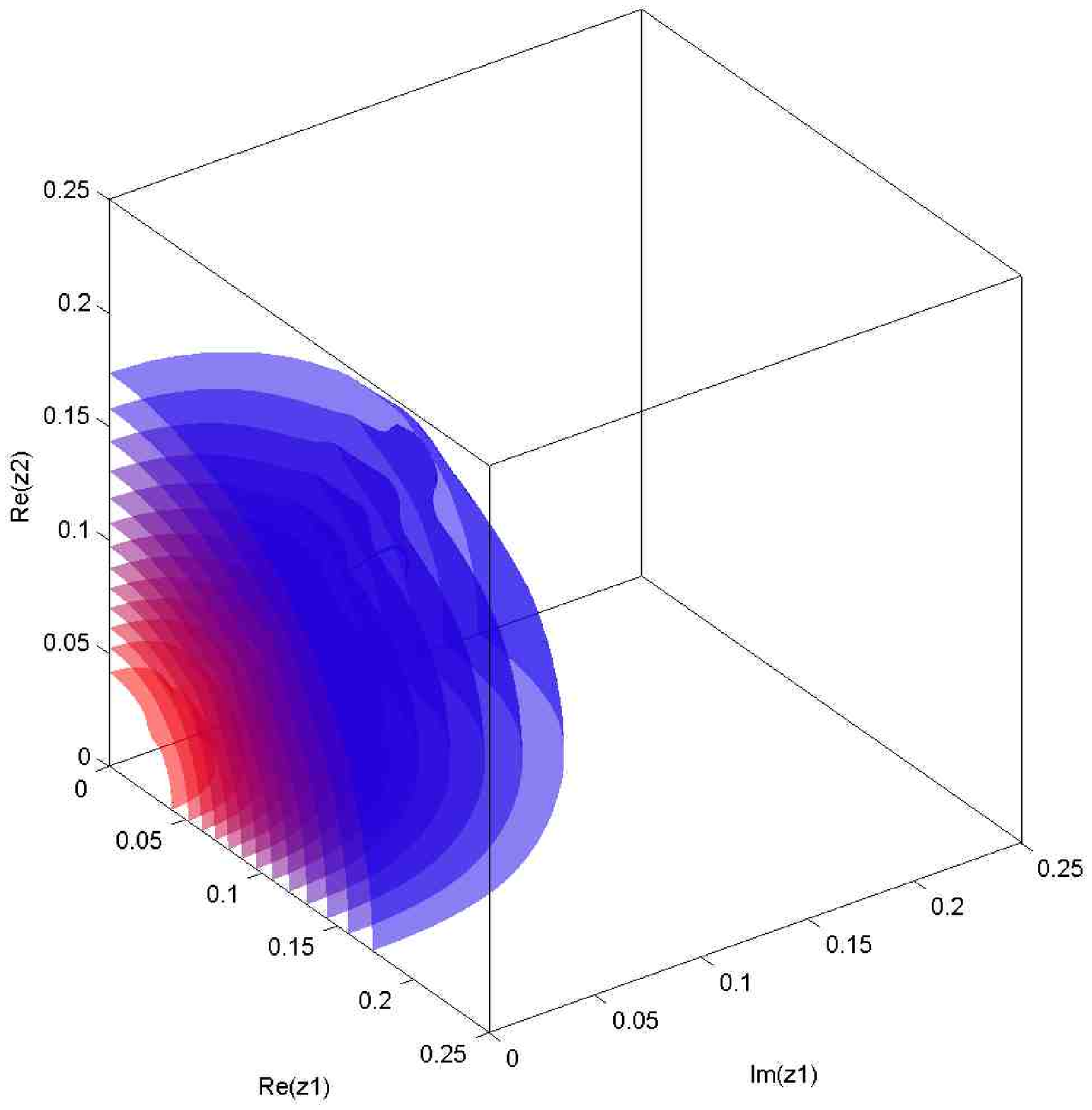,width=3.in}\psfig{file=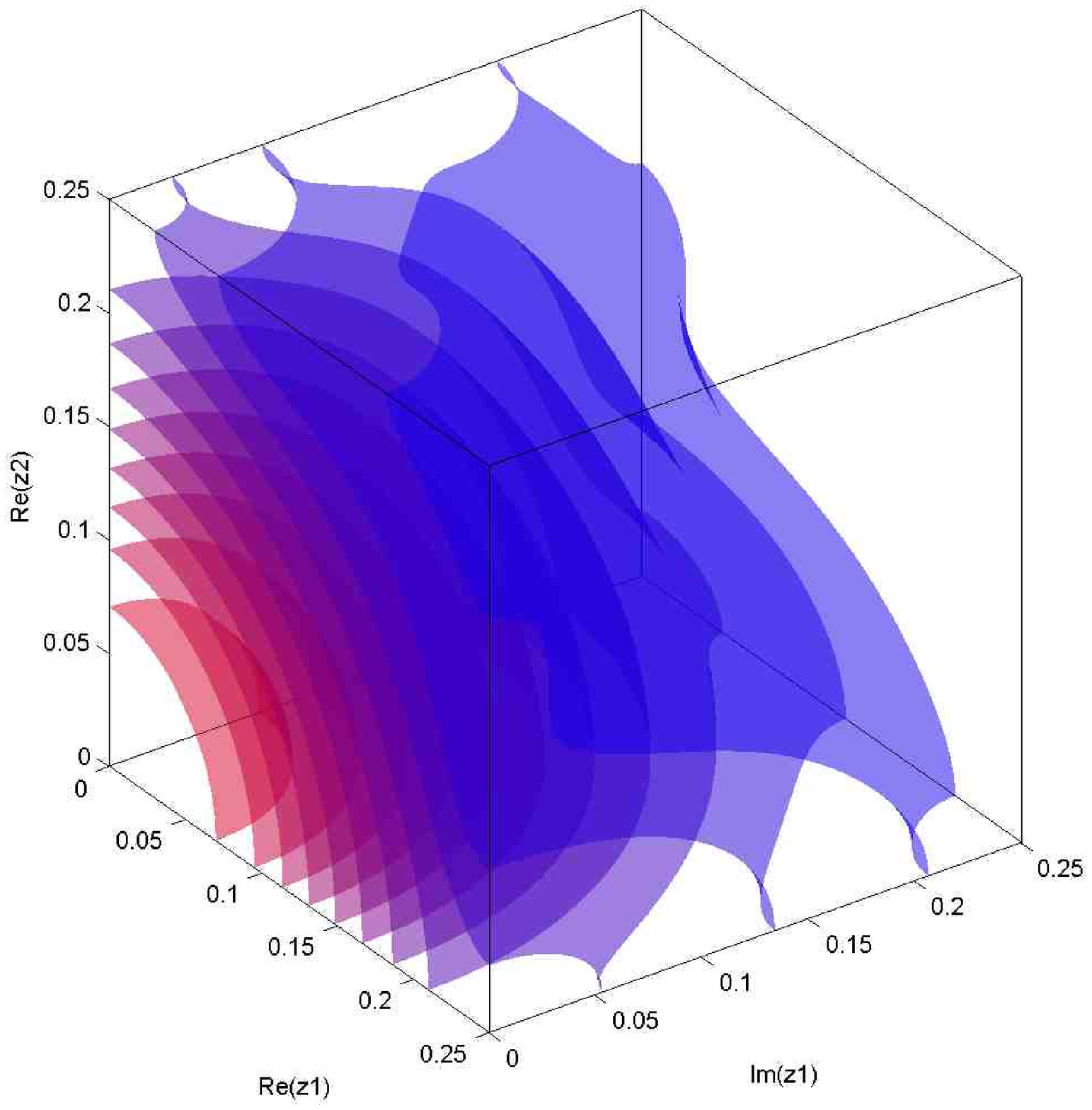,width=3.in}}
\centerline{\psfig{file=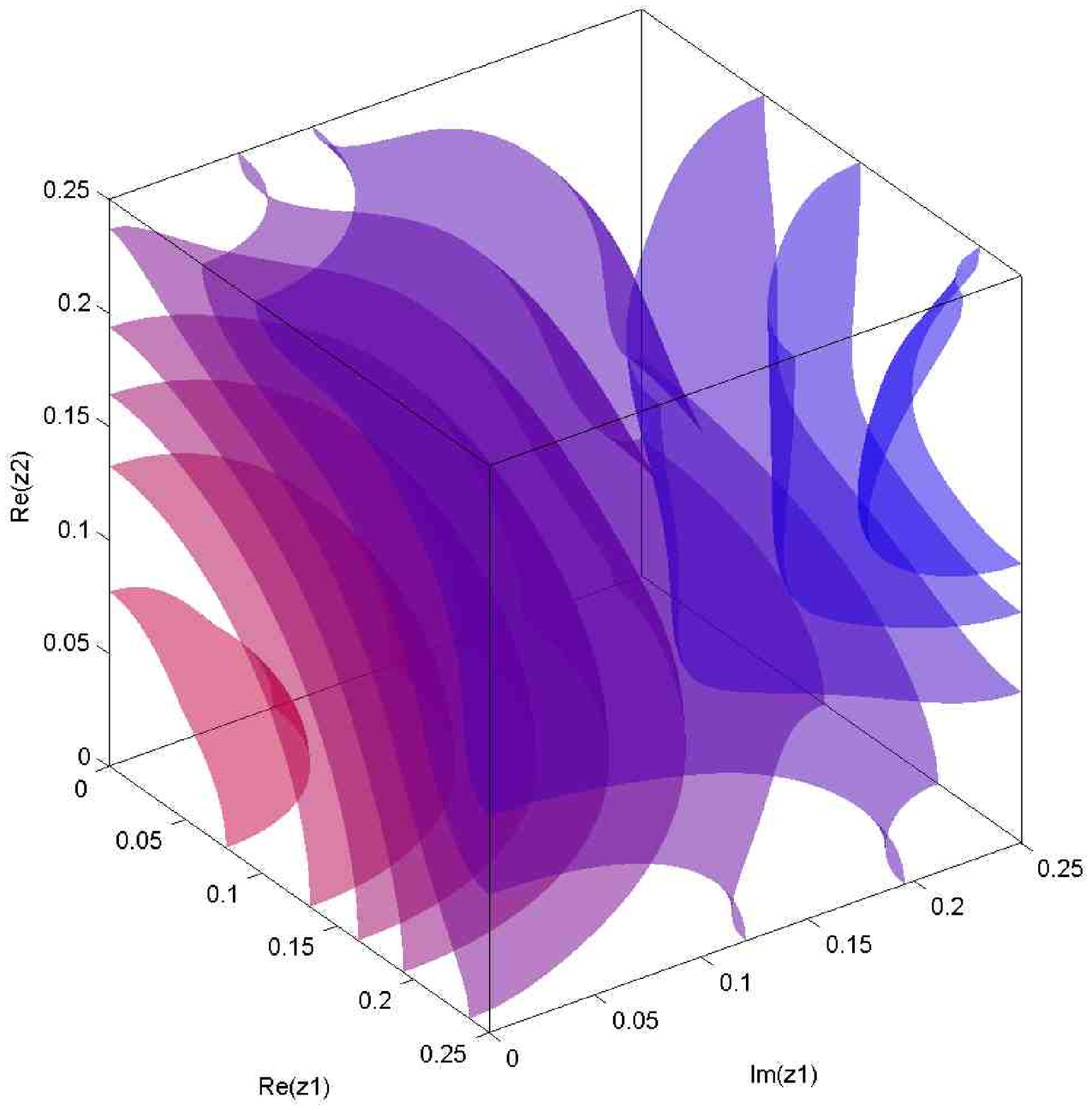,width=3.in}\psfig{file=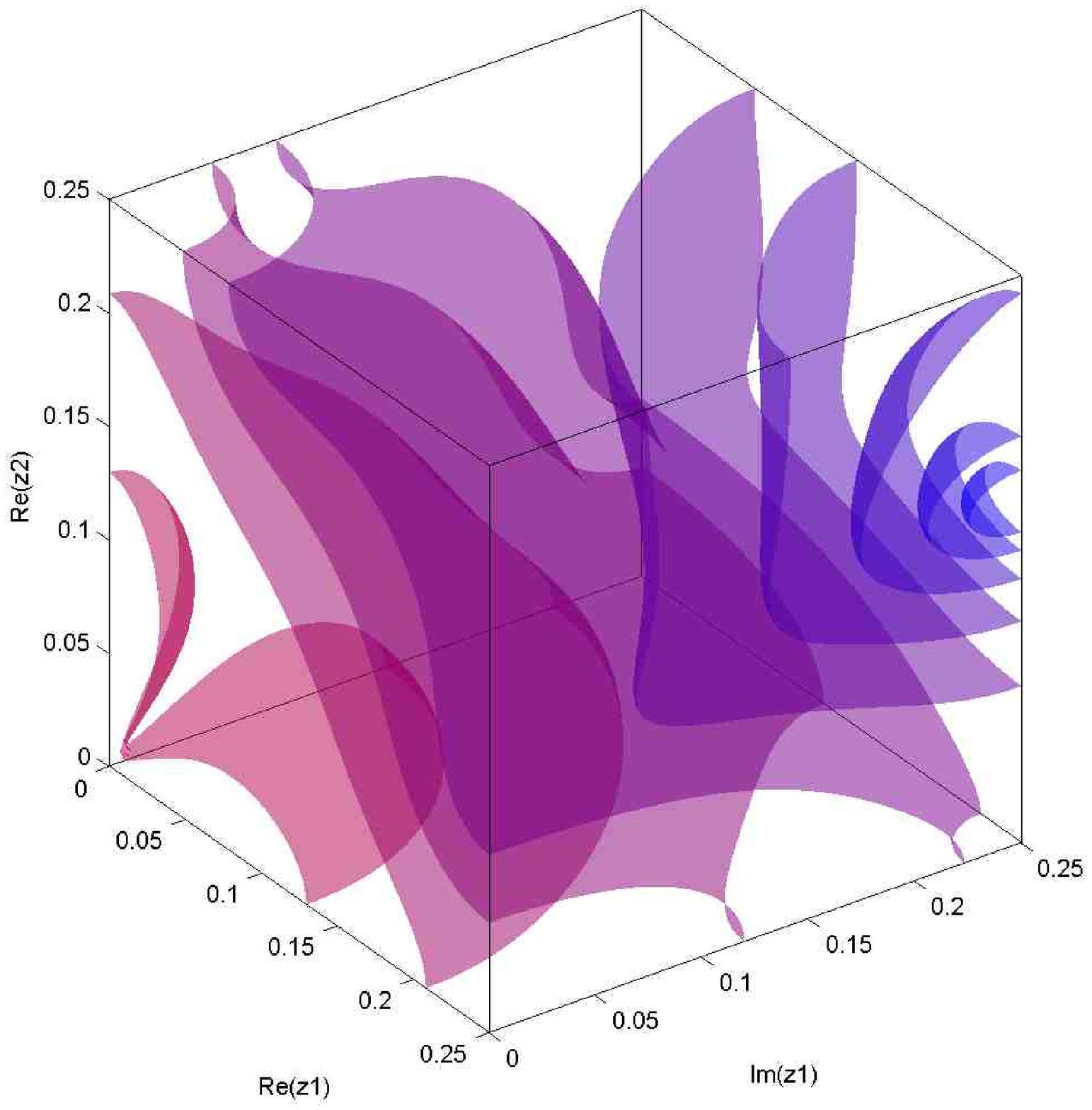,width=3.in}}
\caption{Isosurfaces of the Euler density $\rho$ in the $z^i$ coordinate
system, on the $\I z^2=0$ slice. Only the fundamental domain of the
discrete symmetry group is shown, namely the cubical region $0\le \Rl
z^1,\I z^1,\Rl z^2\le\frac14$ (thus the full $\I z^2=0$ slice, which
is a $T^3/\Z_2$, is 32 times as large as what is shown). Blue
indicates lower curvature, red higher, with surfaces shown at
$\rho=10^{-1.5},10^{-1},\dots,10^{4},10^{4.5}$. Left to right and top
to bottom, $\alpha=0.03, 0.13, 0.28,0.50$.
\label{fig:iso1}
}
}

\FIGURE{
\centerline{\psfig{file=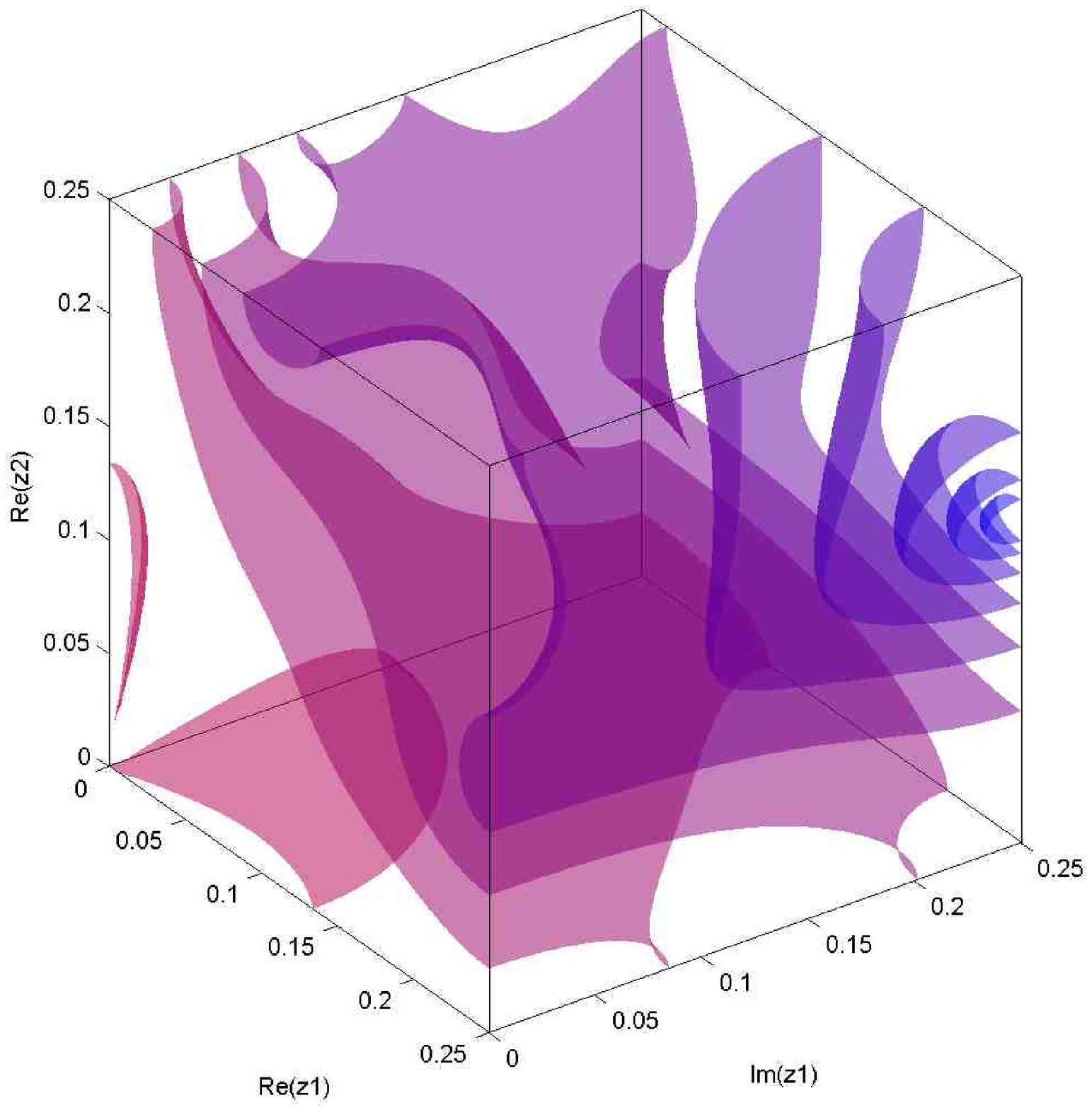,width=3.2in}\psfig{file=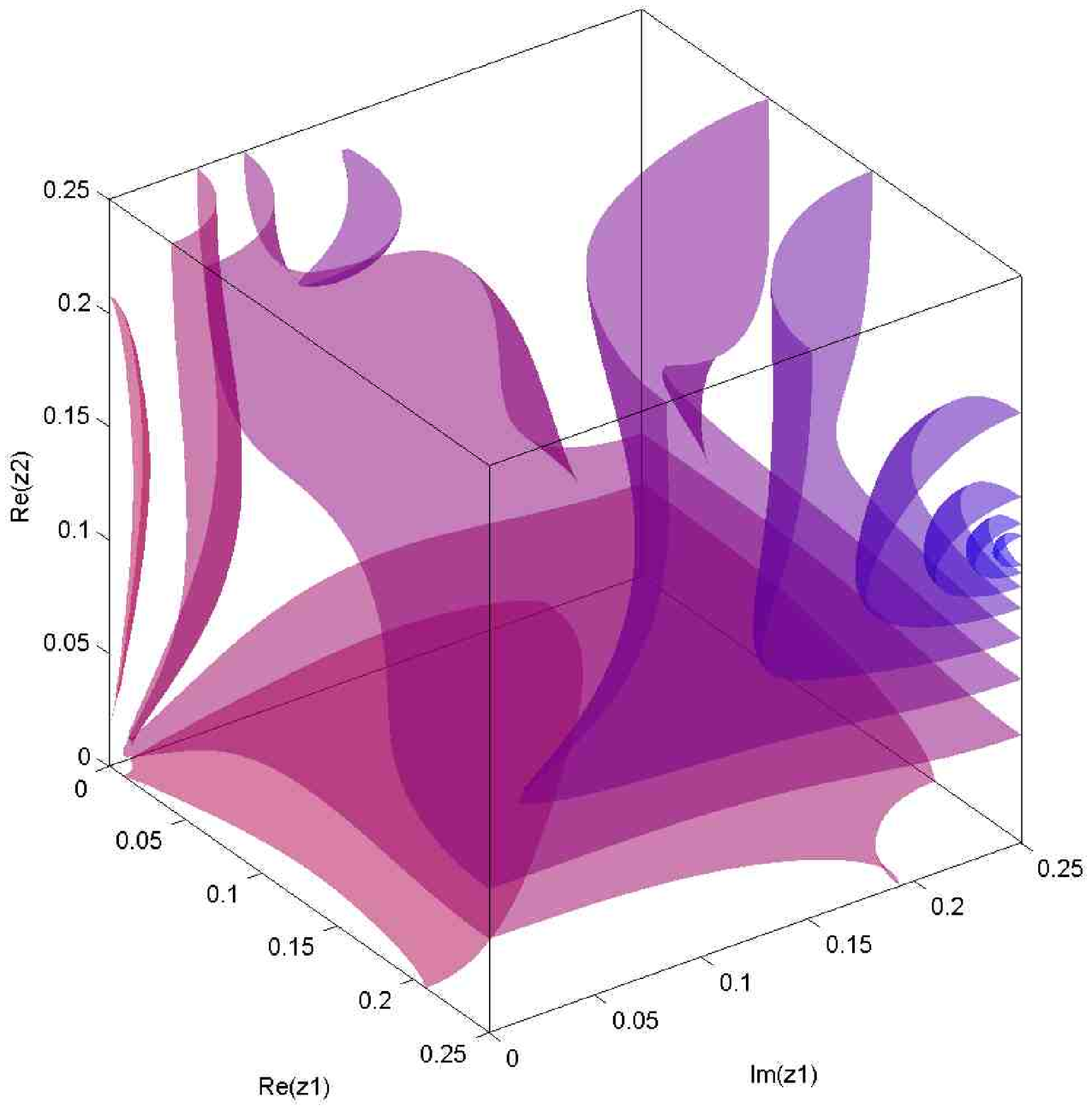,width=3.2in}}
\centerline{\psfig{file=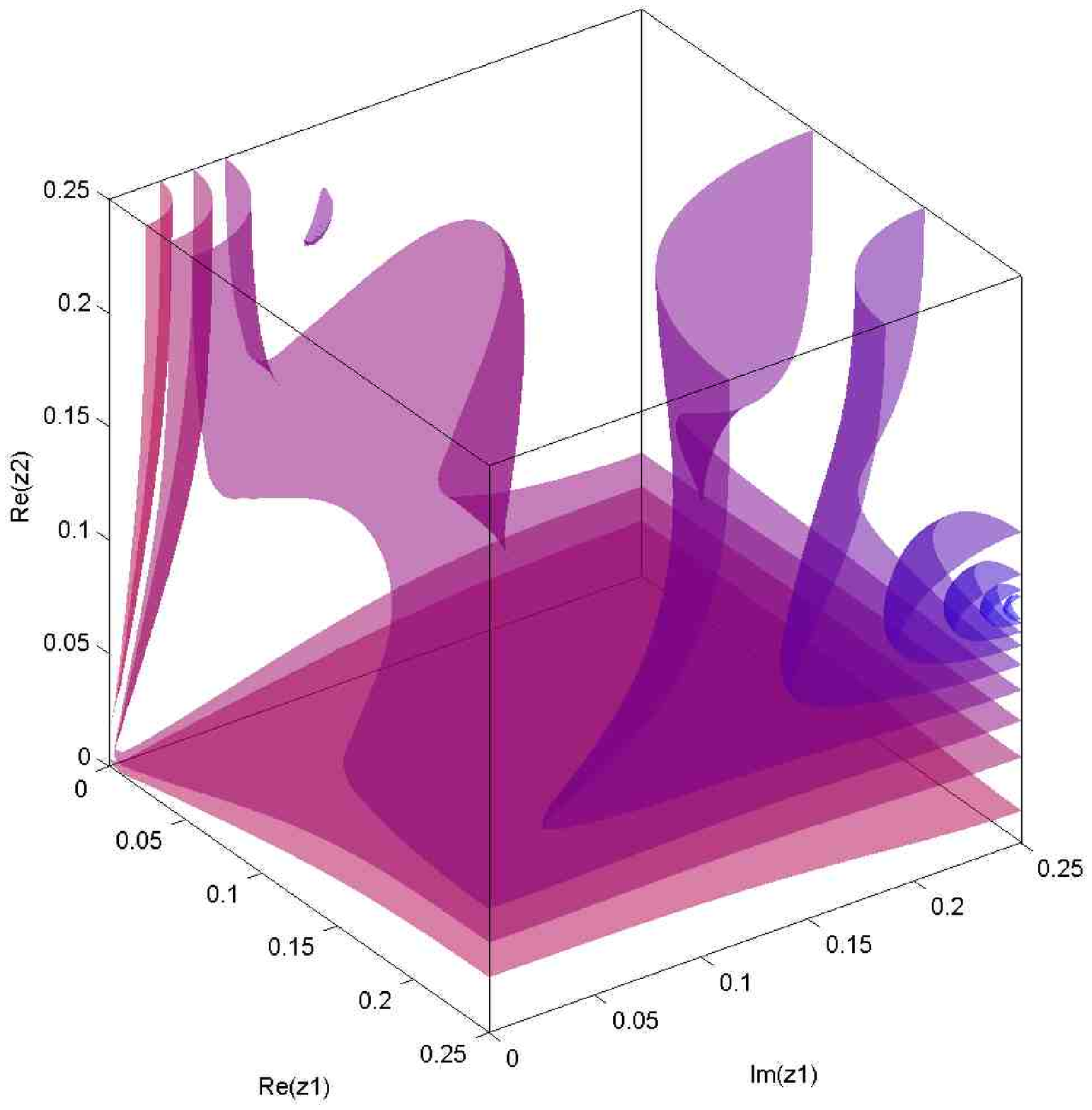,width=3.2in}\psfig{file=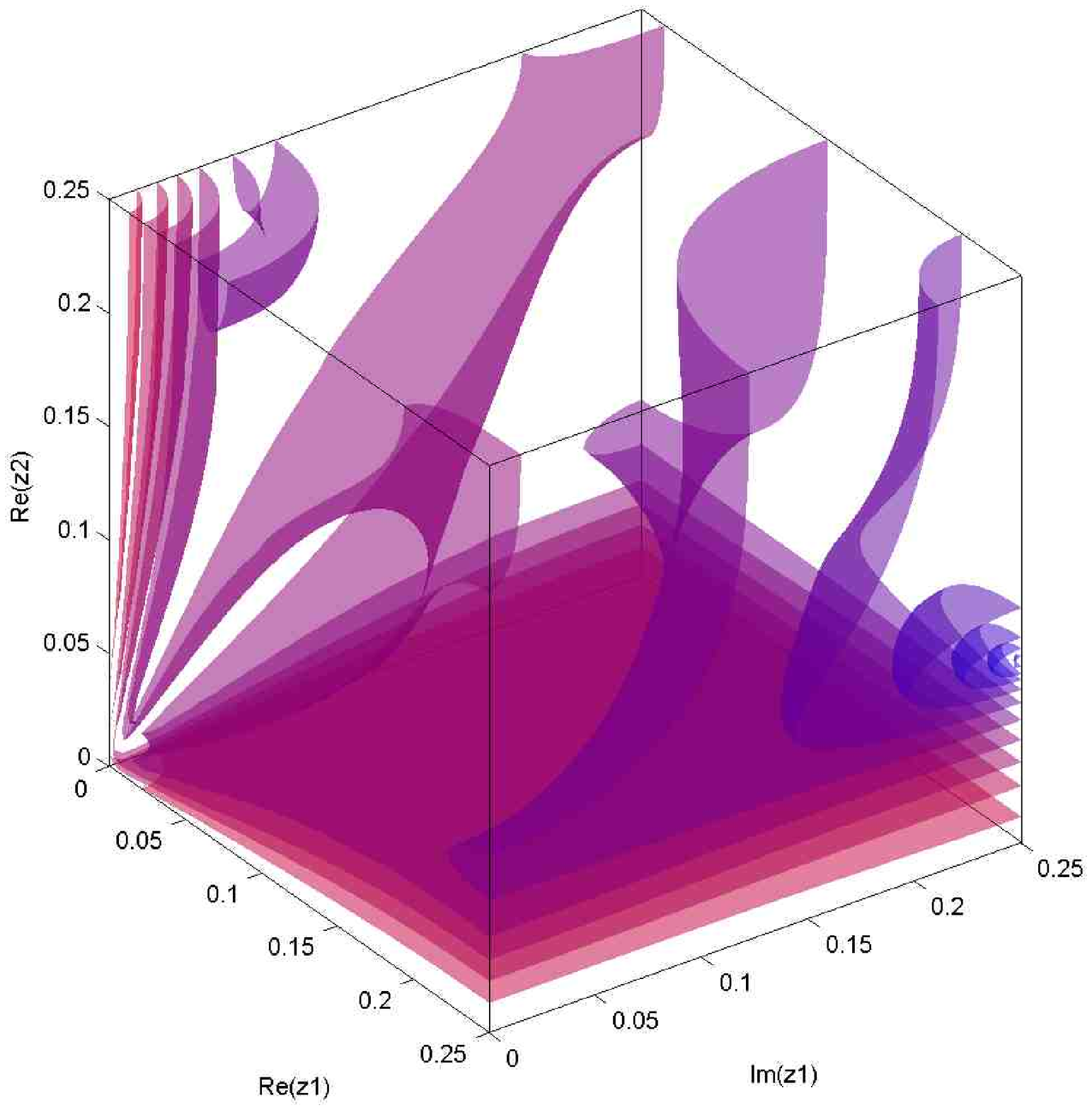,width=3.2in}}
\caption{As in figure \ref{fig:iso1}, but for $\alpha=0.61,0.72,0.85,0.92$.
\label{fig:iso2}
}
}

On a Ricci-flat manifold, the simplest non-trivial curvature invariant one can construct is the square of the Riemann tensor (sometimes referred to as the Kretschmann invariant). In four dimensions, this happens to be proportional to the Euler density $\rho$:
\begin{equation}
\chi=\int\sqrt g\,\rho, \qquad \rho = \frac1{8\pi^2}R_{i\bj k\bar l}R^{i\bj k\bar l}.
\end{equation}
Thus its integral over the manifold is fixed, in this case at
24. Figures \ref{fig:iso1} and \ref{fig:iso2} show surfaces of
constant $\rho$ on the three-dimensional slice $\I z^2=0$ at eight
different values of $\alpha$.\footnote{Animations of these plots as a function of $\I z^2$, showing the entire four-dimensional geometry, are available for download at \href{http://schwinger.harvard.edu/~wiseman/k3/}{\tt http://schwinger.harvard.edu/~wiseman/k3/}.} At the smallest value ($\alpha=0.03$), the curvature is
highly concentrated near the fixed point, and is spherically
distributed; here the metric closely approximates the Eguchi-Hanson
metric (for which the isosurfaces of $\rho$ are spherical in these
coordinates, although the full geometry is only axisymmetric). As we
increase $\alpha$, the curvature spreads out from the fixed
point. However, it does not diffuse uniformly over the
manifold. Instead, starting in the $\alpha=0.72$ figure (third to
last), it gathers along the $z^1=0$ and $z^2=0$ planes. These are the
rational curves, discussed in the last subsection (and referred to as
$\hat c_1$ and $\hat c_2$ in Appendix \ref{homology}), that shrink to
zero size as $\alpha$ approaches 1. By symmetry, the curvature also
accumulates at the 6 other shrinking rational curves, $\{z^1=C\}$ and
$\{z^2=C\}$ for $C=\frac12,\frac i2,\frac12+\frac i2$.

\FIGURE{
\centerline{\psfig{file=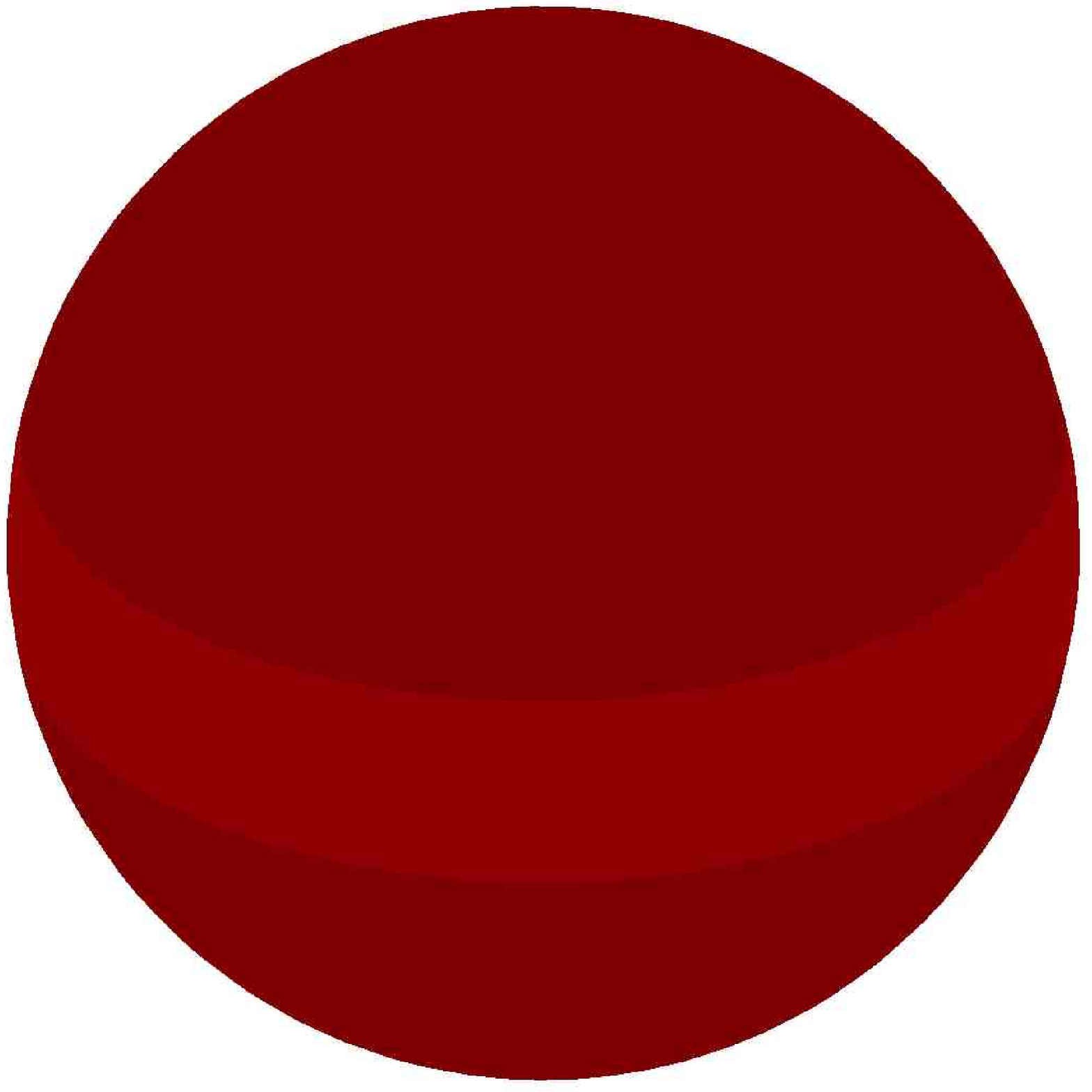,width=1.8in}\psfig{file=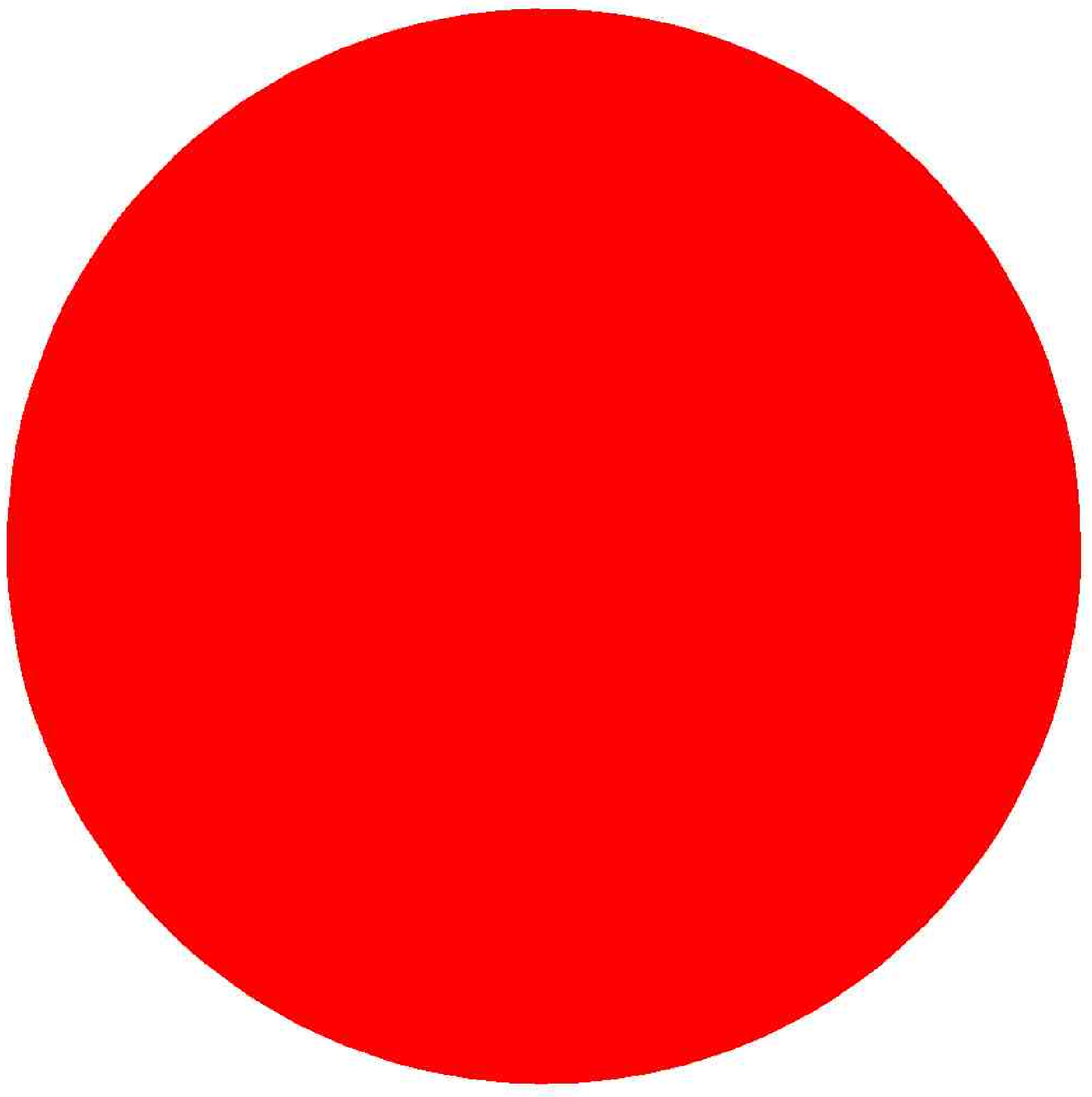,width=1.8in}\psfig{file=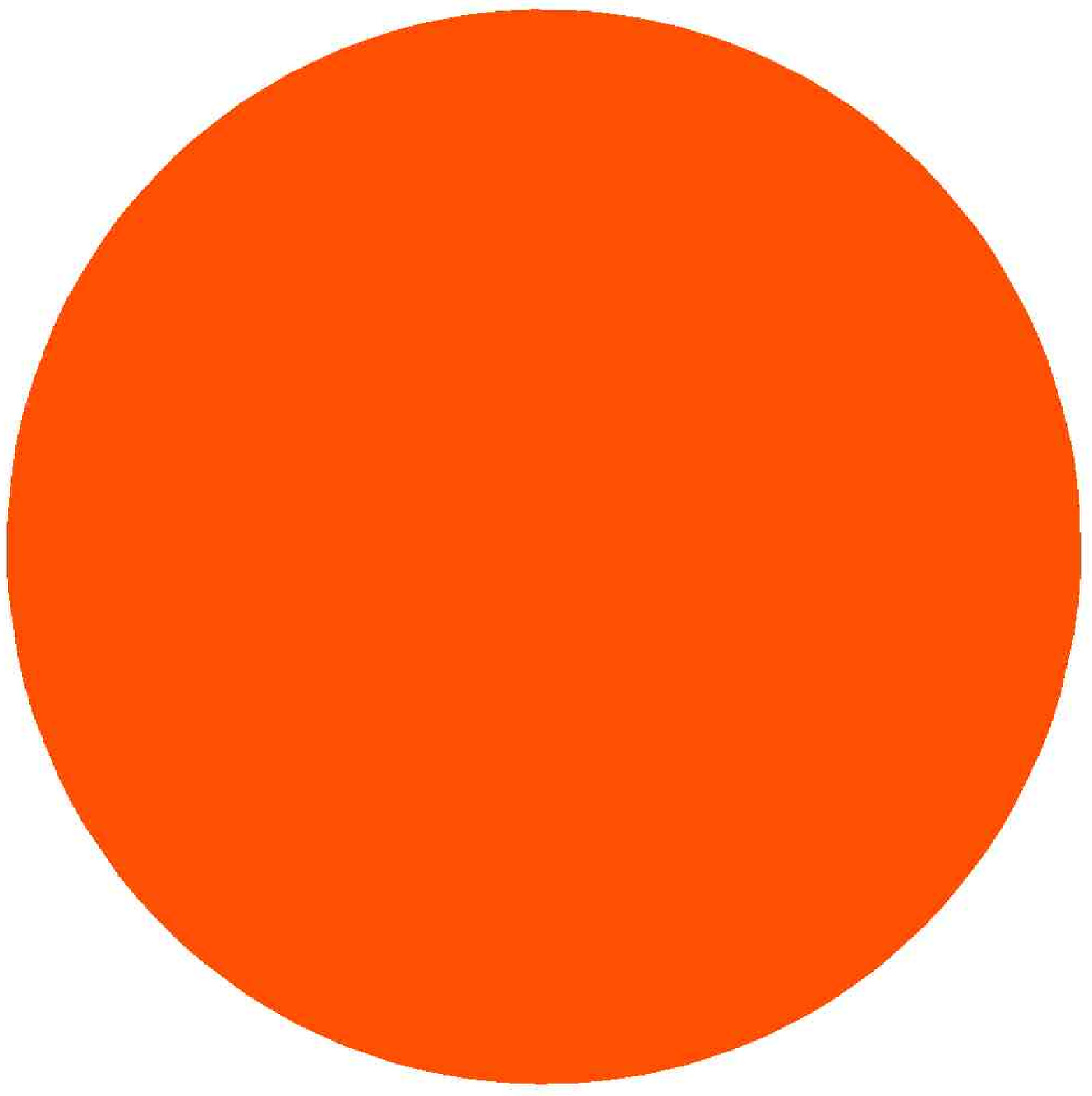,width=1.8in}\psfig{file=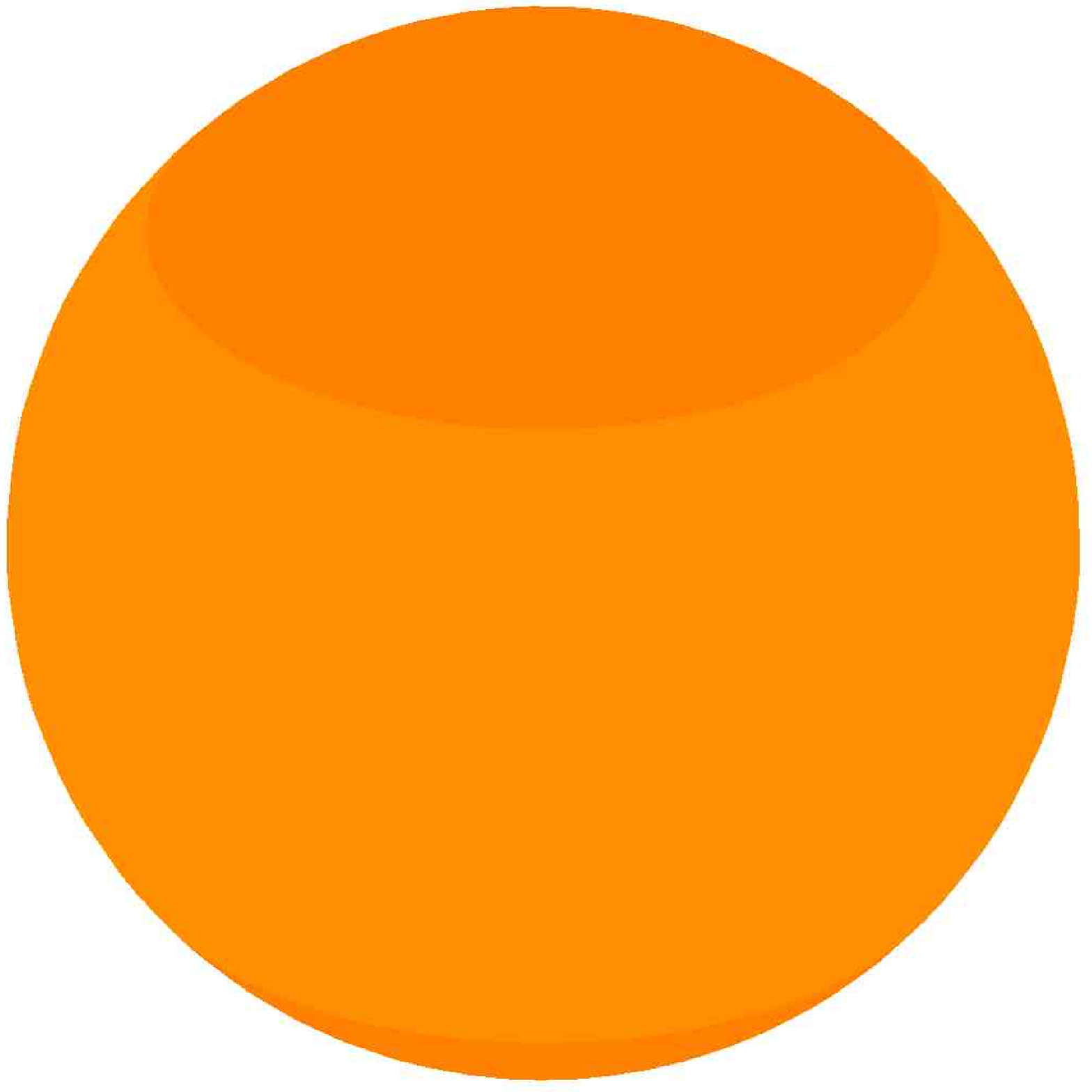,width=1.8in}}
\centerline{
\psfig{file=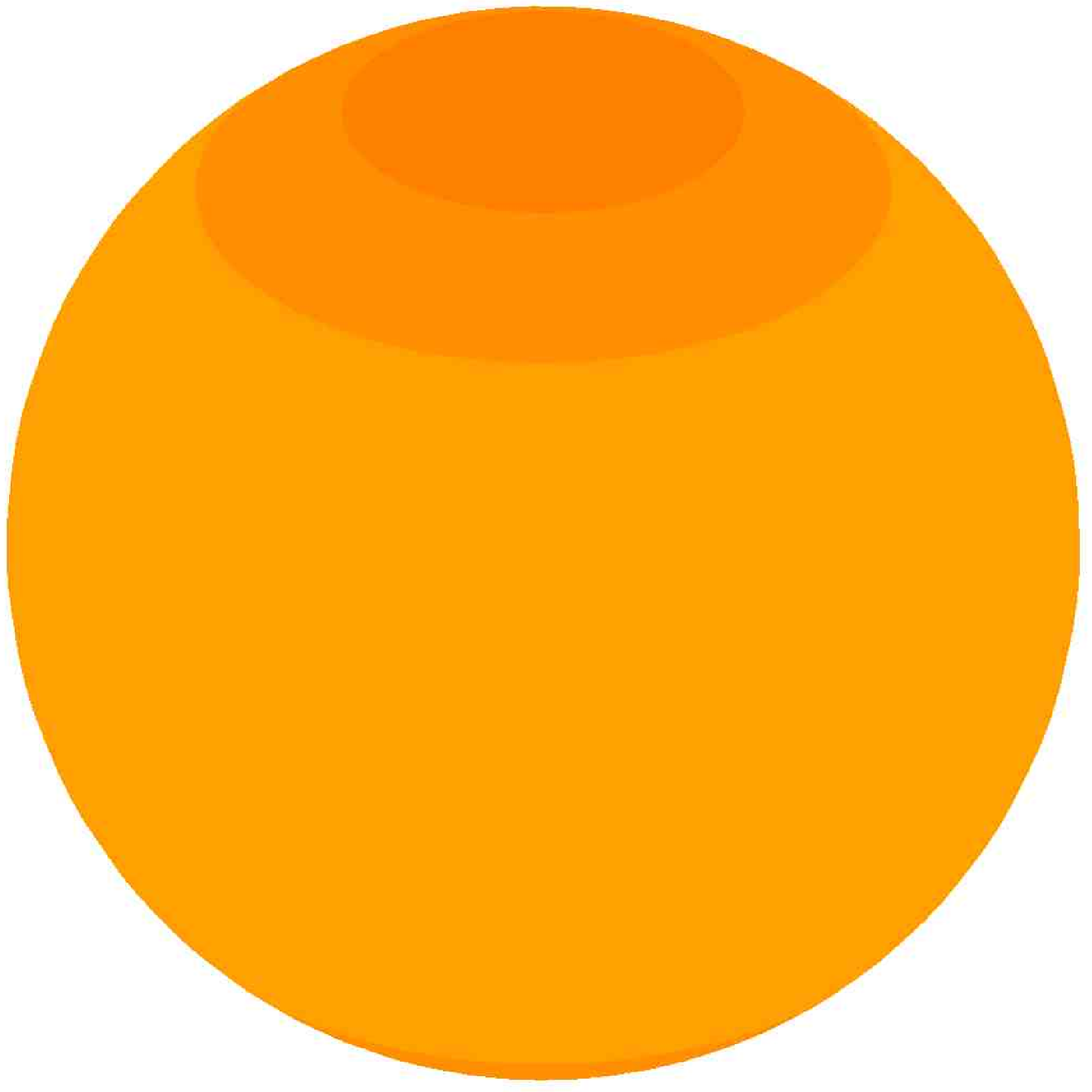,width=1.8in}\psfig{file=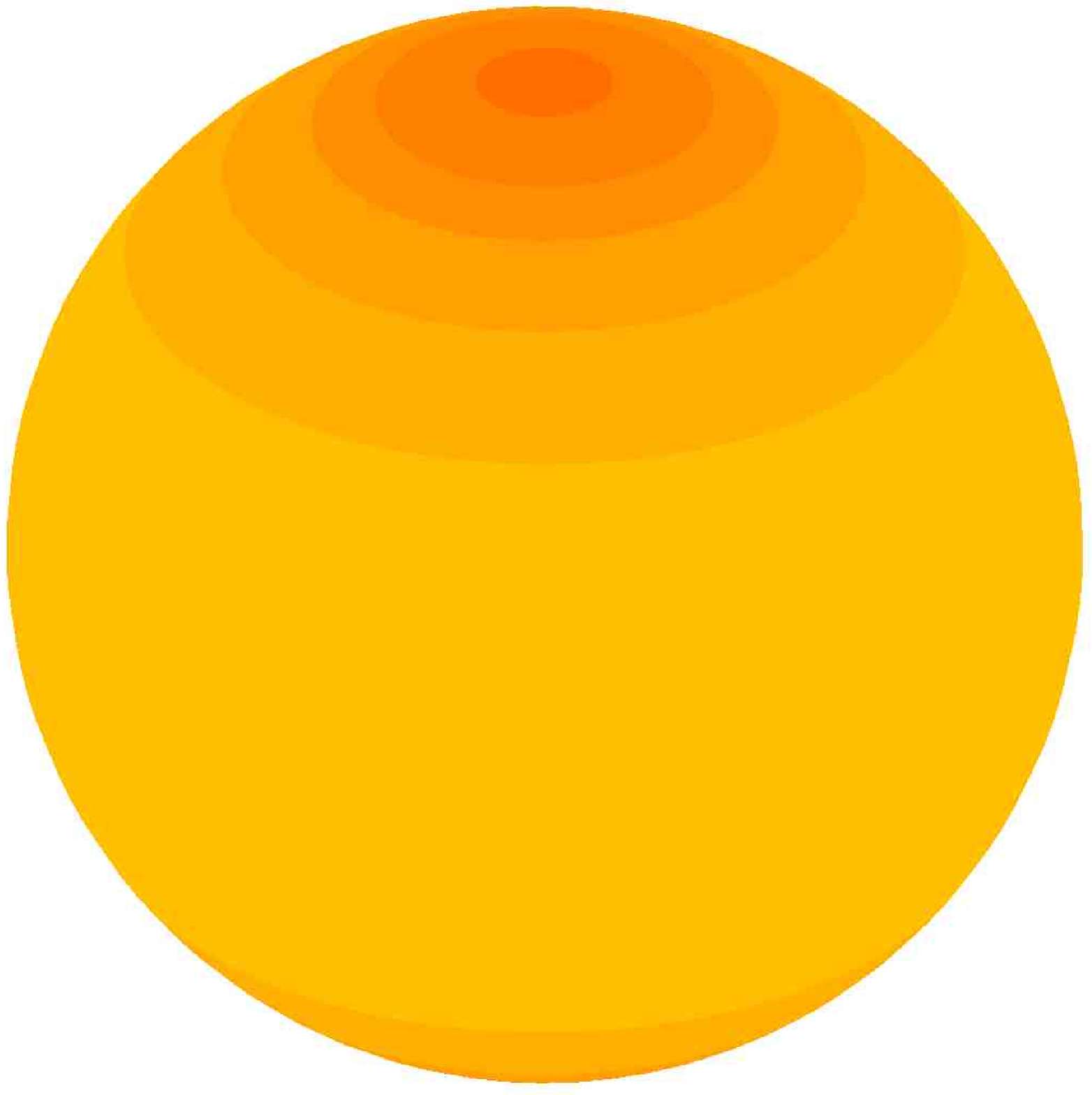,width=1.8in}\psfig{file=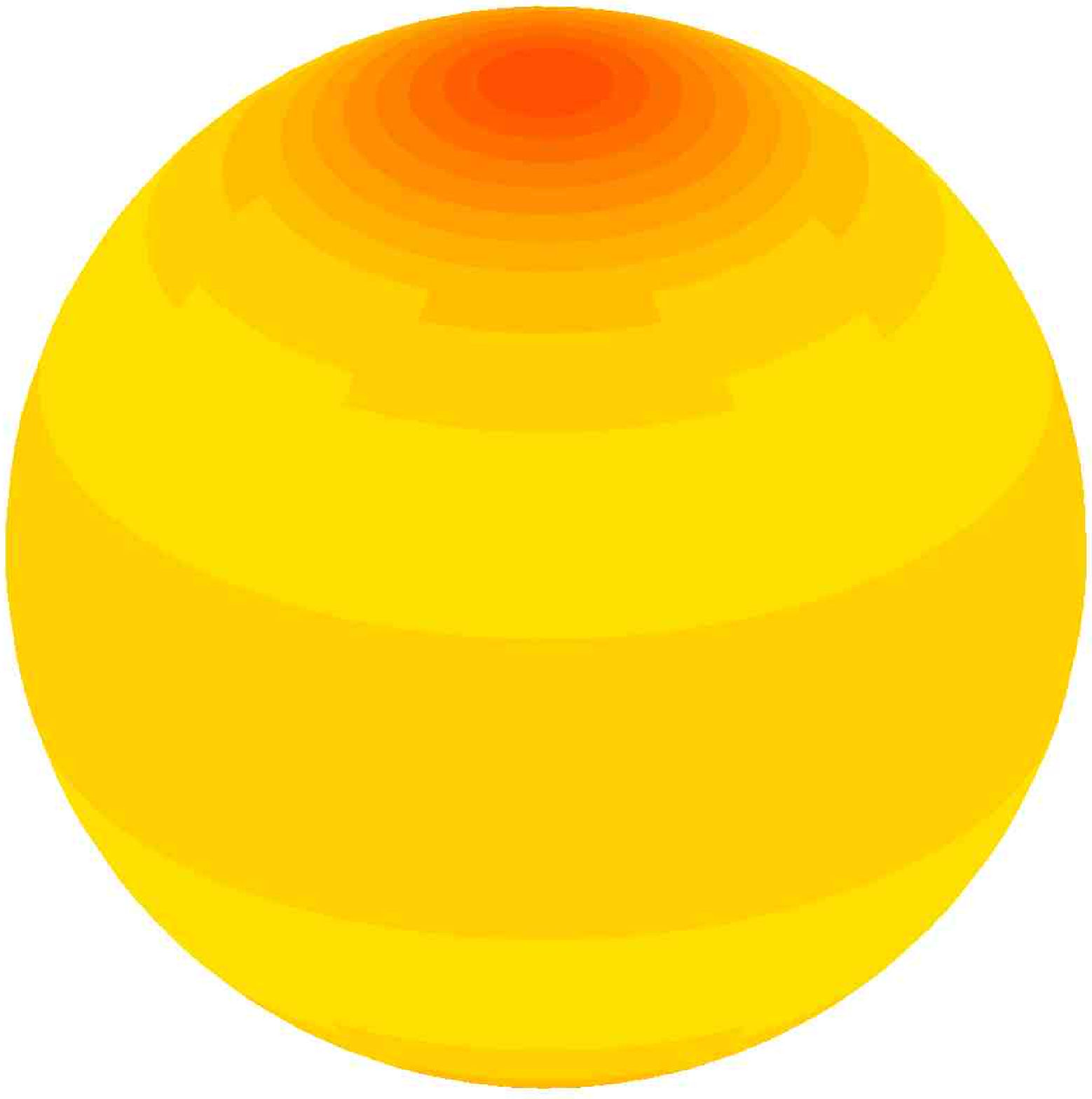,width=1.8in}\psfig{file=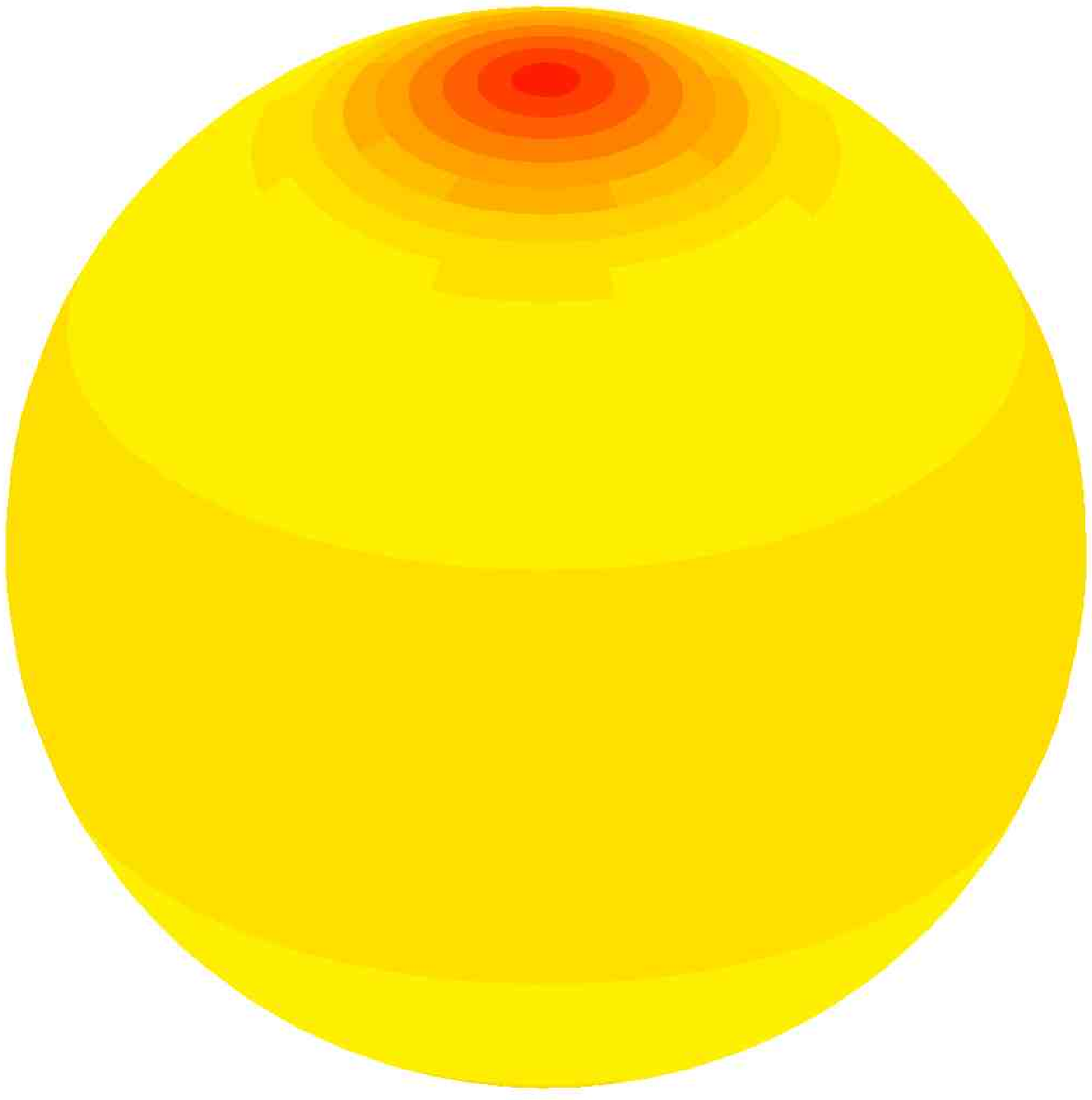,width=1.8in}}
\centerline{ \psfig{file=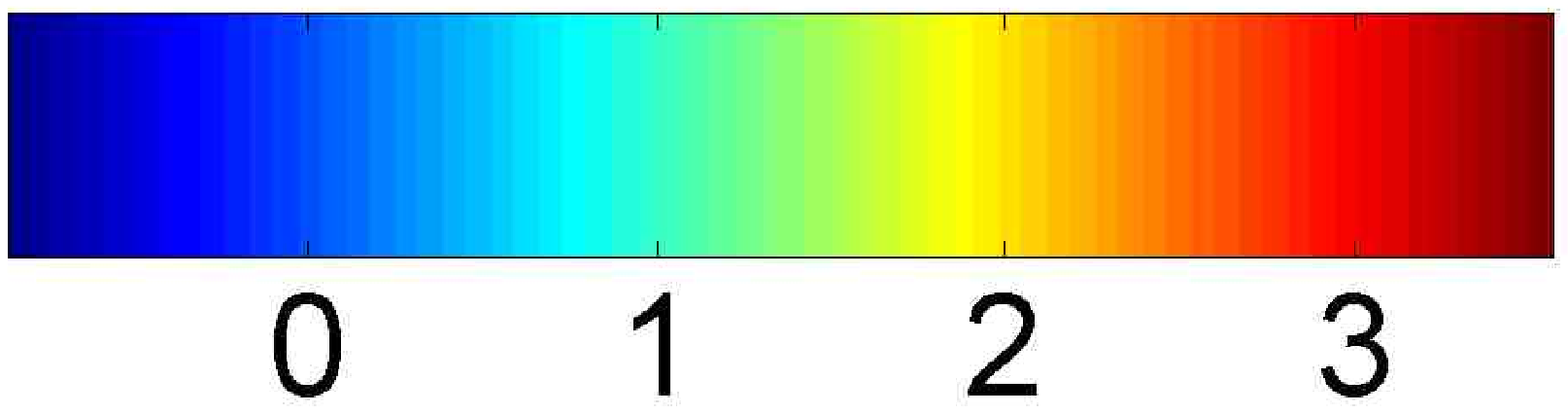,width=2in} }
\caption{Log (base 10) of the Ricci scalar of the induced metric on the exceptional
divisor $w=0$. Points on the sphere are mapped stereographically from their $y$ coordinate, with the north and south poles corresponding to $y=0$ and $y'=0$. Values of $\alpha$ are as in figures \ref{fig:iso1} and \ref{fig:iso2}. The average value of the Ricci scalar over the sphere decreases as its area increases, as is clear from the figure.
\label{fig:EHsphere}
}
}

In the $z^i$ coordinate system used in figures \ref{fig:iso1} and
\ref{fig:iso2}, the exceptional divisor (of the original orbifold) is
represented as a single point, namely the origin. This is misleading,
since in fact it is topologically an $S^2$. It's interesting to study
how its geometry changes as we vary $\alpha$. The induced metric on
the exceptional divisor of the Eguchi-Hanson geometry is that of a
round sphere.  We therefore expect the same to be true for the
exceptional divisor of the Kummer surface at small $\alpha$. In figure
\ref{fig:EHsphere}, we plot the Ricci scalar of the induced metric on
the exceptional divisor. For small values of $\alpha$, it is indeed
uniform. However, as $\alpha$ increases, it becomes non-uniform (its
integral is of course fixed at $8\pi$): the sphere is becoming
prolate. The poles, where the curvature is highest, are at $y=0$ and
$y'=0$, in other words where the exceptional divisor intersects the
planes $z^1=0$ and $z^2=0$ respectively; these are the points that are
closest to neighboring exceptional divisors.

\FIGURE{
\centerline{\psfig{file=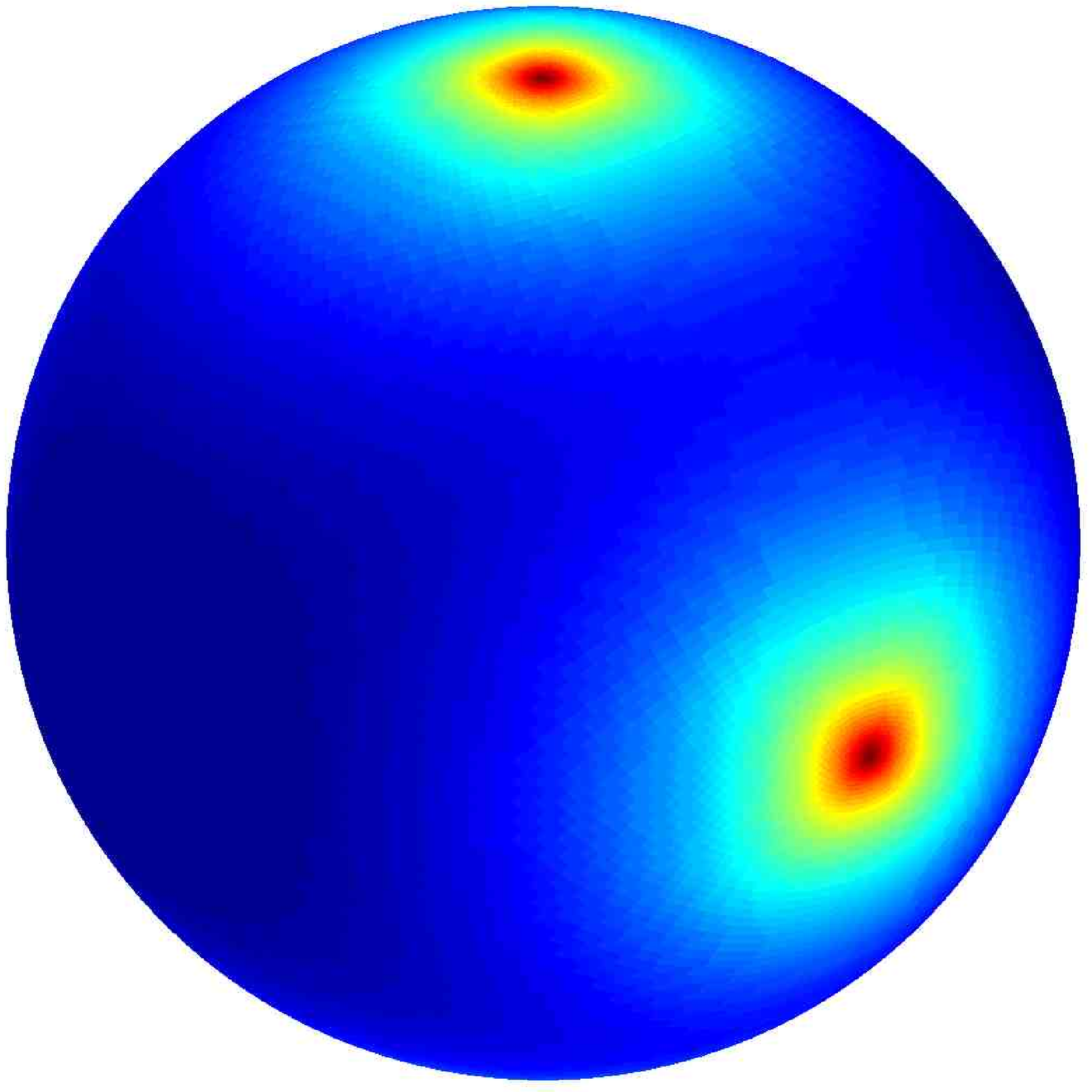,width=1.8in}\psfig{file=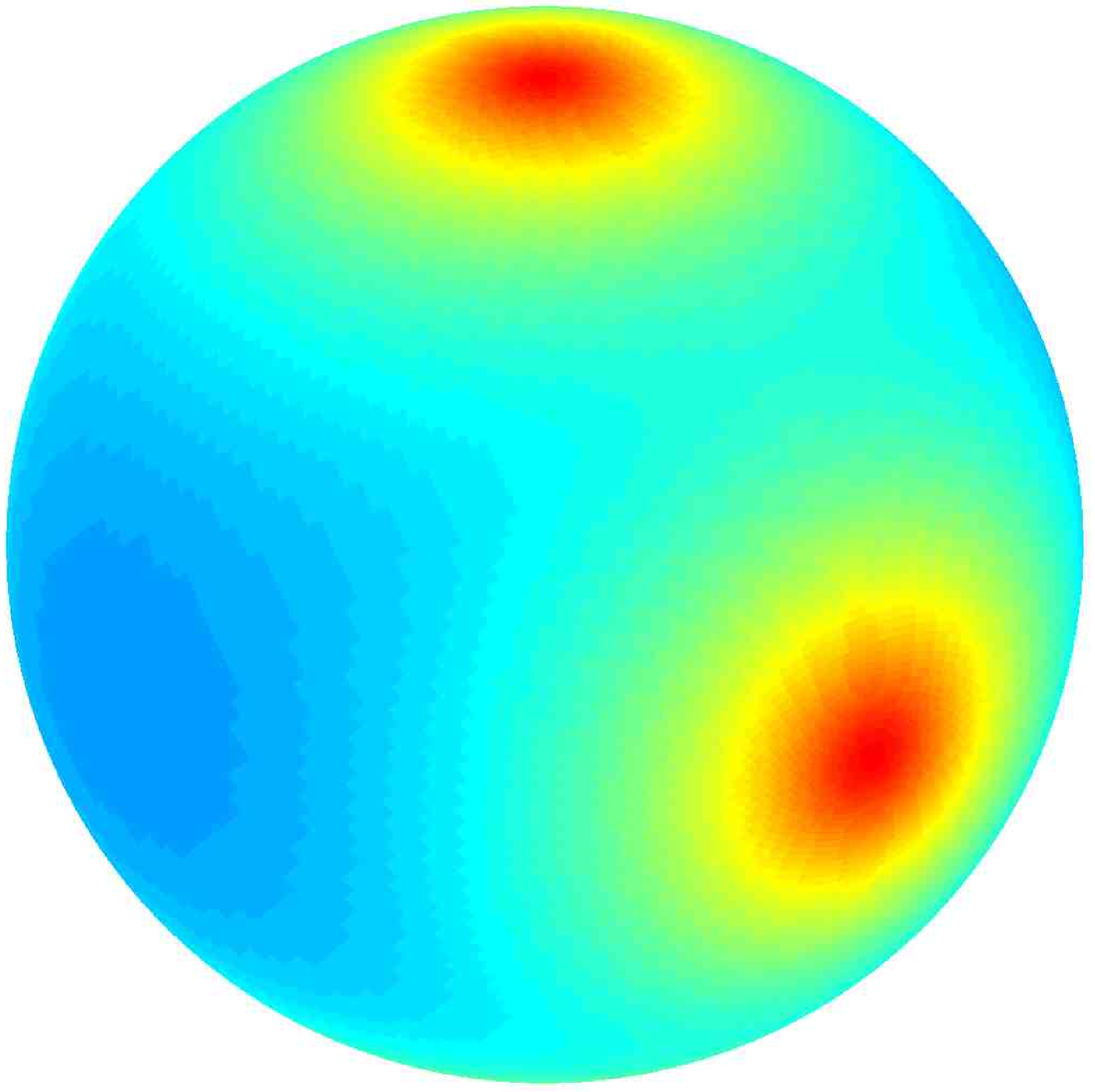,width=1.8in}\psfig{file=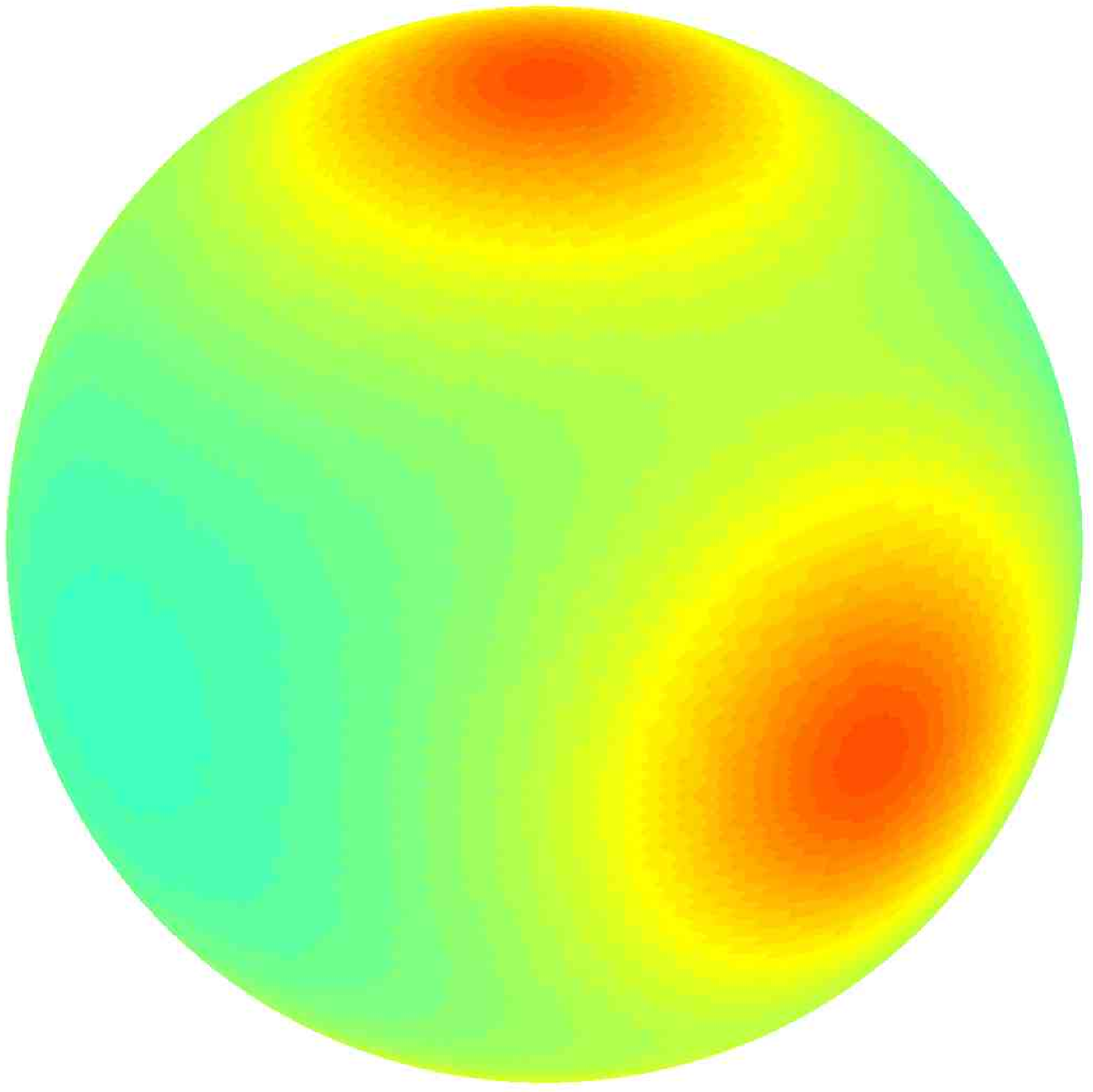,width=1.8in}
\psfig{file=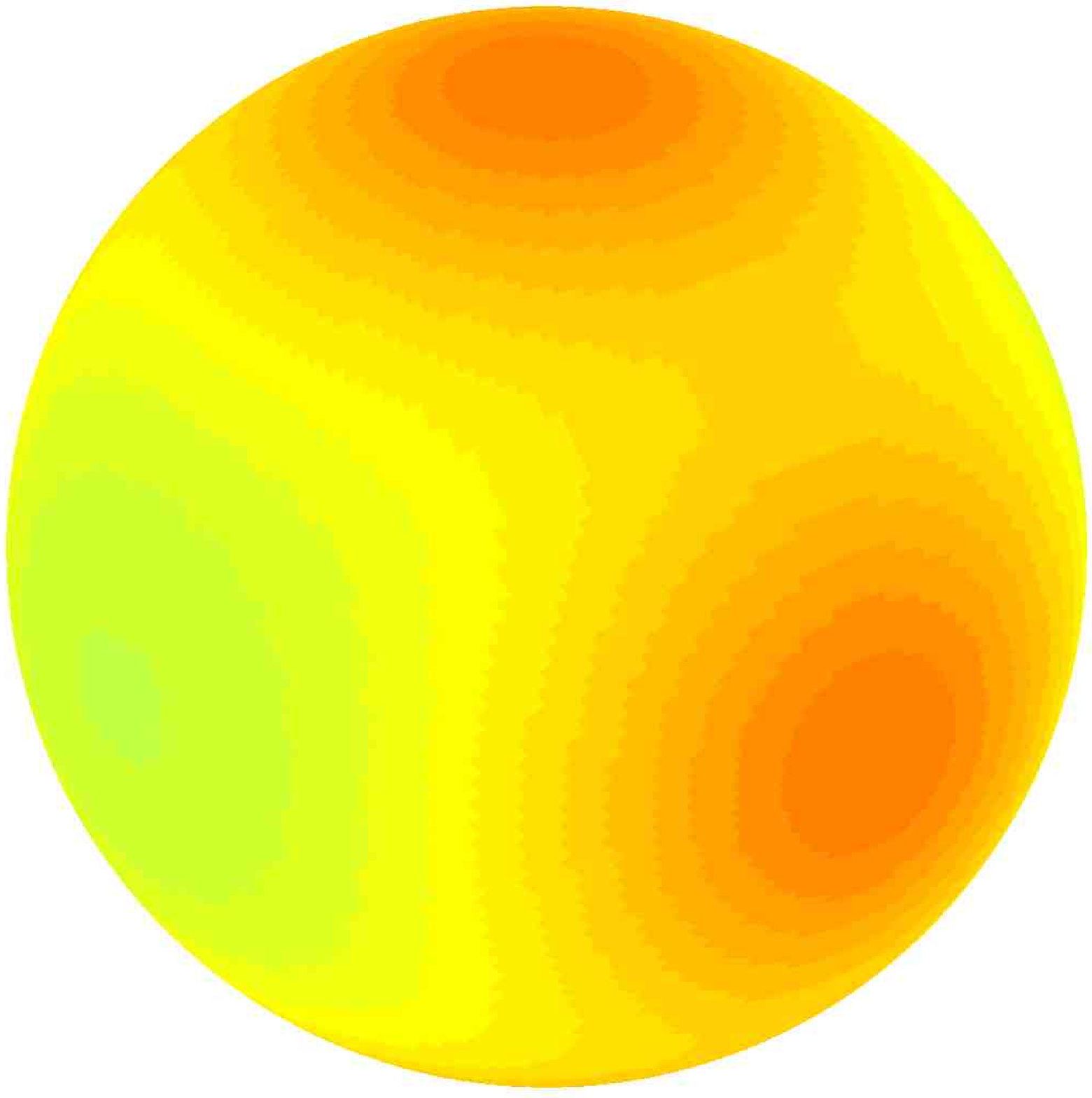,width=1.8in}}
\centerline{\psfig{file=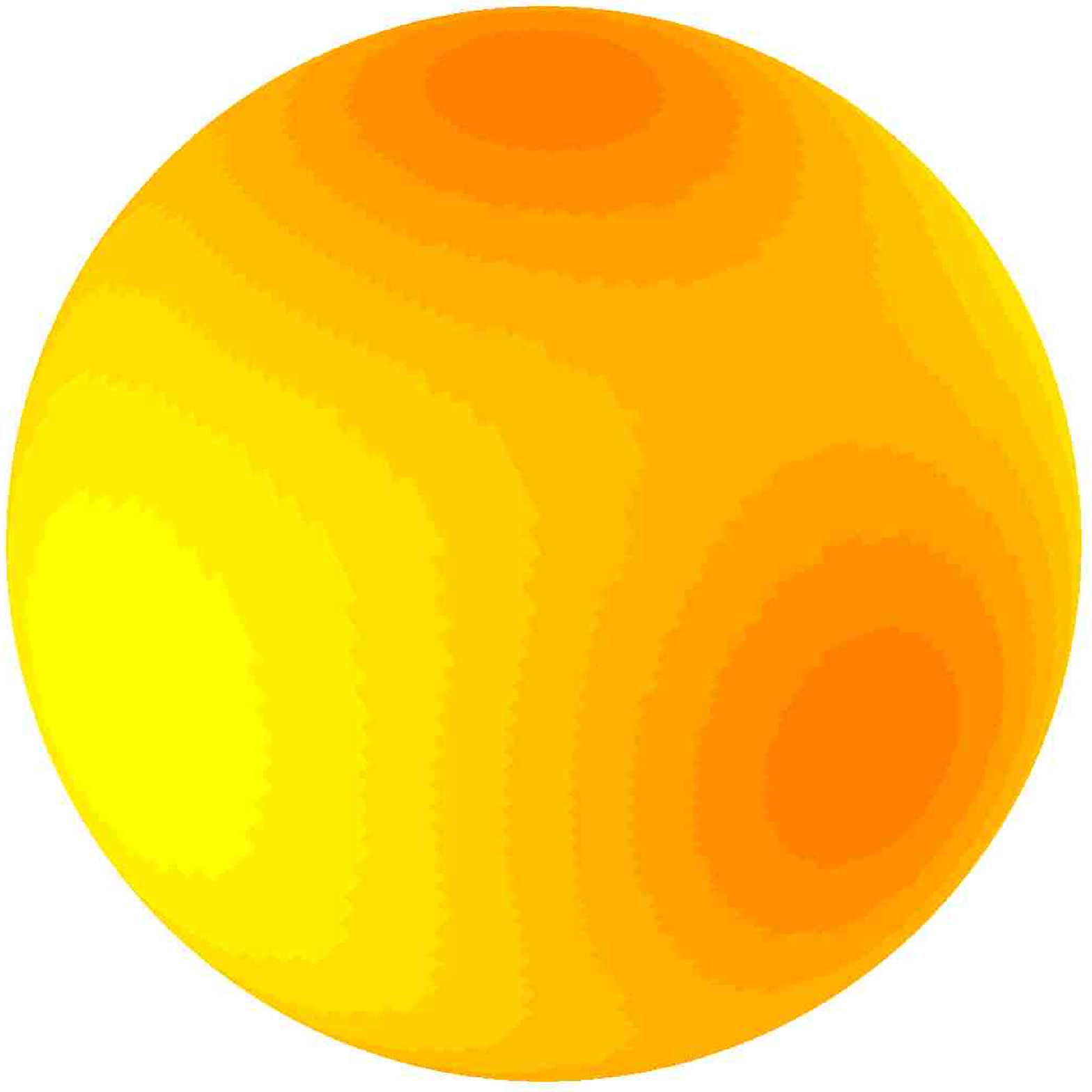,width=1.8in}\psfig{file=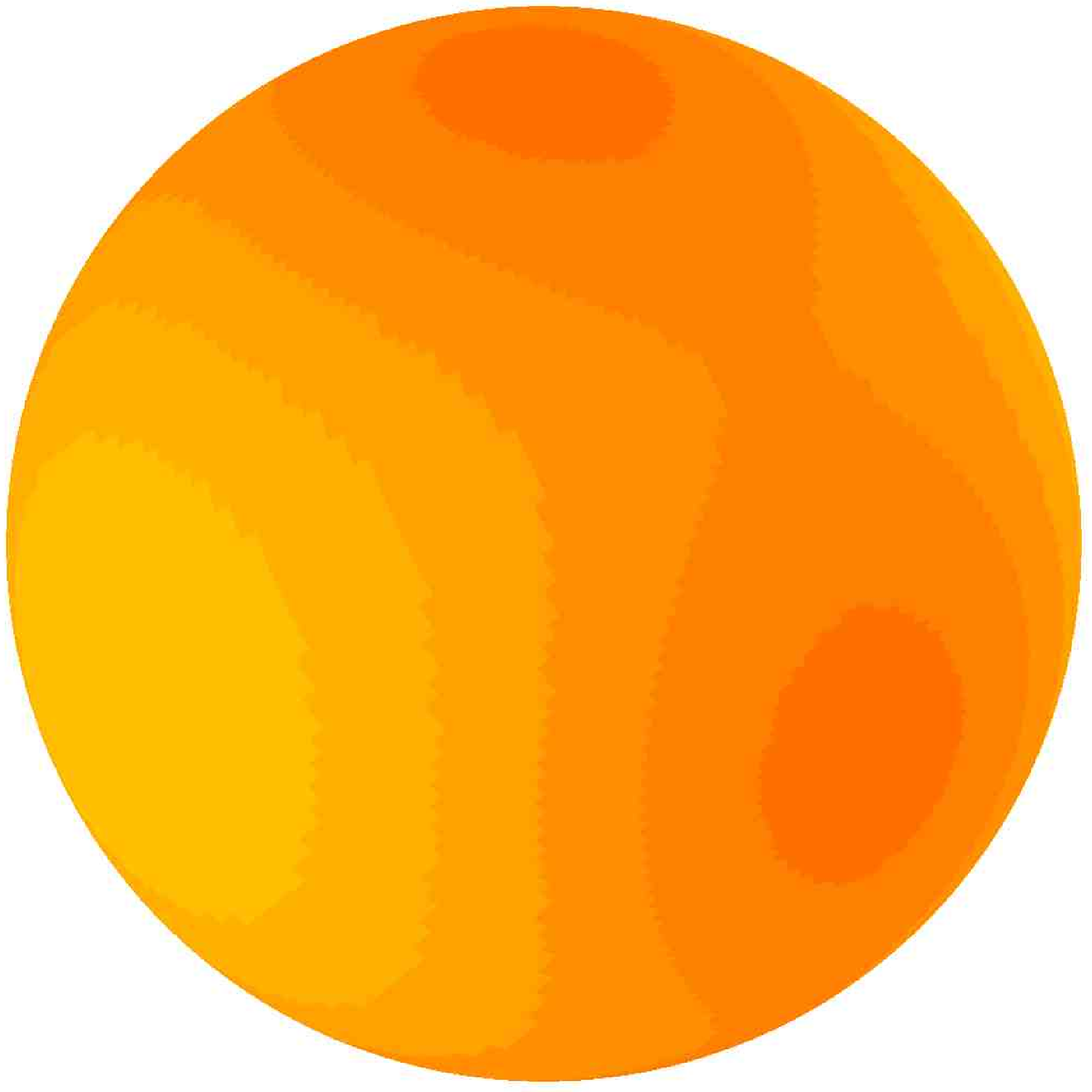,width=1.8in}\psfig{file=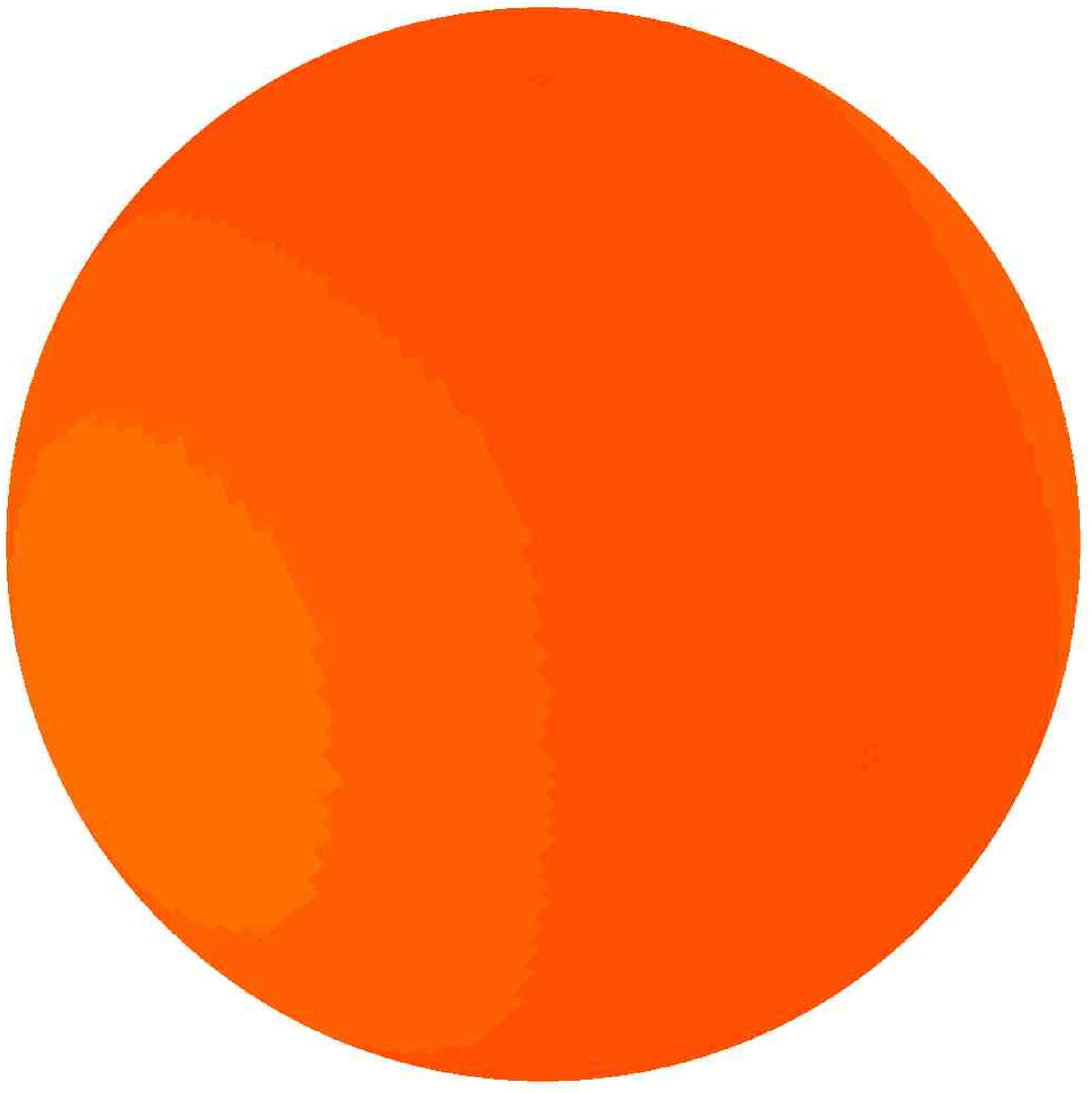,width=1.8in}\psfig{file=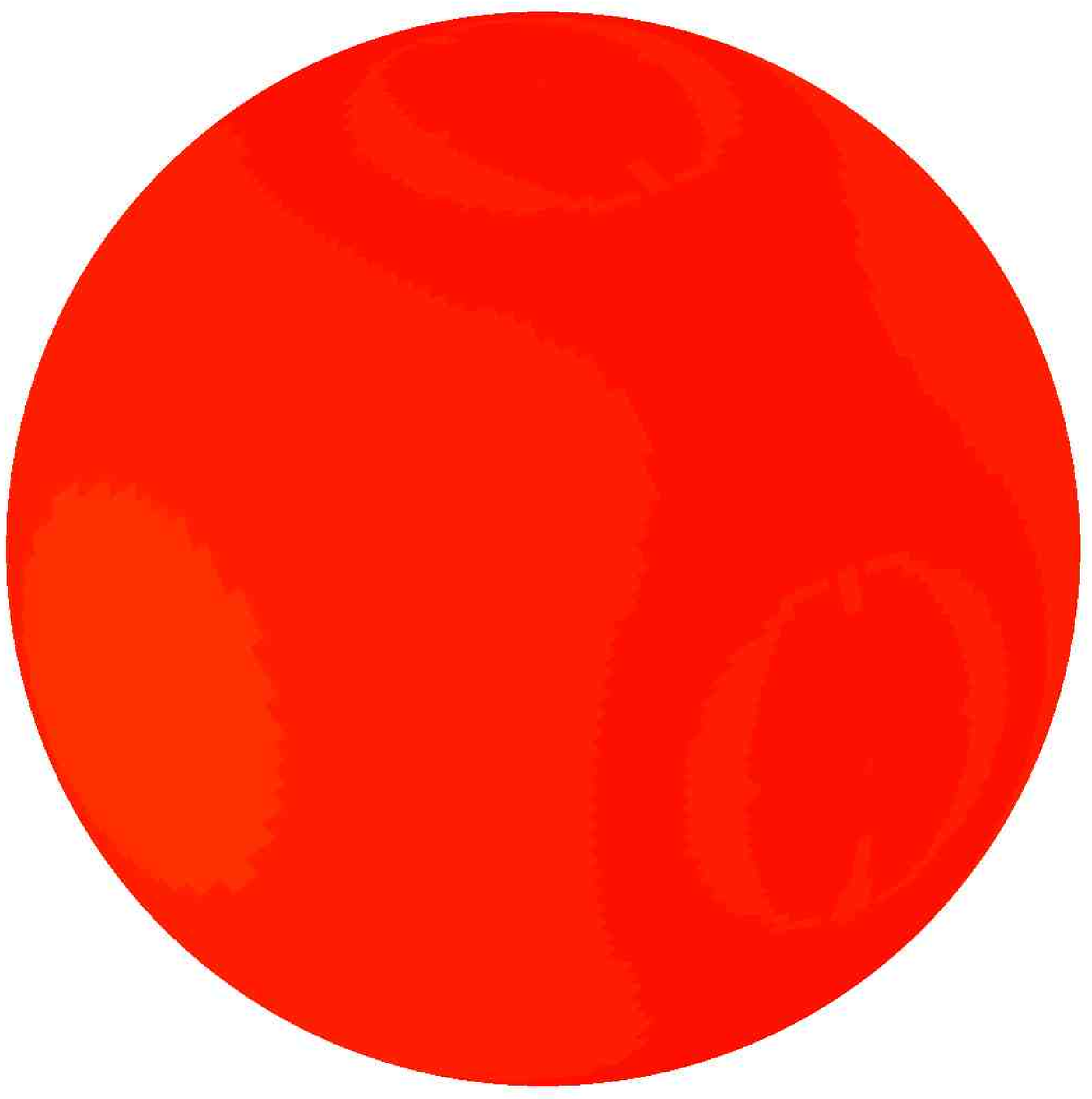,width=1.8in}}
\centerline{ \psfig{file=figs/Colorbar.ps,width=2in} }
\caption{Log (base 10) of the Ricci scalar of the induced metric on
the rational curve $\{z^1=0\}$. (The Ricci scalar is everywhere
positive.) Using the function $\hat y(z^2)=-i\wp(z^2)/\wp(\frac12)$
(with $\wp$ the appropriate Weierstrass elliptic function), points are
mapped from $T^2/\Z_2$ to the complex plane; $\hat{y}$ is then mapped
stereographically to the sphere. The maxima of the Ricci scalar are at
$z^2=0,\frac12,\frac i2,\frac12+\frac i2$. Values of $\alpha$ are as
in figures \ref{fig:iso1} and \ref{fig:iso2}. The average value of the
Ricci scalar over the sphere increases as its area shrinks (see
equation \eqref{neweh}), as is clear from the figure.
\label{fig:Torussphere}
}
}

\FIGURE{
\centerline{\epsfig{file=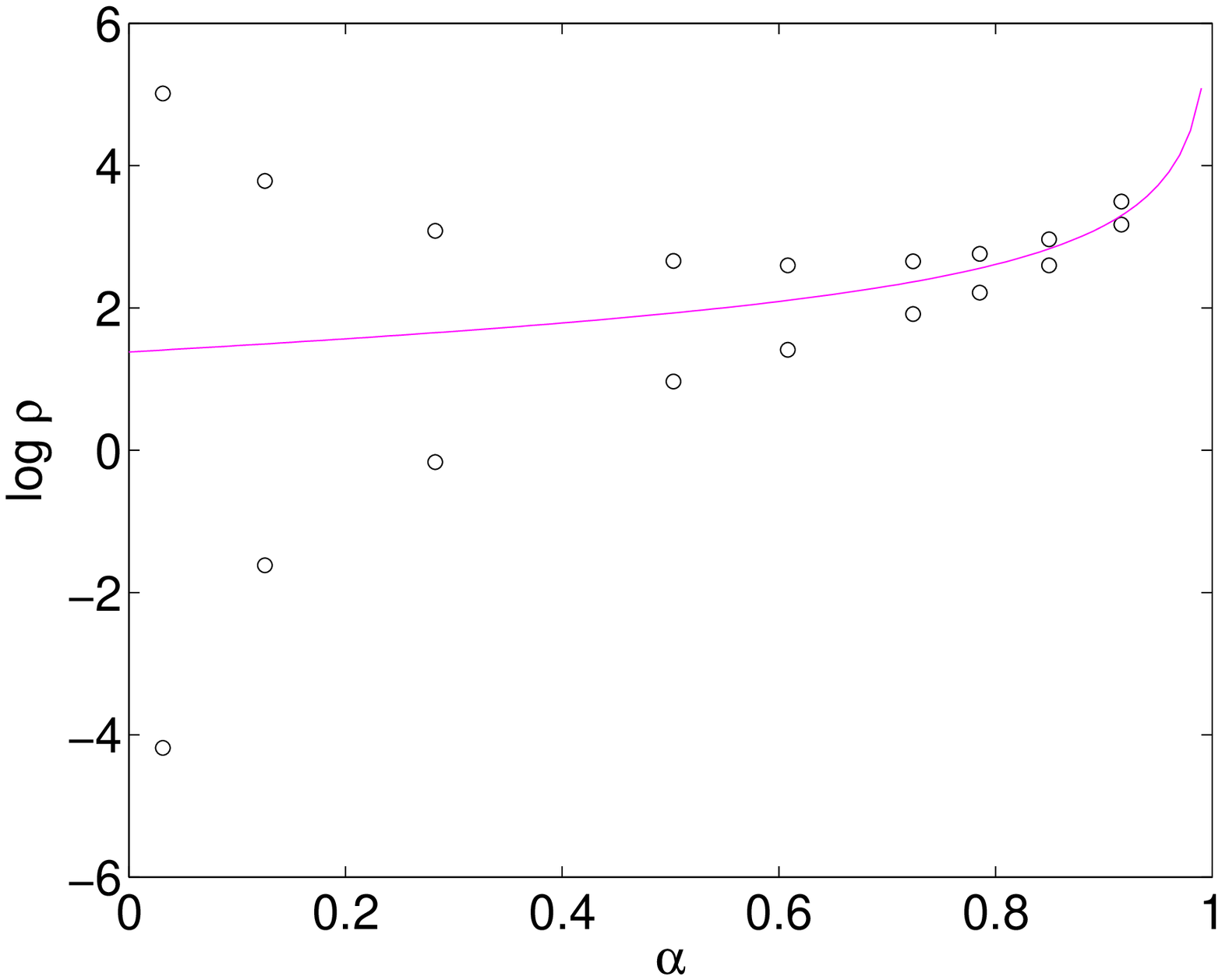,width=4.5in}}
\caption{Maximum and minimum values of the Euler density $\rho$ on the
rational curve $\{z^1=0\}$ as a function of $\alpha$. The solid curve
is the value of $\rho$ on the exceptional divisor of an Eguchi-Hanson
geometry whose exceptional divisor has area $\hat A$.
}
}

Similarly, it is interesting to study the induced geometry of the
shrinking rational curves, for example the curve $\{z^1=0\}$, as shown
in figure \ref{fig:Torussphere}. At $\alpha=0$ its geometry is that of a flat $T^2$
orbifolded by $\Z_2$, which is topologically $S^2$ but with all the
curvature concentrated at the 4 fixed points $z^2=0,\frac12,\frac
i2,\frac12+\frac i2$ (picture a square envelope). As $\alpha$
increases, the curvature spreads out around the sphere, which becomes
rounder and rounder. In the limit $\alpha\to1$, the geometry in the
vicinity of the shrinking rational curve should approach that of an
Eguchi-Hanson metric whose exceptional divisor has area $\hat A=\frac12b^2-2\pi a^2$. Indeed, we see that for values of $\alpha$ approaching 1, the sphere
becomes almost completely round. As another test that the geometry is
approaching Eguchi-Hanson, in figure 4 we plot the maximum and minimum
values of $\rho$ on the curve against $\alpha$. On an Eguchi-Hanson, the Euler density is
constant on the exceptional divisor; the solid curve in that plot is
the value of $\rho$ on the exceptional divisor of an Eguchi-Hanson of the appropriate size.
One sees that as $\alpha$ approaches 1 the maximum
and minimum values of $\rho$ approach each other and the Eguchi-Hanson
value.

\FIGURE{
\centerline{\epsfig{file=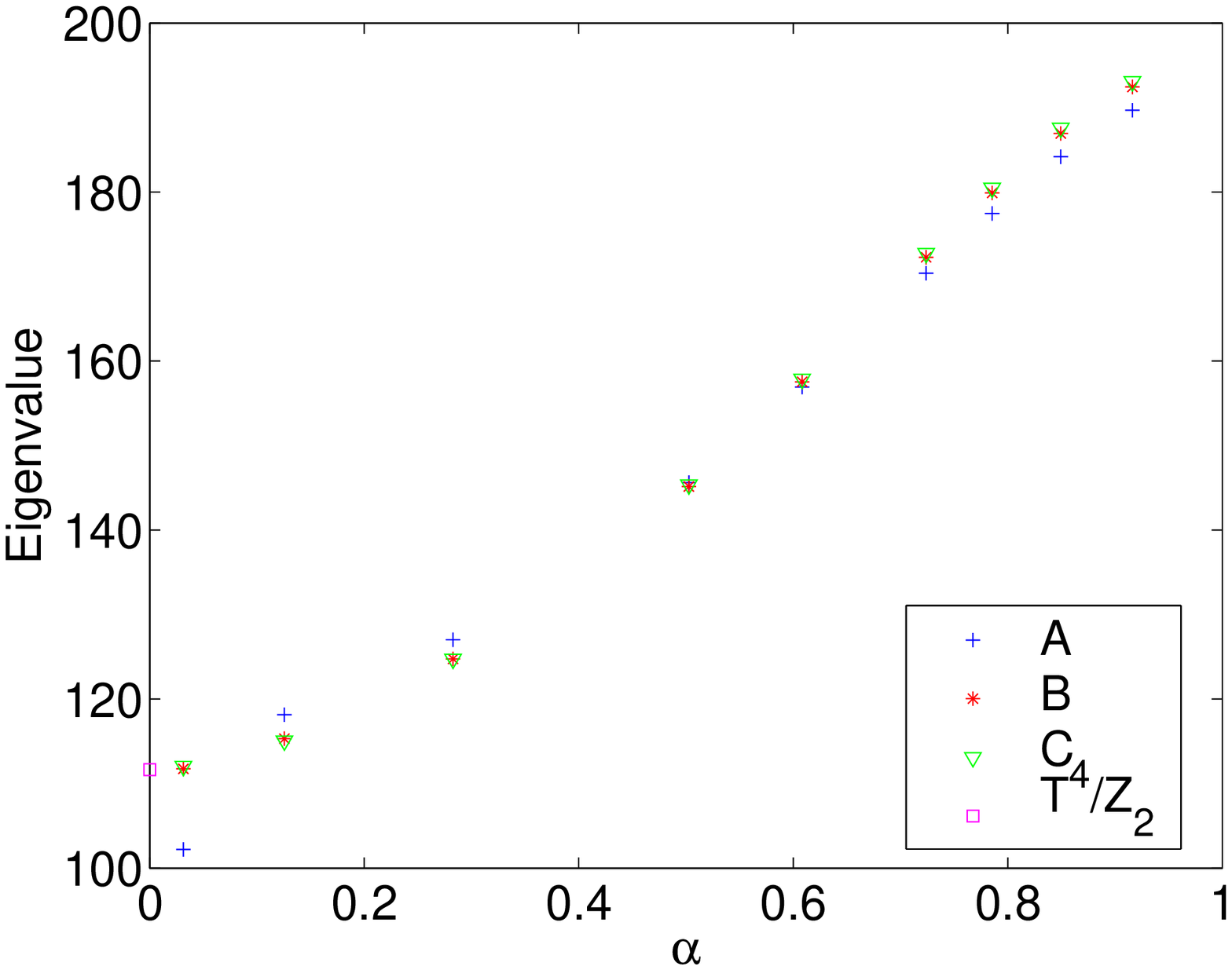,width=4.5in}}
\caption{An eigenvalue of minus the scalar Laplacian as a function of
$\alpha$. Specifically, the lowest eigenvalue whose eigenfunction
is antisymmetric under all 4 generators of the $\Z_2^4$ translation
group of the symmetric Kummer surface. A, B, and C represent increasing lattice
resolutions, as detailed in Appendix \protect\ref{appnum}. The exact
eigenvalue at the orbifold point $\alpha=0$ is also shown.
\label{fig:eval}
}
}

Given the Ricci-flat metric, one can compute the spectrum of various
geometric operators of physical interest, the simplest example being
the scalar Laplacian. As a proof of principle, we used our numerical
metrics to compute a low-lying eigenvalue (and eigenfunction) of it.
The eigenfunctions of the Laplacian on our symmetric Kummer surface
can be classified by their eigenvalues under the $\Z_2^4$ translation
subgroup of its full symmetry group. We calculated the lowest
eigenvalue in the sector with eigenvalue $-1$ under all 4
translations. The results are shown in figure 5. The eigenvalue at the
orbifold point $\alpha=0$, which can be computed exactly, is also
shown for comparison. The fact that the eigenvalue increases with
$\alpha$ can be understood heuristically (at least for small values of
$\alpha$) in the following way. As the rational curves discussed above
shrink, the geodesic distance between a point on the exceptional
divisor, such as $y=w=0$, and its image under each of the discrete
translations decreases, requiring steeper gradients in the
eigenfunction. Finally, we would like to point out that, as we are
studying a long wavelength eigenfunction, even the lowest resolution
(A) computes the eigenvalues to typically within a few percent of the
continuum extrapolated value.

We close this section on a technical note. Careful inspection of the
isosurfaces plotted for the smallest value of $\alpha$ in figure
\ref{fig:iso1} shows a localized deviation from sphericity, caused by
small errors near patch overlaps. Similarly, in figure
\ref{fig:EHsphere} the smallest $\alpha$ sphere is slightly less
uniform than for the next larger value of $\alpha$; in figure
\ref{fig:Torussphere} for the largest value of $\alpha$ we can just
see some ``rings" where small errors are introduced due to coordinate
patch overlaps; and the distance from the extrapolated continuum
values in figure \ref{fig:eval} increases near $\alpha = 0$ or
$1$. Numerical discretization errors are to be expected, and will be
larger where there are higher curvatures. Our solutions containing the
highest curvature regions occur near the two orbifold points $\alpha =
0,1$, explaining why we see the effects mentioned above. Hence, at a
fixed resolution, the global quality of the solutions will not be as
high near these orbifold points, although the local geometry away from
the regions of high curvature should be quite acceptable. Of course,
increasing the resolution, the quality of the solution will improve,
independently of where we are in moduli space. These issues are
discussed in more detail in appendix \ref{continuum}.

\section{Discussion}\label{discussion}

\subsection{Generalizations}\label{generalizations}

We opened this paper with the question of whether it is possible, in
practice, to solve the Einstein equation numerically on a Calabi-Yau
manifold. We have shown that the answer is affirmative when the
Calabi-Yau is a K3 surface with a high degree of discrete symmetry. In
this subsection we will investigate the possibility of generalizing
this accomplishment, first to more general K3 surfaces, then to
Calabi-Yau three-folds, and finally to
\K manifolds with cosmological constant or with matter.

It is clear that in principle our method extends to a general blow-up
of a Kummer surface, and more generally to any K3 surface. It is
merely necessary to implement the topology and complex structure with
complex coordinates defined appropriately on an atlas of charts---for
example one could use patches derived from any algebraic construction
of K3. However, the question remains as to what resolution is
attainable in the absence of large amounts of discrete symmetry. Our
highest resolution, $D$, simulated approximately $2\times10^7\approx80^4$
points. However, due
to the high degree of discrete symmetry our one parameter family of
K3's enjoy, every computed point actually represents $2^9$ points in
the true K3 geometry, and hence we have described the full K3 with an
\emph{effective} resolution of around $400^4$.

On a current high-end desktop computer (with 1 to 2 gigabytes of
memory) one could comfortably increase the total number of points
simulated to $10^8$. For a K3 with no discrete symmetry this would
yield a resolution of $100^4$ for the full K3 geometry.  To estimate
how accuately this could represent the metric, we may compare it to
our intermediate resolution, B. With a linear resolution 4 times lower
than that for D, the effective resolution for the full geometry is
also about $100^4$. As seen in subsection \ref{results} and
appendix \ref{appnum} we find that resolution B adequately reproduces
the geometry and derived properties, such as the low wavelength
eigenmodes of the Laplace operator, provided one is not too close to
the edge of the \K cone. For example, in figure \ref{fig:eval} one
sees that run B computed the Laplacian eigenvalue with an accuracy of
around 1\% compared to the extrapolated continuum value. Near the edge
of the \K cone, where regions of high curvature develop in the
manifold, the best strategy may be to combine numerical with analytic
techniques. For example, at values of the moduli near an orbifold
point, one could patch an analytic Ricci-flat metric, such as the
Eguchi-Hanson metric, into the region where the high curvature is
developing.

If we now wish to move to a Calabi-Yau three-fold, $10^8$ points
translates to a mere 20 points linearly in each direction. This is
over a factor of 2 less in linear resolution than the lowest effective
resolution used in this work (run A, which had an effective resolution of
$50^4$ for the full geometry). This might be acceptable if one were
well away from the \K cone edge, but is certainly rather low.  On the
other hand, if one were to consider a highly symmetric three-fold,
such as a symmetric blow-up of $T^6/\Z_3$, then one would again expect
to attain an effective resolution of around $100$ points linearly in
the full geometry, and thus again expect accuracy comparable to, or
better than, our resolution B.

Moving to the general three-fold appears to be a very challenging
task. As we discuss in Appendix \ref{appnum}, processing time actually
scales rather well with increasing dimension. Instead the problem is
limited by storage. In six real dimensions any appreciable increase in
linear resolution is extremely costly. Thus whilst $20^6$ points would
be possible on a desktop computer, $40^6$ requires 64 times more
memory and is already beyond the abilities of modest clusters. Often a
tough computational problem becomes easy in time, as computer memory
and speed have closely followed Moore's prediction of a doubling every
2 years. However, even assuming that Moore's law continues to hold, it
will require 12 years to increase the linear resolution by a factor of
2 for the three-folds. It is therefore clear that in order to tackle
the general three-fold, one must employ considerably more
sophisticated discretization schemes than we have used here,
presumably adapting the lattice points to regions of high
curvature. Whilst adaptive grids are difficult to implement in the
elliptic context, one can easily use fixed grids that increase
resolution in areas where curvature is expected to be high.

To summarize, we expect that for a general K3 surface one can obtain
very satisfactory results provided one does not wish to probe too near
the edge of the \K cone. For a highly symmetric three-fold similarly
high quality results can be expected. However, moving to the general
three-fold appears tough and new techniques must certainly be
employed.

Having considered the problem of constructing K3's at arbitrary points
in moduli space, we should point out the enormity of the moduli
space itself: with 57 directions to explore it is highly implausible that one
could ever map the entire space of K3's. On the other hand it is
unlikely that one would ever need to map the entire space. One can
imagine wishing to find K3 surfaces with specific properties;
presuming the observables of interest vary smoothly over the moduli
space, it is plausible that one could scan the moduli space for
examples that fit the specific requirements.

Whilst Calabi-Yau's have large numbers of moduli, spaces with Ricci
curvature tend to have fewer of them. An example of this are the four
(real) dimensional del Pezzo surfaces dP${}_n$ ($n=1,\dots,8$),
compact \K spaces that can be constructed by blowing up $n$ points in
$\C P^2$. Del Pezzos have positive first Chern class, and those with
$n\ge 3$ admit \K-Einstein metrics with no \K moduli. The cases of
dP${}_3$ and dP${}_4$ are particularly interesting because they also
have no complex structure moduli. The Einstein equation can again be
reduced to a Monge-Amp\`ere equation for the \K potential. As in our
K3 example, two patches would be required to cover each of the $n$
blown up points, and three to cover the ambient $\C P^2$. Hence we
expect the methods we have applied here for K3 to be applicable, and
the \K-Einstein metrics to be attainable to high resolution on a
desktop computer (certainly comparable to or better than our
resolution C, depending on how many points are blown up in $\C
P^2$). This would be very satisfying for dP${}_3$ and dP${}_4$ as, with
no moduli, one would in principle have constructed these geometries
explicitly and completely.

Other physically interesting geometries that are related to Calabi-Yau
manifolds are supersymmetric flux compactifications in string
theory. A class of type IIB solutions can be constructed as warped
products of flat four-dimensional Minkowski spacetime and a Ricci-flat
Calabi-Yau three-fold, where the warp factor satisfies a Poisson
equation on the Calabi-Yau sourced by fluxes, D-branes, and
orientifold planes (see e.g.\
\cite{deWit:1986xg,Greene:2000gh,Giddings:2001yu,Frey:2003tf}). As we
discussed above, finding the metric on a generic three-fold is
probably out of reach using the methods of this paper. However,
considering fluxes on K3 or a highly symmetric three-fold would likely
be a manageable task. Having found the Ricci-flat metric, solving the
Poisson equation on this geometry is quite simple---indeed even easier
than finding eigenfunctions of the Laplacian as we did earlier in this
paper. Studying the solutions to the Poisson equation on the
Calabi-Yau background would provide a detailed understanding of how
the fluxes backreact on the vacuum geometry, and in particular of how
fluxes on adjacent cycles interact.

\subsection{Lessons for solving general Euclidean geometries}\label{euclidean}

The key simplification in our work has been that of \K geometry. What
are the prospects for constructing general Euclidean geometries?
Without {\K}ity we require many metric components to describe the
geometry, and the Einstein equation becomes complicated. Whilst
this may be technically complicated, in principal one would hope that
using a harmonic gauge condition would allow the system to be locally
solved as an elliptic relaxation problem. Note that one could also use
a gauge fixed Ricci flow, but as discussed eariler, it is more
efficient to solve the elliptic Ricci flatness condition directly
rather than to construct an entire flow when only the endpoint is
required.  In our K3 example, the complex coordinates on the
coordinate patches provide exactly such a local harmonic set of
coordinates. However the most challenging aspect of the problem is to
understand the global issues, such as residual coordinate freedom,
adaptedness of the coordinates, and moduli of the solutions.

Let us for a moment consider the problem of finding the Ricci flat
metric of symmetric K3's as we have done, but ignoring the \K
structure and using only real geometry. We might hope to find harmonic
coordinates on the various patches that make up the topology. This
would require us to solve the harmonic gauge condition (essentially
locally solving Laplace equations) at the same time as the
Einstein equation. Presumably this full system is globally elliptic in
the case of K3, although for more general geometries we should note
that negative modes of the Lichnerowicz operator will exist, as occurs
for the Euclidean Schwarzschild solution, and it is unclear how these
would affect the situation.

Given the link between the complex coordinates of \K geometry and the
harmonic coordinates natural for Euclidean real geometry, we can make
various speculations. We saw that our complex coordinates were well
adapted to the symmetric K3 geometry, and one might hope the same to
be true for more general harmonic coordinates. As explained above, for
the complex coordinates there are no residual holomorphic coordinate
transformations; in real geometry, choosing harmonic coordinates one
again expects only finitely many residual coordinate freedoms on a
compact manifold. Assuming this, one might conclude from our work that
we may see the complex structure moduli of the K3 arise simply from
the global data required to specify the harmonic coordinates, just as
we have fixed the complex structure moduli by taking particular
complex coordinates on the manifold. Then for more general compact
manifolds, one might plausibly associate physical moduli to global
choices when constructing the harmonic coordinates (that are not
simply one of the finite residual coordinate transformations).

Clearly about any Ricci flat real geometry one can always linearize
metric fluctuations and then, in principle, directly determine the
zero modes of the resulting Lichnerowicz operator, and hence determine
all physical moduli of the solution. However, this is obviously very
complicated to imagine doing in practice, and what we really wish to
find is a way to include the moduli as boundary conditions in the
elliptic problem as we have done in our \K example.  Assuming our
presumptions above about the complex structure moduli hold, the key
remaining question is how to understand the \K moduli of K3 as
boundary conditions, without actually making use of the \K structure.
It might be possible to understand this in terms of the volumes of
minimal representatives of the two-cycles. However, whilst for K3 this
approach would work, for the \K-Einstein case of the del Pezzos it
cannot, since the two-cycles present are not associated with any
moduli; hence this approach would not work in general. Thus, finding
new ways to understand the \K moduli and how to actually implement
fixing them whilst solving the real geometry Einstein equations for
K3, look to be important questions. If they can be addressed, it might
allow one to understand the moduli of general real geometries, and
enable explicit metrics to be found in very general Euclidean
geometry-matter systems.

\subsection{Applications}\label{applications}

We now briefly discuss possible applications for the numerical
construction of geometries. We consider first mathematical, then
formal physical, and finally phenomenological applications.

There are various outstanding mathematical conjectures regarding the
geometry of Calabi-Yau manifolds. Most notorious is the
Strominger-Yau-Zaslow conjecture, that Calabi-Yaus with mirrors can be
constructed as toric fibrations and that mirror symmetry acts by
T-duality on the fibers \cite{Strominger:1996it, MR1714827}. A related
conjecture concerns mean curvature flows and special Lagrangian
submanifolds in Calabi-Yau manifolds \cite{MR1957663}. In principle
these conjectures can be tested directly on any Calabi-Yau for which
the Ricci-flat metric is known explicitly.

From the point of view of physics, explicit constructions of
Ricci-flat Calabi-Yau metrics may help us learn more about the sigma
model description of geometry. The Ricci-flat metric on K3 is the
target space of an ${\cal N} = (4,4)$ non-linear sigma model. Due to
the high degree of supersymmetry the classical Ricci flat metric
receives no perturbative or non-perturbative corrections in $\alpha'$
\cite{Aspinwall:1996mn}. Thus the geometries we have constructed in
this paper can, remarkably, be viewed as fully quantum geometries from
the viewpoint of this sigma model. Knowing these metrics then in
principle allows one to compute properties of the quantum sigma
model---for example, the spectra of operators on the target manifold
correspond to conformal weights on the worldsheet.

The Ricci-flat Calabi-Yau three-folds are again target spaces of sigma
models. However, now the classical geometry only gives the leading
$\alpha'$, or large volume, approximation to the true quantum
geometry. Understanding how $\alpha'$ corrections modify the classical
geometry is an important physical issue. Supersymmetry implies that
these corrections preserve the {\K}ity of the metric, and hence they
will appear as higher derivative terms modifying the Monge-Amp\`ere
equation for the \K potential. Each higher derivative term will make
this equation less local, but in principle we may still apply the same
local iterative methods we have used here to solve it. (Presumably on
a compact manifold, one could in principle include infinitely high
derivative terms if their form were known.) From a discretized
viewpoint, if we linearized the equation about some background, we
would find an $N \times N$ operator ($N$ being the total number of
lattice points) and the structure of its matrix representation will no
longer be sparse. However, the terms that fill in the zero components
in the sparse Monge-Amp\`ere case will be small, being down by factors
of $\alpha'$. Hence our Gauss-Seidel iterative methods may still work,
although obviously evaluation of the equation at each point will take
much longer.

Finally, from a phenomenological point of view, being able to compute
metrics is crucial for actually making contact with low energy physics
in string theory. Whilst for simple vacuum Calabi-Yau reductions it is
possible to compute the entire low energy effective action using only
topological data \cite{Strominger:1985ks}, as soon as matter is added
to the compactification manifold this is no longer true. The simplest
example is adding a single brane that is localized in the compact
space. The moduli space metric for this brane, and hence the kinetic
term for its position in the low-energy action, is simply given by the
metric on the internal space \cite{douglas}.  Thus from a physics
standpoint, being able to compute the low energy action of geometric
reductions is a strong motivation to further understand and improve
the numerical geometry methods we are exploring here.

It is worth mentioning that in TeV fundamental scale senarios
\cite{Arkani-Hamed:1998rs,Antoniadis:1998ig} it is possible that in a
few years the LHC might directly probe not just the low energy action,
but also high energy excitations on the internal space, such as
Kaluza-Klein modes. Although the LHC would likely not measure a
sufficient number for any detailed spectroscopy of the internal space,
future colliders would then be able to measure these massive
excitations accurately. If one could extend our methods to the general
three-fold---and surely this would be enough motivation to direct
serious resources to it---then one might scan the moduli space of
Calabi-Yaus, presumably with fluxes, to find candidate geometries
matching the observed resonances.

\acknowledgments

We would like to thank the following people, who have given us
invaluable help in this project: A. Adams, F. Denef, J. Distler,
M. Douglas, D. Gaiotto, S. Gukov, J. Hartle, B. Julia, B. Kors,
J. Minahan, D. Morrison, L. Motl, W. Nahm, A. Neitzke, C. Nunez,
R. Reinbacher, A. Sen, J. Sparks, A. Strominger, P. Tripathy,
B. Wecht, K. Wendland, and S.-T. Yau. M.H. is supported by a
Pappalardo Fellowship, and by the U.S.\ Department of Energy through
cooperative research agreement DF-FC02-94ER40818. T.W. is supported by
the David and Lucile Packard Foundation, grant number 2000-13869A.

\appendix

\section{Details of numerical construction}\label{appnum}

\subsection{Construction of the atlas and initial data}

As discussed in section \ref{construction} our one parameter family of
K3's have many discrete symmetries. In particular these imply we may
take the \K potential to be identical in each of the 16 regions
describing the blow up of the torus fixed points. These regions are
each described in terms of 2 coordinate patches given by $w,y$ and
$w',y'$ and the symmetry $z^1 \leftrightarrow z^2$ implies that the \K
potential may be taken to be identical in each of these. Thus we
reduce our problem to one patch describing the fundamental domain of
the torus using $z^1, z^2$ coordinates, where we orbifold by the
identification $(z^1,z^2) \equiv (-z^1,-z^2)$, and one patch
describing (half of) the blow up of the fixed point contained in that
fundamental domain using coordinates $w, y$. We term these patches the
`Torus patch' and the `Eguchi-Hanson patch'. To begin describing the
geometry of our atlas we define,
\begin{equation}
\sigma = | z^1 |^2 + | z^2 |^2 = 2 | w | \left( 1
+ | y |^2 \right)
\end{equation}
We take our torus patch to cover the coordinate range of the
fundamental domain, but exclude the region near the blown up fixed
point so,
\begin{equation}
- \frac{1}{4} \le \Rl{z^{1,2}} \le \frac{1}{4} \quad , \quad - \frac{1}{4} \le \I{z^{1,2}} \le \frac{1}{4} \quad \mathrm{and} \quad \sigma \ge \sigma_{min}
\end{equation}
Outside the boundaries of this domain we must act with the
translations $z^{1,2} \rightarrow z^{1,2} \pm (1,i)/2$ to map the
point back into the domain. Note that when we do this, we must also
ensure we perform the torus \K transformation derived from
\ref{toruskt4} reduced to this domain. The Eguchi-Hanson patch is
taken to have coordinate range,
\begin{equation}
0 \le | y | \le y_{max} \quad \mathrm{and} \quad \sigma \le \sigma_{max}
\label{ehpatch}
\end{equation}
In order to cover the manifold we must ensure $y_{max} \ge 1$, and
$\sigma_{max} \ge \sigma_{min}$ with $\sigma_{max} < 1/4^2$ to avoid
complicated multiple overlaps. 

In order to evaluate our Monge-Amp\`ere equation at the edge of a
coordinate patch we must necessarily compute derivatives involving \K
potential data from neighbouring coordinate patches. Once we have
finite differenced our patches this will require the \K potential at
some point to be computed from the neighbouring patch and then \K
transformed into the original patch. The coordinate location in the
neighbouring patch will not necessarily fall on a lattice point and
therefore we will need to perform interpolation to compute the desired
\K potential \footnote{Note that whilst extrapolation is more
economical as we would require no patch overlap, it is also
potentially dangerous as it is likely to introduce spurious data, and
since specifiying the correct data is our key concern we have opted to
use interpolation which takes a little more storage (due to the
patches overlapping) but removes the risk of specifying data
incorrectly.}. Thus we require our patches to overlap sufficiently in
order to perform our necessary interpolations.

For example, suppose when we evaluate a derivative at the boundary of
the Eguchi-Hanson patch we require knowing the \K potential at a point
still with $\sigma<\sigma_{max}$ but now $| y | > y_{max}$. Then we
must use the coordinate patch $w',y'$ - which due to the discrete
symmetries is identical to the $w, y$ one. Explicitly, we transform to
the $w', y'$ coordinates where still $\sigma<\sigma_{max}$, but now $|
y' | < y_{max}$, so the point does indeed lie within this $w',y'$
patch. We find the \K potential in this patch using the necessary
interpolation, and then return to our original coordinate patch $w, y$
by performing the necessary \K transformation \ref{ehkt3}. Similarly,
at the large $\sigma$ boundary of the Eguchi-Hanson patch, or small
$\sigma$ boundary of the torus patch we will find the \K potential in
that patch by interpolating from the other and performing the
appropriate \K transformations \ref{ehkt1}, \ref{ehkt2}.

We note that in fact we actually require only one coordinate patch, as
the Eguchi-Hanson patch can quite satisfactorily represent the torus
region. However, the torus patch boundary conditions, essentially
derived from \ref{toruskt1}-\ref{toruskt4} become complicated and
non-local in the $w, y$ coordinates, and we have found it simpler to
use the two patches above, rather than the minimal choice of one
patch.

Even after the reduction to these 2 patches, and the coordinate
domains above, we still have discrete symmetries remaining. The
orbifold symmetry, holomophic isometries $z^j \rightarrow i z^j$, $z^1
\leftrightarrow z^2$ and anti-holomorphic isometry $z^{1,2}
\leftrightarrow \bar{z}^{1,2}$ further reduce the torus coordinate
domain (in addition to $\sigma \ge \sigma_{min}$) to,
\begin{eqnarray}
\label{reducetorus}
0 \le \Rl{z^{1,2}} & \le & \frac{1}{4} \quad , \quad 0 \le \I{z^{1,2}} \le \frac{1}{4} \nonumber \\
\mathrm{and}   \qquad     (z^{1},z^{2}) & \sim & (z^{2},z^{1}) \nonumber \\
                    (z^{1},z^{2}) & \sim & (\bar{z}^{1},\bar{z}^{2})
\end{eqnarray}
In the Eguchi-Hanson patch these isometries allow us to further reduce
\ref{ehpatch} to,
\begin{equation}
0 \le \Rl{w,y} \qquad 0 \le \I{w,y}
\end{equation}
The \K potential does not transform under these discrete isometries,
and therefore if we now require a point to be evaluated outside our
reduced coordinate domains we simply use these discrete symmetries to
map the point back into our reduced coordinate domain.

In the data we present we have chosen a fixed geometry for the patches
and their overlaps. This is independent of the numerical resolution of
the discretization so that for convergence testing we are comparing
like with like. As discussed in section \ref{methods} it also allows
us to directly compare the \K potential in the same overlapping
regions as we vary the resolution which provides a check of numerical
convergence. The parameters we have chosen are,
\begin{equation}
y_{max} = 1.25 \qquad \sigma_{min} = \frac{1}{4^2} \times 0.32 \quad \mathrm{and} \quad \sigma_{max} = \frac{1}{4^2} \times 0.60
\end{equation}

Before we begin relaxing the Monge-Amp\`ere equation we require some
smooth initial data compatible with our choice of \K parameters $a, b$
(although in fact we have found that in some cases even taking
non-smooth initial data the algorithm still converges). We construct
this by taking an initial guess \K potential to have the behaviour of
the Eguchi-Hanson potential near $\sigma = 0$ in the Eguchi-Hanson
patch, and which then interpolates smoothly up to second derivatives
to behave as the flat torus potential for $\sigma > \sigma_{max}$.

The Eguchi-Hanson potential in the coordinate patch $w, y$ is given
by,
\begin{equation}
K^{\mathrm{EH}}_{(y,w)} = \frac{1}{2} \sqrt{\sigma^2 + a^4} + \frac{a^2}{2} \ln \frac{\ln\left(1+|y|^2\right)}{1 + \sqrt{1 + \frac{\sigma^2}{a^4}}}
\end{equation}
and the flat torus potential is simply $K^{\mathrm{torus}} =
\frac{1}{2} \sigma b^2$. For our interpolation, in our Eguchi-Hansen
patch (with $\sigma<\sigma_{max}$) we take,
\begin{eqnarray}
K_{(y,w)} & = & \frac{1}{2} a^2 \log{(1 + |y|^2)} + k_0 + k_2 \sigma^2 + k_4 \sigma^4  \nonumber \\
\nonumber \\
\mathrm{with} \qquad k_0 & = & \frac{1}{16} \left( 6 a^2 + 3 \sigma_{max} b^2 - 8 a^2 \log{\sigma_{max}} \right) \nonumber \\
k_2 & = & \frac{1}{8 \sigma_{max}^2} \left( 3 \sigma_{max} b^2 - 4 a^2 \right) \nonumber \\
k_4 & = & - \frac{1}{16 \sigma_{max}^4} \left( \sigma_{max} b^2 - 2 a^2 \right) \nonumber \\
\end{eqnarray}
and in the torus patch with $\sigma>\sigma_{min}$ we take,
\begin{eqnarray}
\sigma \le \sigma_{max} \qquad K_{(z1,z2)} & = & \frac{1}{2} a^2 \log{\sigma} + k_0 + k_2 \sigma^2 + k_4 \sigma^4  \nonumber \\
\sigma > \sigma_{max} \qquad K_{(z1,z2)} & = & \frac{1}{2} \sigma b^2 \nonumber \\
\end{eqnarray}
The constants above ensure the \K potential has smooth second
derivatives at $\sigma = \sigma_{max}$. However, it is easy to show
that the above does not define a positive \K form over the whole range
$\alpha = 0$ to $1$. As discussed in the main text, one advantage in
solving the Monge-Amp\'ere equation directly, rather than performing
Ricci flow, is that the \K form need not be positive. Hence the simple
interpolation above suffices as initial data.

\subsection{Discretization, memory and time requirements}

We discretize our system in the most naive way. In both patches we
discretize by creating a uniform lattice in the real and imaginary
parts of each complex coordinate. We then use second order finite
differencing to implement the Monge-Amp\'ere equation and third order
accurate interpolation at the patch overlaps.

At each step of our relaxation, we firstly interpolate the values of
the \K potential at the very edges of a patch from the appropriate
neighbouring patches, and secondly perform one iteration of the
Gauss-Seidel generalizated to our Monge-Amp\`ere equation. When
updating each point, as a by-product we may quickly compute the value
of $\det{g}$ at that point. During the Gauss-Siedel update, we keep a
running average of this quantity over all lattice points, and then use
this as the value of $\tilde{\lambda}$ for the next step. This ensures
our Gauss-Seidel iterations asymptote to a fixed solution.

In this paper we have used 4 different resolutions each differing in
linear resolution by a factor of 2. In the the torus patch we
discretize with equal lattice spacing $dz$ in the real and imaginary
$z^1$ and $z^2$ directions. In the Eguchi-Hanson patch we discretize
with spacing $dw$ in the real/imaginary $w$ directions and $dy$ in the
real/imaginary $y$ directions. We label the various resolutions $A-D$,
and they are specified as,
\begin{equation} \label{resolutions}
\begin{array}{rclrclrcl}
A \qquad \qquad  dz & = & 0.025   \qquad \qquad  & dw & = & 0.003125  \qquad \qquad  & dy & = & 0.25  \\
B \qquad \qquad  dz & = & \frac{1}{2} 0.025 \qquad \qquad  & dw & = &  \frac{1}{2} 0.003125 \qquad \qquad  & dy & = & \frac{1}{2} 0.25  \\
C \qquad \qquad  dz & = & \frac{1}{2^2} 0.025  \qquad \qquad  & dw & = & \frac{1}{2^2} 0.003125  \qquad \qquad  & dy & = & \frac{1}{2^2} 0.25  \\
D \qquad \qquad  dz & = & \frac{1}{2^3} 0.025 \qquad \qquad  & dw & = & \frac{1}{2^3} 0.003125  \qquad \qquad  & dy & = &  \frac{1}{2^3} 0.25  \\
\end{array}
\end{equation}
Remembering that the coordinates $z$ range from $0\rightarrow0.25$ for
our reduced domain in the torus patch, this yields 80 points along a
side of the torus in resolution $D$.

In the torus patch we implement the 2 identifications in
\ref{reducetorus} imperfectly by simply only storing points with
$\I(z^2)>\max{\left(\Rl(z^1),\I(z^1)\right)}$. This is rather
convenient, but doesn't fully take advantage of these discrete
symmetries. Asymptotically this yields a reduction of 3 rather than
the optimal value 4 for these 2 identifications. With this minor
imperfections in mind, the total number of points stored in computer
memory for each resolution to represent the manifold is then,
\begin{equation}
\begin{array}{rcl}
\mbox{number of points} \quad A  & = & 6 \times 10^3 \\
B  & = & 7 \times 10^4 \\
C  & = & 1 \times 10^6 \\
D  & = & 2 \times 10^7 \\
\end{array}
\end{equation}

The Gauss-Seidel scheme is implemented by solving the discrete
Monge-Amp\`ere equations at each point in the lattice. We found that
under-relaxation was not required for stability and the scheme
converged stably. Interestingly we found that we could not over-relax
the equation with any appreciable over-relaxation rendering the scheme
unstable. Presumably this is an effect associated with the patch
boundaries rather than their interiors which we expect to behave in an
analogous manner to the Poisson equation. Thus in principle it might
be possible to include an over-relaxation parameter that varied over
the patch to be one at the edges, but larger than one in the interior.

We measure the distance from convergence by computing the maximum
update of the \K potential in a given Gauss-Seidel iteration over the
whole lattice. This is equivalent to computing the maximum violation
of the discretized Monge-Amp\`ere equation.

When this number falls below $10^{-12}$ we classify the solution as
having relaxed. Certainly for any quantity we have computed, there is
no further change if the solutions are subjected to further iterations
of the Gauss-Seidel scheme. The time taken for our implementation of
the algorithm from initial guess to the relaxed condition is shown in
figure \ref{fig:timing} for the various resolutions $A-D$ on a
standard desktop computer (3Ghz Pentium, 500 Mb). After a little
relaxation, as for the Poisson equation, the number of Gauss-Seidel
iterations required to improve the discretized error by a factor of 10
quickly tends to a constant. This is also shown in the same figure.

\FIGURE{
\centerline{\psfig{file=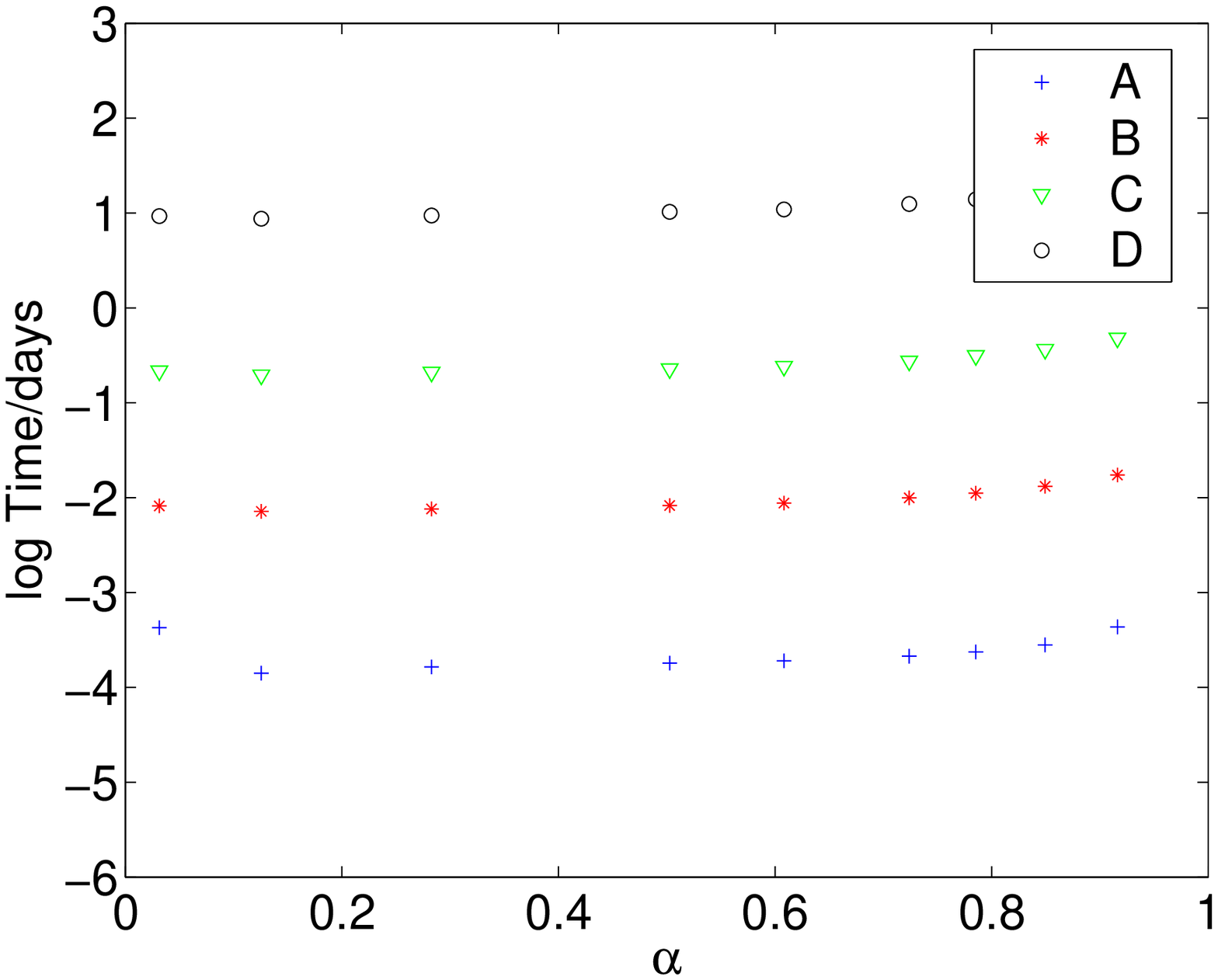,width=3.5in}\psfig{file=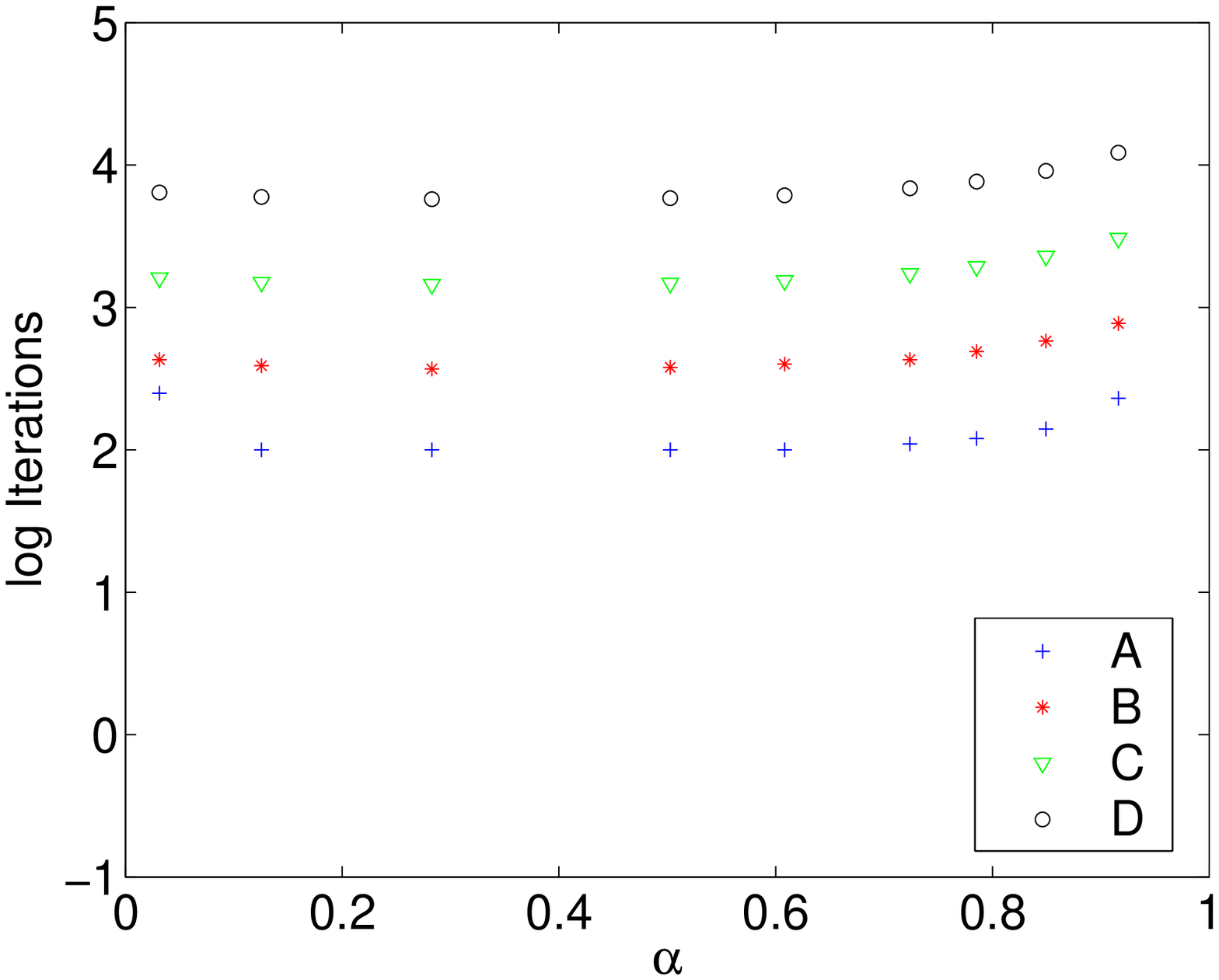,width=3.5in}}
\caption{ The lefthand plot shows the log (base 10) total time
measured in days for the various resolutions $A-D$ to converge to our
required degree. The righthand plot shows the number of Gauss-Seidel
iterations to improve the maximum error in the discretized
Monge-Amp\`ere equation by a factor of 10. 
\label{fig:timing}
}
}

We see that both these quantities exhibit only a weak dependence on
the position in moduli space that we choose. The lowest resolution $A$
relaxes in seconds while our highest resolution $D$ requires up to a
week.

The time taken to relax using the local Gauss-Seidel scheme can be
estimated as going as $N^{1+2/d}$ in $d$ real dimensions where $N$ is
the total number of lattice points. Every iteration takes a time of
order $N$, and the total number of steps can be estimated as $N^{2/d}$
by considering the spectral radius of the linearized Laplace operator.
We see this scaling is certainly consistent with the results of figure
\ref{fig:timing}.

At every step in the iteration we must also perform an interpolation
of points at the edge of each coordinate patch from its neighbours. We
only require the edge points to be interpolated that are required by
our second order differenced Monge-Amp\`ere equation. Thus the work
involved scales as a codimension one quantity, namely as
$N^{(d-1)/d}$. In practice our third order interpolation actually
takes considerable time. Asymptotically at large $N$, however, it will
obviously become subdominant to the Gauss-Seidel iteration time which
scales as $N$.

It is very interesting to note that as we move to higher dimensions,
the total relaxation time more and more closely approaches $N$. Thus
the advantage in using highly non-local schemes such as multi-grid to
improve convergence times (typically to $N \log{N}$) becomes
considerably reduced. In this sense we may claim that the problem of
extending our methods to Calabi-Yau 3-folds is storage limited rather
than speed limited.

In a memory limited problem, it is rather natural to move to higher
order methods. Our implementation uses second order finite
differencing on the Monge-Amp\'ere equation. However, we might hope
for improved convergence to the continuum if we were to use $4^{th}$
order discretization of the Monge-Amp\`ere equation (and also for
interpolating between patches). We did try this in our case of K3, but
the additional time required by each iteration slowed the total
convergence time to approximately the same as the next higher
resolution using second order differencing. In order to procede to our
highest resolution we decided to stay with second order
differencing. Bear in mind that whilst a higher order method
approaches the continuum more quickly, in order to resolve short
length scales one requires high resolutions, so to get accurate
results near the orbifold regions of our moduli space, we require the
highest resolutions possible.

With a particular physics or maths question in mind, one might attempt
computations using more resources and improved stamina than we have
used here. Then the fundamental problem of limited storage should
probably be tackled by both a combination of more efficient
discretization, and also higher order methods. The increased cost in
processor time could certainly be ameliorated by some form of
parallelisation - which is well suited to this problem which,
afterall, naturally divides into coordinate patches.

The eigenfunction of the scalar Laplace operator presented in section
\ref{results} was computed using a naive iterative scheme. The initial
guess for the eigenfunction $\psi_{(\alpha)}$ with odd parity under
the $Z_2^4$ translation isometry ($z^{i} \rightarrow z^{i} \pm 1/2$)
was taken to be that for the torus orbifold, $\alpha = 0$, where,
\begin{equation}
\psi_{(0)} = \cos{2 \pi \Rl z^1} \cos{2 \pi \I z^1} \cos{2 \pi \Rl z^2} \cos{2 \pi \I z^2}
\end{equation}
giving an eigenvalue $4 (2 \pi/b)^4$. The eigenvalue equation was
solved using Gauss-Seidel iteration everywhere but at one point, with
the eigenvalue being determined dynamically from the condition that
the eigenfunction be smooth at that point. Of course, explicit
independence of the actual point chosen was checked. This method is
very simple to implement, but is really only suited to finding the
lowest eigenfunction in a parity sector. More sophisticated (but
standard) methods would be required to computer higher eigenfunctions.

\subsection{Convergence tests} \label{continuum}

We now breifly present data that demonstrates increasing numerical
resolution improves observed quantities in a manner consistent with a
second order approach to the continuum.

Firstly as discussed in the main text, we compute a numerical
$\tilde{\lambda}$ rather than using the analytic value $\lambda$. This
ensures the numerical Monge-Amp\`ere equations converges to a static
end point. We may then compare this numerical determination of
$\lambda$ with the true analytic value. In figure \ref{fig:detg} we
plot the difference between these values for the various resolutions
as a function of position in moduli space. We see the error is
greatest near the orbifold point at $\alpha = 1$ as expected. We
clearly see that the errors improve with increasing resolution
consistent with second order scaling - we remind the reader that the 4
resolutions $A-D$ each differ by a linear resolution factor of 2.

\FIGURE{
\centerline{\psfig{file=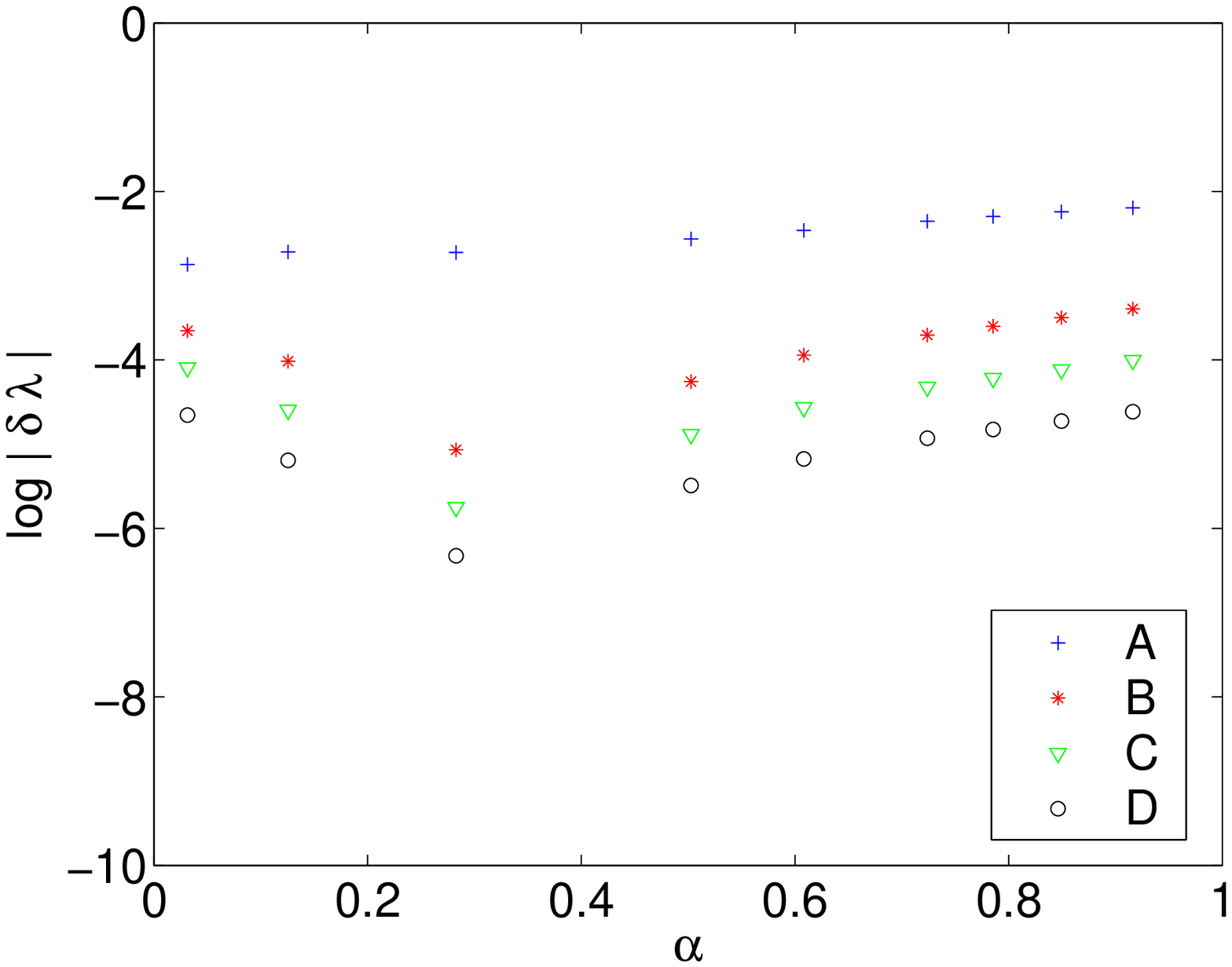,width=3.5in}}
\caption{ Plot of the log (base 10) absolute difference between the
numerically determined $\tilde{\lambda}$ and the true analytic value
$\lambda$ for the various resolutions $A-D$. 
\label{fig:detg}
}
}

In figure \ref{fig:euler} we plot the integrated Euler density for the
resolutions $B-D$. The result, the Euler number, has true value 24. We
clearly see here that increasing resolution does indeed improve the
numerical determination of this quantity as we would hope for. The
lowest resolution gives a very poor estimate of the Euler number and
we have not included it here. Resolution $B$ still provides a rather
poor estimate. We see the error in the Euler number is practically
quite small for resolution $C$ provided we are not too near either
orbifold point, and resolution $D$ gives errors of less than $0.1\%$
over most of the moduli space.

\FIGURE{
\centerline{\psfig{file=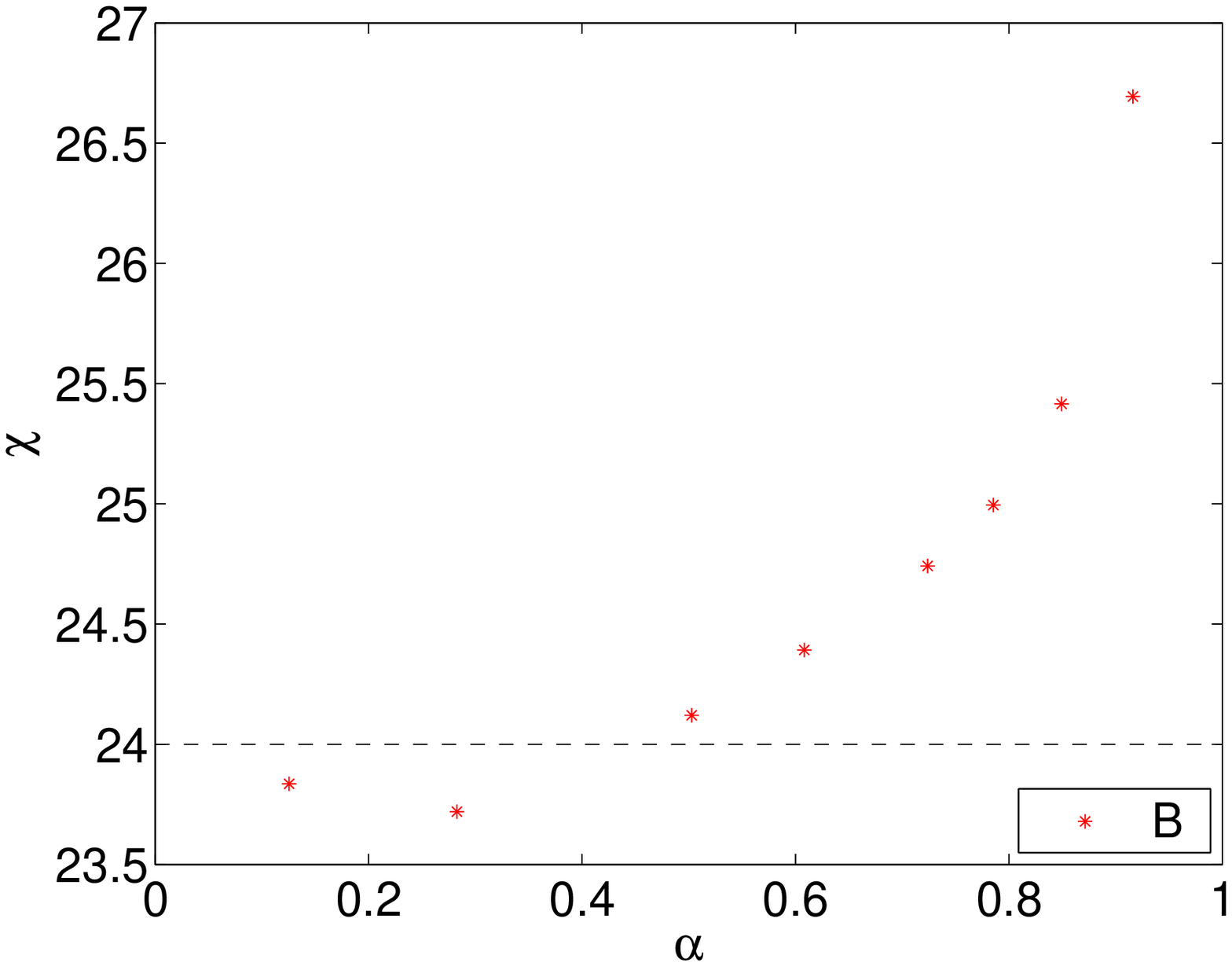,width=3.5in}\psfig{file=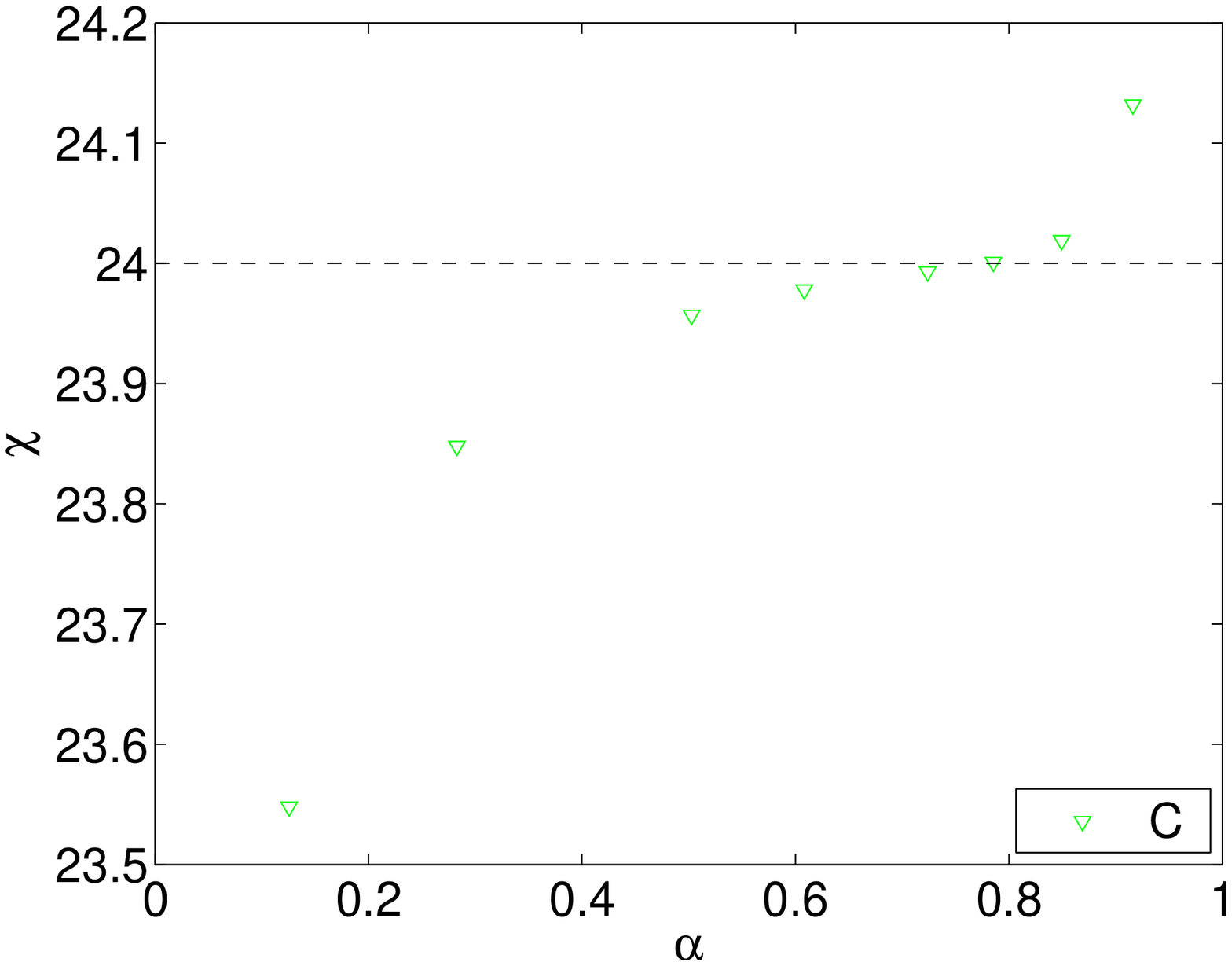,width=3.5in}}
\centerline{\psfig{file=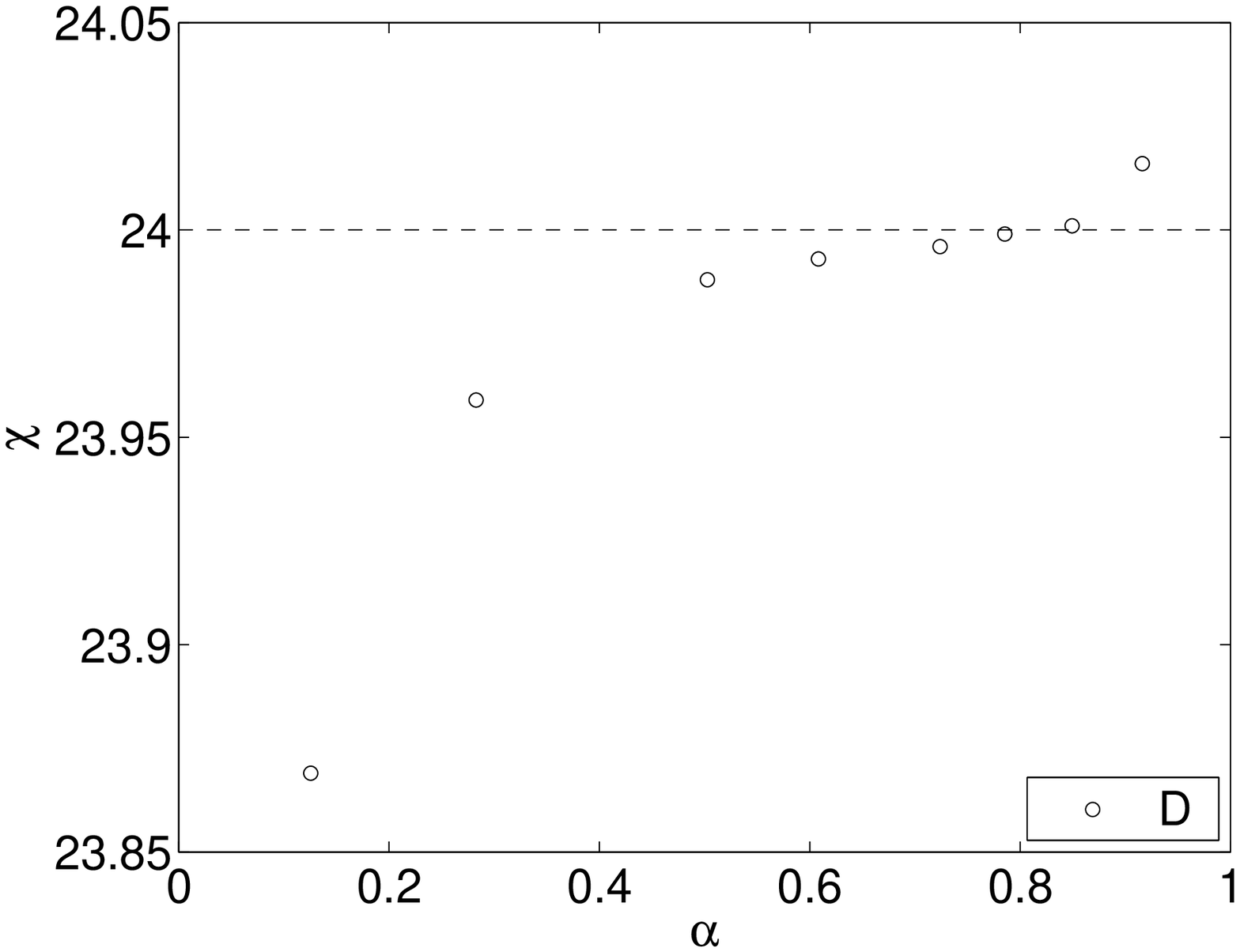,width=3.5in}}
\caption{ 
The Euler number integrated from the Euler density for the
resolutions $B$, $C$, $D$.
\label{fig:euler}
}
}

Finally our patches geometrically overlap in coordinate
space. Therefore as discussed in the main text, once we have found a
solution, we can test its quality by computing the error in the \K
potential in the overlapping regions (obviously taking into account
the relevent \K transformation between the 2 patches). In figure
\ref{fig:overlap} we show the maximum error found by comparing 2
different patch overlaps; the overlap of the Eguchi-Hanson patch with
itself, and then its overlap with the torus patch. We see that this
maximum error again decreases consistent with second order scaling as
resolution is increased.

\FIGURE{
\centerline{\psfig{file=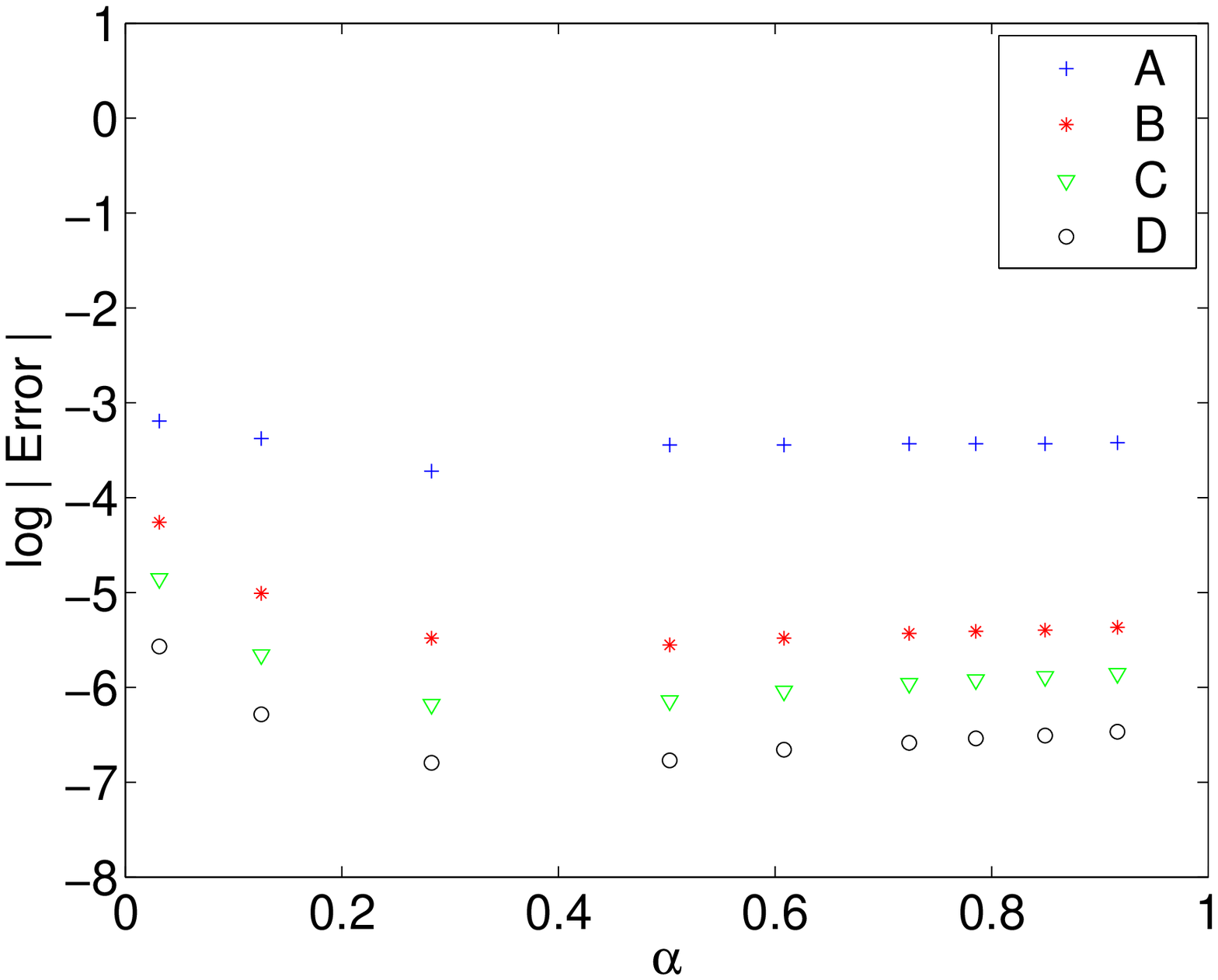,width=3.5in}\psfig{file=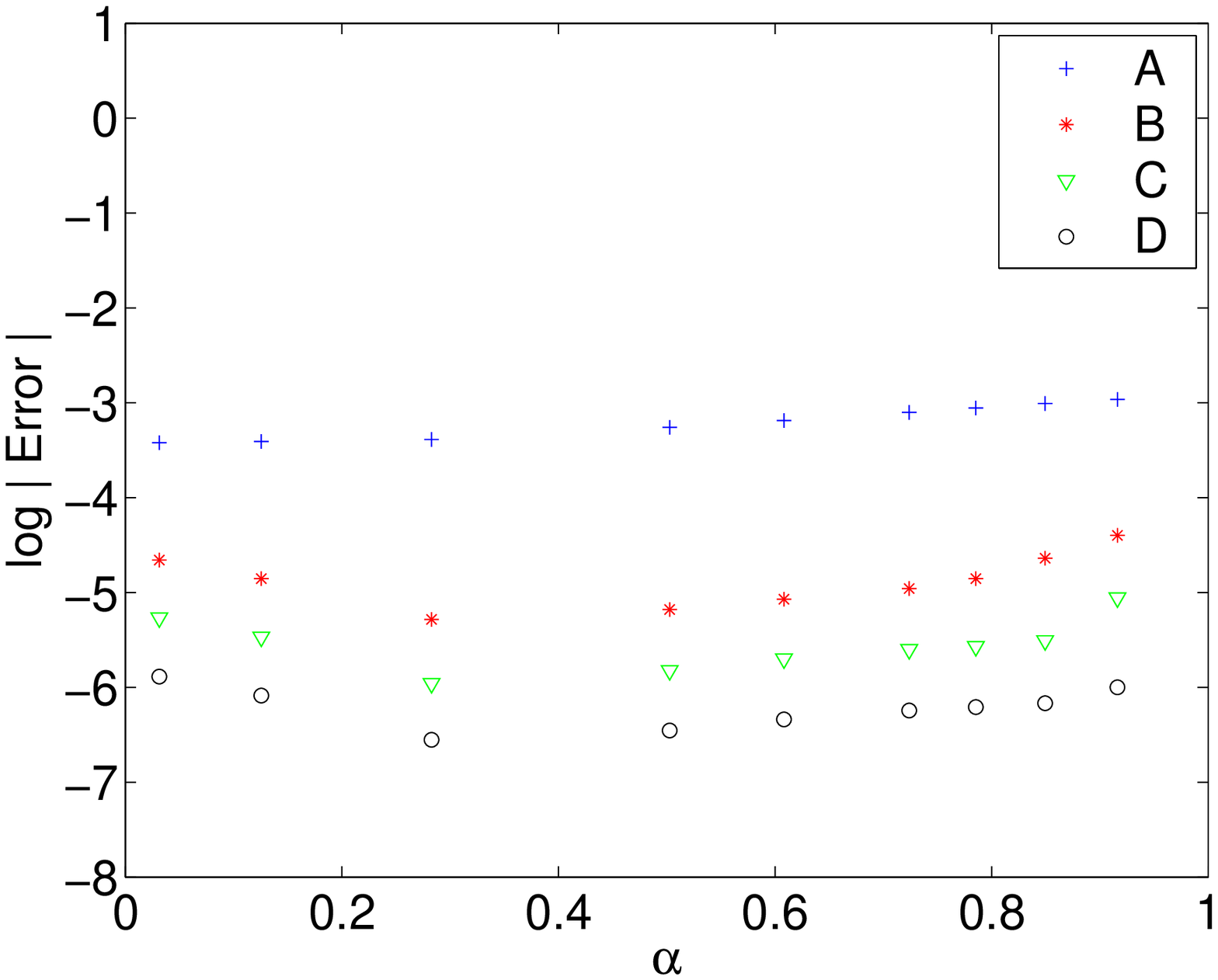,width=3.5in}}
\caption{ Lefthand plot shows the log (base 10) absolute maximum error
between \K potentials on the overlap of the Eguchi-Hanson patch with
itself. The righthand plot shows the maximum error on the overlap
between the Eguchi-Hanson and torus patches.  
\label{fig:overlap}
}
}

\section{Homology of Kummer surfaces}\label{homology}

In this appendix we derive some properties of the second homology
group of Kummer surfaces that were used in the main text. We also show
how to relate the natural basis for the homology to the standard basis
for the integral homology of K3. As far as we know this explicit
relation is new.

K3 has second Betti number $b_2=22$. In the Kummer construction, 6
two-cycles are inherited from $T^4$, and the other 16 are the
exceptional divisors at the blown up fixed points. The second homology
of $T^4$ is generated by the 2 holomorphic curves $\{z^1=C\}$ and
$\{z^2=C\}$, as well as the 4 non-holomorphic curves $\{\Rl
z^1=C^1,\Rl z^2=C^2\}$, $\{\Rl z^1=C^1,\I z^2=C^2\}$, $\{\I
z^1=C^1,\Rl z^2=C^2\}$, and $\{\I z^1=C^1,\I z^2=C^2\}$. When taking
the orbifold, one may choose the constant(s) in a such a way that the
curve either passes through or avoids the fixed points (e.g.\ the
holomorphic curves pass through four fixed points if $C=0$ but avoids
them if $C=1/4$). In the latter case, one must include the image under
the orbifold, e.g.\ $\{z^1=C\}\cup\{z^1=-C\}$, to obtain a two-cycle
in the orbifold. (The cycles that pass through the fixed points are
linear combinations of those that don't and the exceptional divisors.)
For our purposes, it is simplest to take as a basis the 6 cycles that
avoid the fixed points, which we will refer to as ``torus cycles",
along with the 16 exceptional divisors. We refer to these cycles as
$c_I$, where $c_{1,2}$ are the holomorphic torus cycles, $c_{3,4,5,6}$
are the non-holomorphic torus cycles, and $c_{7,\dots,22}$ are the
exceptional divisors (which are holomorphic).

It is straightforward to write down the intersection matrix in this
basis. The torus cycles intersect each other in pairs, with
intersection number 2. For example $c_1$ intersects $c_2$ at two
points, $(C,C)$, and $(C,-C)$ (of course we don't count $(-C,-C)$ and
$(-C,C)$ separately). By construction, none of the torus cycles
intersect the exceptional divisors. The latter do not intersect each
other, but have self-intersection $-2$ (since they are topologically
$\C P^1$'s). All in all, we find the following block diagonal
intersection matrix:
\begin{equation}
h_{IJ} = \#(c_I,c_J) = 2
\begin{bmatrix} U & & & \\ & U & & \\ & & U & \\ & & & -I_{16} & \end{bmatrix}, \qquad
U=\begin{bmatrix} 0 & 1 \\ 1 & 0 \end{bmatrix}.
\end{equation}

From the \K transformations (\ref{toruskt1}--\ref{ehkt3}) the periods
of the \K form $j_I = \int_{c_I}J$ may be computed for the symmetric
Kummer surfaces:
\begin{equation}
j_I = \begin{cases} b^2, & I=1,2 \\ 0, & I=3,4,5,6 \\ \pi a^2, & I = 7,\dots,22 \end{cases}.
\end{equation}
In the case of the holomorphic curves ($I=1,2,7,\dots,22$) these
periods are their areas. In terms of the periods the volume is
\begin{equation}
V = \frac12\int J\wedge J = \frac12h^{IJ}j_Ij_J = \frac12b^4-4\pi^2a^4,
\end{equation}
where $h^{IJ}$ is the inverse intersection matrix,
\begin{equation}\label{inverseim}
h^{IJ} = \frac12
\begin{bmatrix} U & & & \\ & U & & \\ & & U & \\ & & & -I_{16} & \end{bmatrix}.
\end{equation}

Consider now the holomorphic curve $\hat{c}_1=\{z^1=0\}$. This
intersects $c_2$ at one point, $(0,C)$, but none of the other torus
cycles. It also intersects, at one point each, 4 of the exceptional divisors, namely
those located at $(0,0)$, $(0,\frac12)$, $(0,\frac i2)$,
$(0,\frac12+\frac i2)$, which we will call $c_{7,8,9,10}$. Using the
intersection matrix, we have
\begin{equation}
\hat{c}_1 = \frac12\left(c_1 - c_7-c_8-c_9-c_{10}\right).
\end{equation}
Hence the area of this curve is
\begin{equation}
\hat A = \int_{\hat c_1} J = \frac12\left(j_1-j_7-j_8-j_9-j_{10}\right) = \frac12b^2-2\pi a^2,
\end{equation}
as claimed in subsection \ref{construction}.

Now let us return to the general Kummer surface. The two-cycles $c_I$
are obviously integral. However, since the inverse intersection matrix
\eqref{inverseim} is not composed of integers, they do not form a
basis for the integral homology $H_2(\text{K3},\Z)$, but only a sublattice
of it. The standard basis $\{c'_M\}$ for $H_2(\text{K3},\Z)$ is
defined to have intersection matrix
\begin{equation}\label{k3im}
h'_{MN} = \#(c'_M,c'_N)=
\begin{bmatrix}
U &  & & & \\
 & U &  & & \\
 &  & U & & \\
 & & & -E_8 & \\
 & & & & -E_8
\end{bmatrix},
\end{equation}
where $E_8$ represents the Cartan matrix of that group,
\begin{equation}
-E_8 =
\begin{bmatrix}
-2 & 1  &     &    &     &     &     & \\
1  & -2 &  1 &    &     &     &     & \\
    & 1  & -2 & 1 &     &     &     & \\
    &     & 1 & -2 &  1 &     &     & \\
    &     &    &  1 & -2 &  1 &  1 &  \\
    &     &    &     &  1 & -2 &    & \\
    &     &    &     &  1 &    & -2 & 1 \\
    &     &    &     &     &    &  1 & -2
\end{bmatrix}.
\end{equation}
The change of basis relating the $\{c_I\}$ to the $\{c'_M\}$ is as follows:
\begin{equation}
c'_M = {M_M}^Ic_I,
\end{equation}
where
\begin{multline}\label{basischange}
2{M_M}^I = \\
\left[
\begin{array}{cccccccccccccccccccccc}
 40 & 20 & 11 & 2 & -14 & 0 & -13 & 4 & -8 & -1 & -8 & 1 & 11 & -20 & -20 & -20 & 2 & 0 & 0 & 0 & 0 & 2 \\
 46 & 23 & 11 & 2 & -15 & 0 & -14 & 4 & -10 & 0 & -9 & 1 & 13 & -23 & -22 & -24 & 2 & 0 & 0 & 0 & 0 & 2 \\
 8 & 4 & 0 & 0 & 0 & 0 & -2 & 0 & -2 & 0 & -2 & 0 & 2 & -4 & -4 & -4 & 0 & 0 & 0 & 0 & 0 & 0 \\
 6 & 3 & -1 & -1 & 0 & 0 & -1 & 0 & -1 & 1 & -1 & -1 & 2 & -3 & -3 & -3 & -1 & 0 & 1 & 0 & 0 & 0 \\
 0 & 0 & 0 & 0 & 2 & 0 & 0 & 0 & 0 & 0 & 0 & 0 & 0 & 0 & 0 & 0 & 0 & 0 & 0 & 0 & 0 & 0 \\
 -2 & -1 & 1 & 1 & 1 & 1 & 1 & 0 & 0 & -1 & 0 & 0 & -1 & 1 & 1 & 1 & 0 & 0 & -1 & 0 & 0 & 1 \\
 -4 & -1 & 0 & 0 & 1 & 0 & 1 & 1 & 0 & 0 & 0 & 0 & -1 & 1 & 2 & 2 & 0 & 0 & 0 & 0 & 0 & 0 \\
 0 & -1 & 1 & 0 & -1 & 0 & 0 & 0 & 0 & 0 & 1 & 1 & -1 & 1 & 0 & 0 & 0 & 0 & 0 & 0 & 0 & 0 \\
 4 & 2 & -1 & 0 & 0 & 0 & -1 & 0 & 0 & 1 & -2 & -1 & 1 & -2 & -2 & -2 & 0 & 0 & 0 & 0 & 0 & 0 \\
 2 & 1 & 1 & 0 & 1 & 0 & 0 & 0 & -1 & -1 & 0 & 0 & 1 & -1 & -1 & -1 & 0 & 0 & 0 & 0 & -1 & 1 \\
 0 & 0 & 0 & 0 & 0 & 0 & 0 & 0 & 0 & 0 & 0 & 0 & 0 & 0 & 0 & 0 & 0 & 0 & 0 & 0 & 2 & 0 \\
 14 & 7 & 3 & 0 & -5 & 0 & -4 & 1 & -3 & 0 & -3 & 0 & 4 & -7 & -7 & -7 & 0 & 0 & 0 & 0 & -1 & -1 \\
 -22 & -11 & -5 & 0 & 5 & 0 & 6 & -1 & 5 & 0 & 5 & 0 & -6 & 11 & 11 & 11 & 0 & 0 & 0 & 0 & -1 & -1 \\
 -7 & -3 & -2 & -1 & 3 & 0 & 2 & -2 & 1 & 1 & 1 & 0 & -2 & 3 & 3 & 4 & -1 & 0 & 0 & 0 & 0 & 0 \\
 2 & 1 & 0 & 0 & -1 & 0 & 0 & 0 & 0 & 0 & -1 & 1 & 1 & -1 & -1 & -1 & 0 & 0 & 1 & -1 & 0 & 0 \\
 0 & 0 & 0 & 0 & 0 & 0 & 0 & 0 & 0 & 0 & 0 & 0 & 0 & 0 & 0 & 0 & 0 & 0 & 0 & 2 & 0 & 0 \\
 -4 & -2 & 1 & 0 & 0 & 0 & 1 & 0 & 1 & 0 & 1 & 0 & -1 & 2 & 2 & 2 & 1 & -1 & -1 & -1 & 0 & 0 \\
 0 & 0 & 0 & 0 & 0 & 0 & 0 & 0 & 0 & 0 & 0 & 0 & 0 & 0 & 0 & 0 & 0 & 2 & 0 & 0 & 0 & 0 \\
 2 & 1 & -1 & 0 & 1 & 0 & 0 & 0 & 0 & 0 & 0 & 0 & 0 & -2 & -1 & -1 & -1 & -1 & 0 & 0 & 0 & 0 \\
 -2 & 0 & 0 & 0 & 0 & 0 & 0 & 0 & 0 & 0 & 0 & 0 & 0 & 2 & 0 & 0 & 0 & 0 & 0 & 0 & 0 & 0 \\
 -2 & -2 & 0 & 0 & 0 & 0 & 0 & 0 & 0 & 0 & 0 & 0 & 0 & 2 & 2 & 2 & 0 & 0 & 0 & 0 & 0 & 0 \\
 -15 & -7 & -4 & -1 & 5 & 0 & 5 & -2 & 3 & 0 & 3 & -1 & -5 & 7 & 7 & 7 & 0 & 0 & 1 & 0 & 0 & -1
\end{array}
\right].
\end{multline}
(Thus one has $h'_{MN}={M_M}^I{M_N}^Jh_{IJ}$.) In deriving this change of basis we made use of results in \cite{Nahm:1999ps}. To our knowledge its explicit form is new.

\bibliography{ref} 

\providecommand{\href}[2]{#2}\begingroup\raggedright\begin{thebibliography}{10}

\bibitem{Yau:1977ms}
S.-T. Yau, {\it Calabi's conjecture and some new results in algebraic
  geometry},  {\em Proc. Nat. Acad. Sci.} {\bf 74} (1977) 1798--1799.

\bibitem{Calabi}
E.~Calabi, {\it On {K}\"ahler manifolds with vanishing canonical class},  in
  {\em Algebraic geometry and topology. A symposium in honor of S. Lefschetz},
  pp.~78--89.
\newblock Princeton University Press, Princeton, N. J., 1957.

\bibitem{Candelas:1985en}
P.~Candelas, G.~T. Horowitz, A.~Strominger, and E.~Witten, {\it Vacuum
  configurations for superstrings},  {\em Nucl. Phys.} {\bf B258} (1985)
  46--74.

\bibitem{Morrison}
D.~Morrison.
\newblock private communication (2004).

\bibitem{Cao}
H.~D. Cao, {\it Deformation of {K}\"ahler metrics to {K}\"ahler-{E}instein
  metrics on compact {K}\"ahler manifolds},  {\em Invent. Math.} {\bf 81}
  (1985), no.~2 359--372.

\bibitem{Candelas:1987is}
P.~Candelas, {\it Lectures on complex manifolds}, . IN *TRIESTE 1987,
  PROCEEDINGS, SUPERSTRINGS '87* 1-88.

\bibitem{MR2061425}
B.~Chow and D.~Knopf, {\em The {R}icci flow: an introduction}, vol.~110 of {\em
  Mathematical Surveys and Monographs}.
\newblock American Mathematical Society, Providence, RI, 2004.

\bibitem{MR2112626}
S.-C. Chang, B.~Chow, and S.-C. Chu, eds., {\em Geometric evolution equations},
  vol.~367 of {\em Contemporary Mathematics}, (Providence, RI), American
  Mathematical Society, 2005.

\bibitem{Garfinkle:2003an}
D.~Garfinkle and J.~Isenberg, {\it Critical behavior in {R}icci flow},
  \href{http://xxx.lanl.gov/abs/math.dg/0306129}{{\tt math.dg/0306129}}.

\bibitem{Rubinstein}
J.~H. Rubinstein and R.~Sinclair, {\it Visualizing {R}icci flow of manifolds of
  revolution},  \href{http://xxx.lanl.gov/abs/math.dg/0406189}{{\tt
  math.dg/0406189}}.

\bibitem{MR2115754}
D.~Garfinkle and J.~Isenberg, {\it Numerical studies of the behavior of {R}icci
  flow},  in {\em Geometric evolution equations}, vol.~367 of {\em Contemp.
  Math.}, pp.~103--114.
\newblock Amer. Math. Soc., Providence, RI, 2005.

\bibitem{Aspinwall:1996mn}
P.~S. Aspinwall, {\it K3 surfaces and string duality},
  \href{http://xxx.lanl.gov/abs/hep-th/9611137}{{\tt hep-th/9611137}}.

\bibitem{Nahm:1999ps}
W.~Nahm and K.~Wendland, {\it A hiker's guide to {K}3: Aspects of {N} = (4,4)
  superconformal field theory with central charge c = 6},  {\em Commun. Math.
  Phys.} {\bf 216} (2001) 85--138,
  [\href{http://xxx.lanl.gov/abs/hep-th/9912067}{{\tt hep-th/9912067}}].

\bibitem{Inose}
H.~Inose, {\it On certain {K}ummer surfaces which can be realized as
  non-singular quartic surfaces in {$P\sp{3}$}},  {\em J. Fac. Sci. Univ. Tokyo
  Sect. IA Math.} {\bf 23} (1976), no.~3 545--560.

\bibitem{Wendland}
K.~Wendland.
\newblock private communication (2005).

\bibitem{Wendland:2003ma}
K.~Wendland, {\it On superconformal field theories associated to very
  attractive quartics},  \href{http://xxx.lanl.gov/abs/hep-th/0307066}{{\tt
  hep-th/0307066}}.

\bibitem{Gibbons:1979xn}
G.~W. Gibbons and C.~N. Pope, {\it The positive action conjecture and
  asymptotically euclidean metrics in quantum gravity},  {\em Commun. Math.
  Phys.} {\bf 66} (1979) 267--290.

\bibitem{Bozhkov}
Y.~D. Bozhkov, {\it A construction of almost anti-self-dual connections on
  {K}ummer surfaces},  {\em Serdica} {\bf 14} (1988), no.~3 283--290.

\bibitem{deWit:1986xg}
B.~de~Wit, D.~J. Smit, and N.~D. Hari~Dass, {\it Residual supersymmetry of
  compactified d = 10 supergravity},  {\em Nucl. Phys.} {\bf B283} (1987) 165.

\bibitem{Greene:2000gh}
B.~R. Greene, K.~Schalm, and G.~Shiu, {\it Warped compactifications in {M} and
  {F} theory},  {\em Nucl. Phys.} {\bf B584} (2000) 480--508,
  [\href{http://xxx.lanl.gov/abs/hep-th/0004103}{{\tt hep-th/0004103}}].

\bibitem{Giddings:2001yu}
S.~B. Giddings, S.~Kachru, and J.~Polchinski, {\it Hierarchies from fluxes in
  string compactifications},  {\em Phys. Rev.} {\bf D66} (2002) 106006,
  [\href{http://xxx.lanl.gov/abs/hep-th/0105097}{{\tt hep-th/0105097}}].

\bibitem{Frey:2003tf}
A.~R. Frey, {\it Warped strings: Self-dual flux and contemporary
  compactifications},  \href{http://xxx.lanl.gov/abs/hep-th/0308156}{{\tt
  hep-th/0308156}}.

\bibitem{Strominger:1996it}
A.~Strominger, S.-T. Yau, and E.~Zaslow, {\it Mirror symmetry is {T}-duality},
  {\em Nucl. Phys.} {\bf B479} (1996) 243--259,
  [\href{http://xxx.lanl.gov/abs/hep-th/9606040}{{\tt hep-th/9606040}}].

\bibitem{MR1714827}
D.~R. Morrison, {\it The geometry underlying mirror symmetry},  in {\em New
  trends in algebraic geometry (Warwick, 1996)}, vol.~264 of {\em London Math.
  Soc. Lecture Note Ser.}, pp.~283--310.
\newblock Cambridge Univ. Press, Cambridge, 1999.

\bibitem{MR1957663}
R.~P. Thomas and S.-T. Yau, {\it Special {L}agrangians, stable bundles and mean
  curvature flow},  {\em Comm. Anal. Geom.} {\bf 10} (2002), no.~5 1075--1113.

\bibitem{Strominger:1985ks}
A.~Strominger, {\it Yukawa couplings in superstring compactification},  {\em
  Phys. Rev. Lett.} {\bf 55} (1985) 2547.

\bibitem{douglas}
M.~R. Douglas, {\it Two lectures on d-geometry and noncommutative geometry},
  \href{http://xxx.lanl.gov/abs/hep-th/9901146}{{\tt hep-th/9901146}}.

\bibitem{Arkani-Hamed:1998rs}
N.~Arkani-Hamed, S.~Dimopoulos, and G.~R. Dvali, {\it The hierarchy problem and
  new dimensions at a millimeter},  {\em Phys. Lett.} {\bf B429} (1998)
  263--272, [\href{http://xxx.lanl.gov/abs/hep-ph/9803315}{{\tt
  hep-ph/9803315}}].

\bibitem{Antoniadis:1998ig}
I.~Antoniadis, N.~Arkani-Hamed, S.~Dimopoulos, and G.~R. Dvali, {\it New
  dimensions at a millimeter to a fermi and superstrings at a {TeV}},  {\em
  Phys. Lett.} {\bf B436} (1998) 257--263,
  [\href{http://xxx.lanl.gov/abs/hep-ph/9804398}{{\tt hep-ph/9804398}}].

\end{thebibliography}\endgroup
\bibliographystyle{JHEP}

\end{document}